\documentclass[a4paper,fleqn,usenatbib]{mnras}

\usepackage{newtxtext,newtxmath}
\usepackage{natbib}
\usepackage[usenames]{color}
\usepackage{multirow}

\usepackage[T1]{fontenc}
\usepackage{ae,aecompl}

\usepackage{graphicx}	
\usepackage{amsmath}	
\usepackage{amssymb}	

\newcommand{\HI}{\ion{H}{i}}

\newcommand{\kms}{\mbox{km~s$^{-1}$}}
\newcommand{\Msol}{\mbox{M$_\odot$}}
\newcommand{\surm}{\mbox{M$_\odot$ pc$^{-2}$}}

\newcommand{\siggas}{\mbox{$\Sigma_{\rm gas}$}}
\newcommand{\sigsfr}{\mbox{$\Sigma_{\rm SFR}$}}
\newcommand{\sigstar}{\mbox{$\Sigma_{*}$}}
\newcommand{\sighi}{\mbox{$\Sigma_{\rm HI}$}}
\newcommand{\sightwo}{\mbox{$\Sigma_{\rm H_2}$}}

\newcommand{\coj}{\mbox{$^{12}$CO ($J=1\rightarrow0$)}}
\newcommand{\um}{\mbox{$\micron$}}
\newcommand{\ac}{\mbox{$\arcsec$}}
\newcommand{\Htwo}{\mbox{H$_2$}}

\title[Volumetric Star Formation Prescriptions]{Volumetric Star Formation Prescriptions in Vertically Resolved Edge-on Galaxies}

\author[K. Yim et al. ]{
Kijeong Yim,$^{1}$\thanks{E-mail: kyim@kasi.re.kr or kijeong.yim@gmail.com}
Tony Wong,$^{2}$
Richard J. Rand,$^{3}$
Eva Schinnerer$^{4}$
\\
$^{1}$Korea Astronomy and Space Science Institute, 776 Daedeok-daero, Yuseong-gu, Daejeon 34055, Korea\\
$^{2}$Department of Astronomy, University of Illinois, 1002 West Green Street, Urbana, IL 61801, USA\\
$^{3}$Department of Physics and Astronomy, University of New Mexico, 1919 Lomas Blvd NE, Albuquerque, NM 87131-1156, USA\\
$^{4}$Max Planck Institute f\"{u}r Astronomie, K\"{o}nigstuhl 17, D-69117, Heidelberg, Germany\\
}

\date{Accepted 2020 April 03. Received 2020 April 02; in original form 2019 January 22}

\pubyear{2020}

\begin{document}
\label{firstpage}
\pagerange{\pageref{firstpage}--\pageref{lastpage}}
\maketitle

\begin{abstract}
\label{abs}
We measure the gas disc thicknesses of the edge-on galaxy NGC 4013 and the less edge-on galaxies (NGC 4157 and 5907) using CO (CARMA/OVRO) and/or \HI\ (EVLA) observations. 
We  also estimate the scale heights of stars and/or the star formation rate (SFR) for our sample of five galaxies using $Spitzer$ IR data (3.6 \um\ and 24 \um). We derive the average volume densities of the gas and the SFR using the measured scale heights along with radial surface density profiles. Using the volume density that is more physically relevant to the SFR than the surface density, we investigate the existence of a volumetric star formation law (SFL), how the volumetric SFL is different from the surface-density SFL, and how the gas pressure regulates the SFR based on our galaxy sample. 
We find that the volumetric and surface SFLs in terms of the total gas have significantly different slopes, while the volumetric and surface SFLs in terms of the molecular gas do not show any noticeable difference.  The volumetric SFL for the total gas has a flatter power-law slope of 1.26 with a smaller scatter of 0.19 dex compared to the slope (2.05) and the scatter (0.25 dex) of the surface SFL. The molecular gas SFLs have similar slopes of 0.78 (volume density) and 0.77 (surface density) with the same rms scatter.
We show that the interstellar gas pressure is strongly correlated with the SFR but find no significant difference between the correlations based on the volume and surface densities. 
\end{abstract}




\section{Introduction}
\label{intro}
Stars and the interstellar medium (ISM) are essential ingredients of galaxies, which significantly affect the morphology and dynamics of galaxies. For that reason, it is very important to probe what regulates the star formation rate (SFR) in galaxies to better understand galaxy formation and evolution. 
It has been a long time since \citet{1959ApJ...129..243S} theorized a power-law relationship between the SFR and the gas volume densities by comparing the distribution of \HI\ and young stars in the Milky Way. Afterward, Kennicutt (\citeyear{1989ApJ...344..685K}, \citeyear{1998ApJ...498..541K}) supported the relationship observationally by demonstrating a power-law correlation between the SFR and the gas surface densities integrated over the discs of normal and starburst galaxies, so the power-law correlation is referred to as the Kennicutt-Schmidt (K-S) law or star formation law (SFL). Following their works, a number of studies (e.g.,\citealt{2002ApJ...569..157W};  \citealt{2008AJ....136.2782L};  \citealt{2008AJ....136.2846B}) examined the power-law correlations between \sigsfr\ and \sightwo, \sighi, or \siggas\ (= \sightwo\ + \sighi) using nearby galaxies and they showed  that the SFR is strongly correlated with the molecular gas while the SFR is not closely related with the atomic gas (see also \citealt{2011AJ....142...37S}). In terms of \sigsfr\ versus \siggas, \citet{2008AJ....136.2846B} showed that the relationship between the SFR and the total gas is not clear in the \HI\  dominant environments and concluded that the relationship between \sigsfr\ and \siggas\ is not a universal Schmidt law. 
One may notice that the relationships by the  previous works are based on surface densities unlike the relationship based on volume densities by \citet{1959ApJ...129..243S}.
 Most studies use the mass surface density instead of the volume density that is more directly and physically relevant to the SFR, since it is hard to measure the disc thickness which is involved in the volume density. \citet{2012ApJ...745...69K} supported the importance of the volumetric star formation law based on comparisons between theoretical models and observational data. In addition, the SFL applied to numerical simulations of galaxy formation and evolution is based on the volume densities (e.g., \citealt{1997A&A...325..972G}; \citealt{2008MNRAS.383.1210S}).
When the disc thickness of a galaxy is constant, the surface density, which is simply proportional to the volume density by a constant scale height, is applicable for the K-S law. However, the scale heights of gas and stars are not constant but varying with radius. Recent observations have revealed that the scale heights for gas and stars increase as a function of radius (e.g., \citealt{2010A&A...515A..62O}; \citealt{2014AJ....148..127Y}). 

As an alternative probe of the volumetric SFL,  \citet{2008ARep...52..257A} inferred the gas disc thickness by assuming the hydrostatic equilibrium for the gas and using the Jeans equation for the stellar component, but in the case of edge-on galaxies,  we can measure the gas disc thickness directly. Recently, we have measured the scale height from the edge-on spiral galaxies NGC 891, 4157, 4565, and 5907 (\citealt{2011AJ....141...48Y}, \citeyear{2014AJ....148..127Y}). In this paper, we add one more edge-on galaxy NGC 4013 ($i = 90\degr$; \citealt{1995A&A...295..605B}; \citealt{2015ApJ...808..153Z}) located at a distance of 14.5 Mpc, which is adopted based on the velocity of 1060 \kms\ with respect to the cosmic microwave background \citep{1996ApJ...473..576F} and  
$H_0$ = 73 \kms\ Mpc$^{-1}$. We have observed  NGC 4013 using the Combined Array for Research in Millimeter-wave Astronomy (CARMA) and the Owens Valley Radio Observatory (OVRO) for \coj\ and the Expanded Very Large Array (EVLA) for \HI. In addition, we have carried out the EVLA \HI\ observations in B configuration toward NGC 4157 and NGC 5907 to increase the spatial  resolution of the existing data for better resolving the \HI\ disc. Using the high-angular resolution data, we will investigate how the volumetric SFL is different from the ``general'' SFL based on surface densities.

Another possible prescription for the SFR is the interstellar gas pressure.
\citet{1993ApJ...411..170E} suggested that the hydrostatic midplane pressure is responsible for regulating the molecular to atomic gas ratio. Later, \citet{2002ApJ...569..157W} and \citet{2006ApJ...650..933B} found a power-law  correlation  between the hydrostatic pressure and the gas ratio (as measured by the ratio of CO to \HI\ intensity) based on observations. 
The tight correlation between the pressure and the ratio implies that the pressure is one of the key variables for the SFR since both the SFR and the gas pressure are positively correlated with the gas density. 
However, the previous studies assumed that the vertical velocity dispersion involved in the hydrostatic pressure is constant, contrary to recent observational results,  demonstrating that the velocity dispersions in CO and \HI\ decrease with radius (e.g., \citealt{Boomsma_2008}; \citealt{2009AJ....137.4424T}; \citealt{2014AJ....148..127Y}). \cite{2009AJ....137.4424T} derived the vertical velocity dispersion from the second moment map assuming isotropy of the velocity dispersion, using galaxies that have an inclination less than 50$\degr$.
In previous studies (\citealt{2011AJ....141...48Y}, \citeyear{2014AJ....148..127Y}), we obtained the hydrostatic midplane pressure using an assumed constant velocity dispersion and the turbulent interstellar pressure using the varying velocity dispersions in order to compare the pressures in controlling the gas ratio. We found that the power-law correlation between the ratio and the hydrostatic pressure is still valid for the relation between the interstellar gas pressure and the ratio based on the volume density.  
Here, we will investigate how the gas pressure predicts the SFR by directly comparing the two quantities (on volume density basis) estimated from the edge-on galaxy sample. 

This paper is organized as follows. Section \ref{obs} describes the CO, \HI, and IR (3.6 and 24 \um) observations and the reduction process of NGC 4013 (Section \ref{obs_4013}) and the observations and the reduction of \HI\ for NGC 4157 and 5907 (Section \ref{obs_4157_5907}). We present the position-velocity (p-v) diagrams of CO and \HI\ for NGC 4013 and derive the rotation curve using the p-v diagrams in Section \ref{kinematics}. Section \ref{radialprofile} explains how we obtain the surface mass density profiles for the gas, stars (3.6 \um), and the SFR (24 \um) of NGC 4013. In Section \ref{verticalprof}, we measure the scale heights as a function of radius for the gas, stars, and the SFR and determine the vertical velocity dispersions with radius for CO, \HI, and stars. In Section \ref{SFprescriptions}, using all the information, we investigate how the SFR volume density is correlated with the gas volume density and how the volumetric SFL is different from the surface-density SFL, and compare the volumetric and surface SFEs for the gas, \Htwo, and \HI. In addition, we examine the role of the interstellar gas pressure in regulating the SFR. Finally, we summarize and conclude our results in  Section \ref{sum}.  

\section{Observations and Data Reduction}
\label{obs}

\subsection{NGC 4013}
\label{obs_4013}

\begin{figure*}
\begin{center}
\includegraphics[width=0.95\textwidth,angle=0]{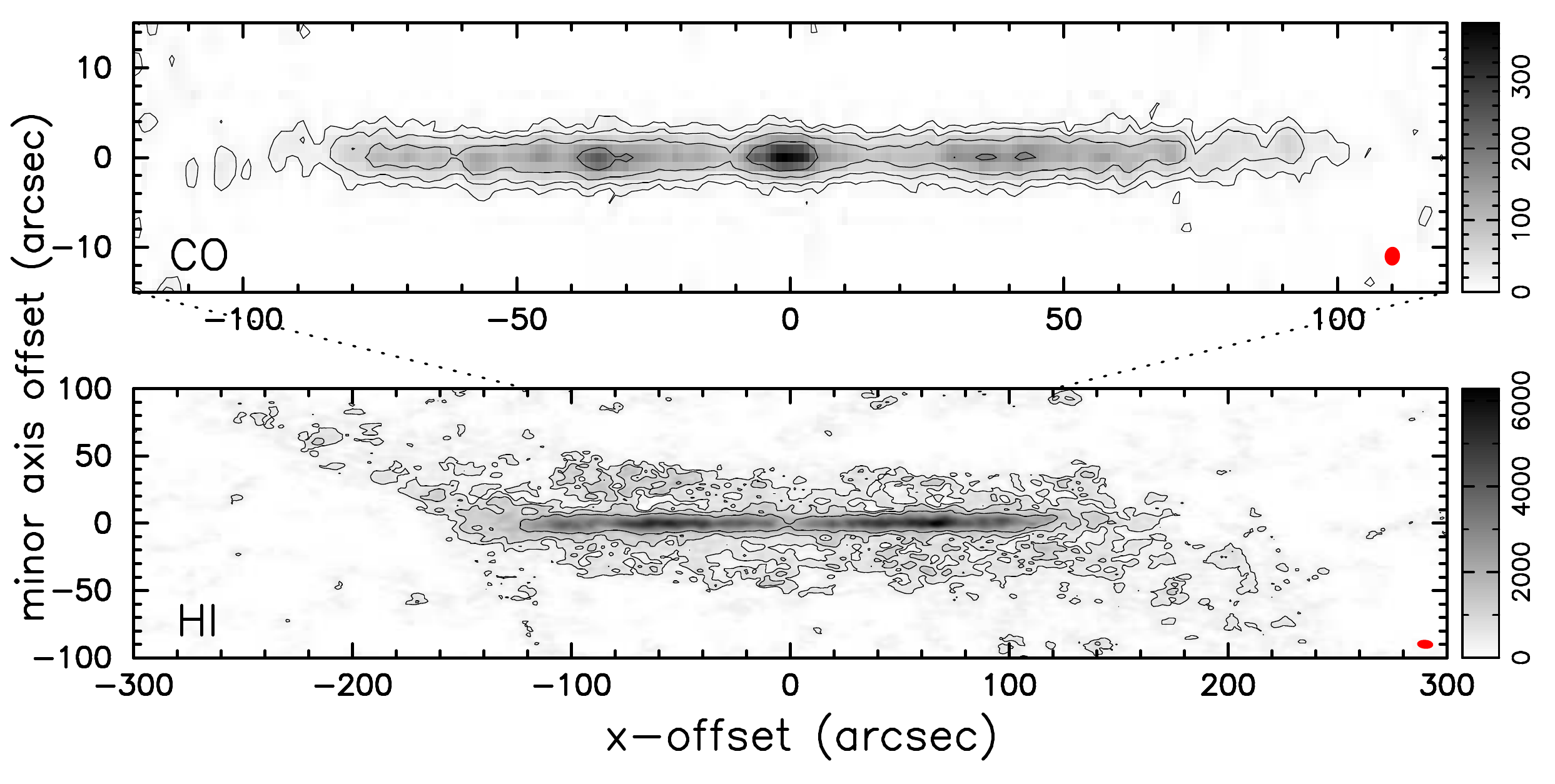}
\caption{CO (top) and \HI\ (bottom) integrated intensity maps of NGC 4013. Contour levels are $15.0 \times 2.2^n$ K \kms, with n=0, 1, 2, 3 for CO and $450.0 \times 2.0^n$ K \kms, with n=0, 1, 2  for \HI. The lowest contour level is $\sim$3$\sigma$. The synthesized beam is shown in the lower right corner of each panel. 
\label{n4013maps}}
\end{center}
\end{figure*}

We observed \coj\ toward NGC 4013 using CARMA in C and D configurations with a velocity coverage of 446 \kms\ in 2008 and reduced them using the MIRIAD package: the task MFCAL for bandpass and gain calibration and the task BOOTFLUX for flux calibration. 
The calibrated data were combined with OVRO observations in the C, L, and H (U) configurations using a 7-point mosaic  \citep{2004IAUS..217..166S} via the MIRIAD task INVERT to increase the sensitivity and image fidelity. We obtained the average rms noise of 0.2 K from line-free edge channels of the combined cube. 
The spatial and velocity resolutions of the combined image are 2.80\ac\ $\times$ 2.08\ac\ (natural weighting) and 10 \kms, respectively. The systemic velocity ($V_{\rm sys}$) is 835 \kms \citep{1995A&A...295..605B}. We measured a position angle of 65$\degr$ from the $Spitzer$ 3.6 \um\ map using the MIRIAD task IMFIT. The integrated intensity map (rotated by 25$\degr$) is shown in the top panel of Fig. \ref{n4013maps}. The map is masked to increase a signal-to-noise ratio (S/N) by blanking regions where signal is below 3$\sigma$ in a smoothed map (to 6\ac\ resolution). 
The galactic centre of the map ($\alpha$ = \rm{11$^h$58$^m$31$\fs$34 and $\delta$ = 43$\degr$56$\arcmin$50.72\ac}) is placed at the $x$-offset=0 and the minor axis offset=0. The CO total flux of the masked integrated intensity map is 880 Jy \kms.

The \HI\ data were obtained from the VLA in D configuration in 2003 (project AS0750; PI: E. Schinnerer) and EVLA (C and B configurations) in 2010 -- 2011 \citep{2015ApJ...808..153Z}. 
Some bad data in the B and C arrays were excluded (tracks from 2010 October 24 and 2011 March 21) and flagged using the task FLAGCMD of the Common Astronomy Software Applications (CASA) package \citep{2007ASPC..376..127M}. 
All the B, C, and D data were calibrated separately using the CASA tasks SETJY, BANDPASS, GAINCAL, FLUXSCALE, and APPLYCAL. The calibrated data were combined using the task CONCAT and cleaned using CLEAN. A continuum subtraction was carried out on the combined image using IMCONTSUB. The final cleaned image with Briggs weighting and a robustness of 0.5 has a resolution of 7.51\ac $\times$ 6.58\ac. 
Fig. \ref{n4013maps} (bottom panel) shows the masked integrated intensity \HI\ map rotated by 25$\degr$. A strong warp is evident on the upper left and lower right sides of the figure, consistent with \cite{2015ApJ...808..153Z}.
The \HI\ total flux of the masked map is 46 Jy \kms\ and the rms noise per channel is 1.4 K.

\begin{figure*}
\begin{center}
\includegraphics[width=0.95\textwidth,angle=0]{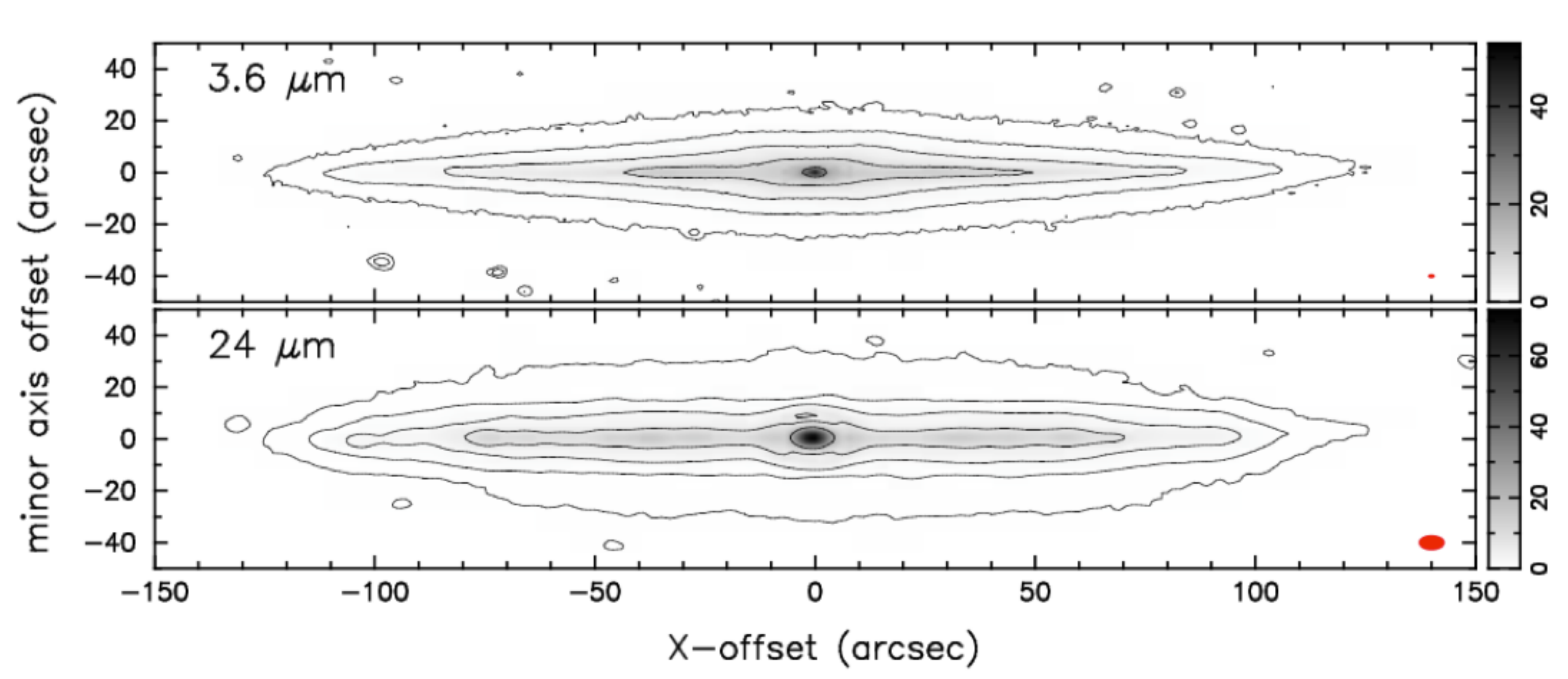}
\caption{The $Spitzer$ 3.6 \um\ (top) and 24 \um\ (bottom) images of NGC 4013. Contour levels are $0.2 \times 3.25^n$ MJy sr$^{-1}$, with n=0, 1, 2, 3, 4.  The point-spread function of the images is shown in the lower right corner of each panel: 1.66\ac\ $\times$ 1.66\ac\ for 3.6 \um\ and 5.90\ac\ $\times$ 5.90\ac\ for 24 \um.
\label{irmaps}}
\end{center}
\end{figure*}

We obtained the $Spitzer$ IRAC 3.6 \um\ (Program ID 215; PI: G. Fazio) and MIPS 24 \um\ (Program ID 30562; PI: J. C. Howk) Basic Calibrated Data (BCD) from the Spitzer Heritage  Archive. The BCD data were mosaicked after background matching using MOPEX (Mosaicking and Point Source Extraction). Before mosaicking, the instrumental artefacts of the 24 \um\  map were removed via Image Reduction and  Analysis Facility (IRAF) tasks. The mosaicked images of the 3.6 \um\ and 24 \um\ data are shown in Fig.  \ref{irmaps}. The resolutions of the maps are 1.66\ac\ $\times$ 1.66\ac\ for 3.6 \um\ and 5.9\ac\ $\times$ 5.9\ac\ for 24 \um. As stated in previous studies (e.g., \citealt{2014AJ....148..127Y}), the 24 \um\ image appears to be similar to the CO image. 

\subsection{NGC 4157 and NGC 5907}
\label{obs_4157_5907}

We presented the CO and \HI\ images of NGC 4157 and 5907 in the previous study by  \cite{2014AJ....148..127Y}. The  angular resolutions of the VLA \HI\ (C and D arrays) images are about 15\ac ($\sim$ 1 kpc), which is not sufficient for comparison with the CO data ($\sim 3.5\ac$; 245 pc) and to resolve the vertical structure of the disc. For that reasons, we observed \HI\ emission with EVLA in B configuration toward NGC 4157 for 24 hours and NGC 5907 for 26 hours in 2012. 
In the case of NGC 4157, we used only 16 hours of data since the data taken on June 8 and July 5 were not good enough to use even after careful flagging. 
Once the data were flagged and calibrated using the CASA package, we combined the visibility data with the VLA (C and D) calibrated data using the CASA task CONCAT.  
For imaging the combined data, we used the task CLEAN with Briggs weighting and a robustness of 0.5, which resulted in a spatial resolution of 4.47\ac\ $\times$ 4.42\ac\ ($\sim$ 280 pc) for NGC 4157 and 5.03\ac\ $\times$ 4.42\ac\ ($\sim$ 230 pc) for NGC 5907. Continuum emissions were subtracted in both cleaned images using the task IMCONTSUB. 

\begin{figure*}
\begin{center}
\includegraphics[width=0.95\textwidth,angle=0]{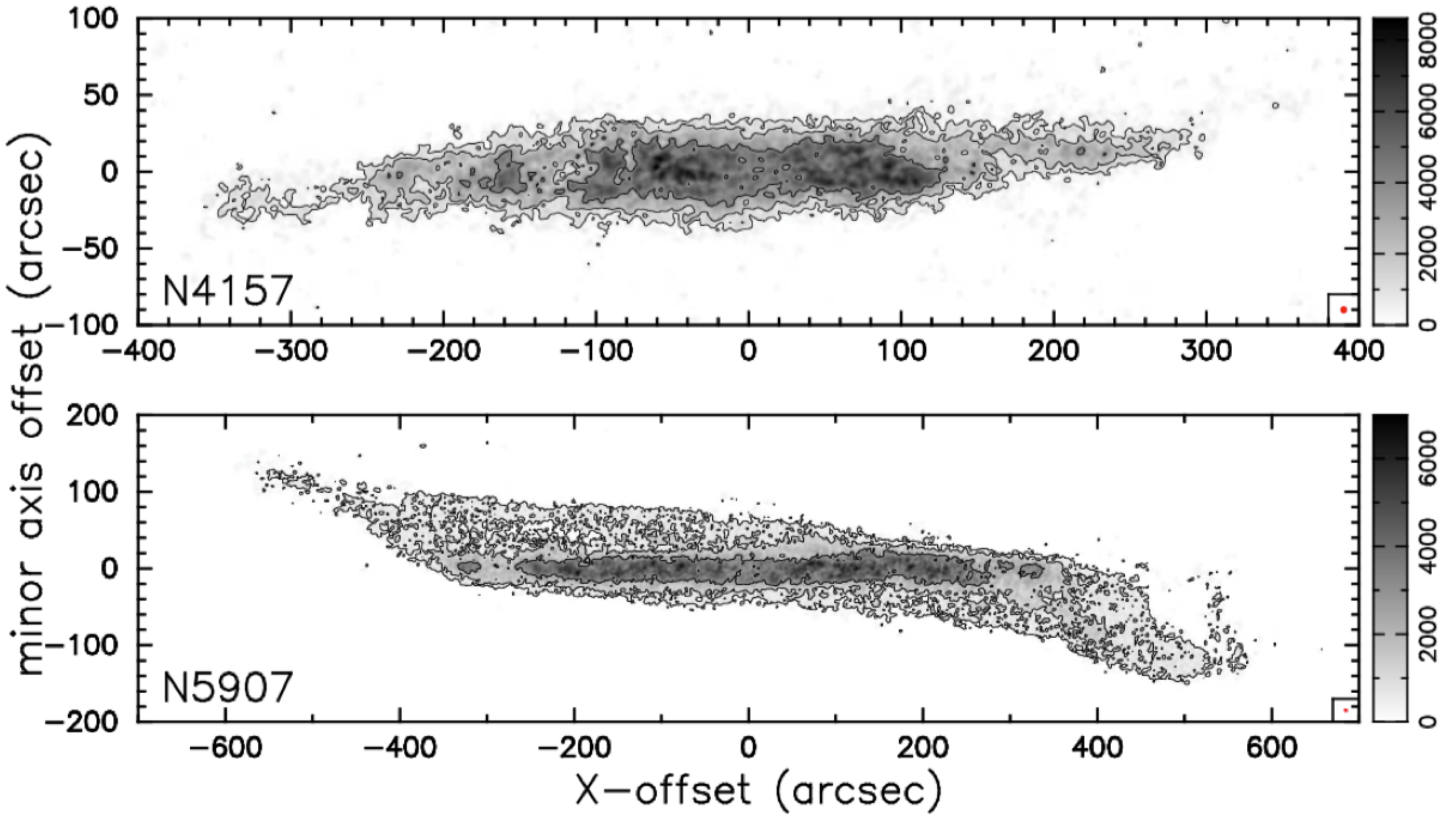}
\caption{\HI\ integrated intensity maps of NGC 4157 (top) and NGC 5907 (bottom). Contour levels are $750 \times 2.3^n$ K \kms, with n=0, 1, 2 for NGC 4157 and $360 \times 2.7^n$ K \kms, with n=0, 1, 2 for NGC 5907.  The lowest contour level is 3$\sigma$. The synthesized beam is shown in the lower right corner of each panel. 
\label{h1maps}}
\end{center}
\end{figure*}

Fig. \ref{h1maps} shows the masked integrated intensity maps of the combined cubes. Each map is rotated such that the blue-shifted signal is on the left side and the red-shifted signal is on the right side along the major axis. The galactic centres are placed in the centre of the offset system. The maps  better resolve small scales compared to the old maps using only  C and D configurations.  The strong warp in NGC 5907 and the weak warp in NGC 4157 shown in \cite{2014AJ....148..127Y} are more evident in the new maps.  

\section{Kinematics of NGC 4013}
In order to derive the surface density profiles of CO and \HI\ using the PVD method \citep{2011AJ....141...48Y}, we first made the position-velocity (p-v) diagrams (Fig. \ref{pv4013}) and obtained the rotation curves (Fig. \ref{vrot}). Since the method uses the Position-Velocity Diagram to derive the radial distribution, we defined it as the PVD method in \cite{2011AJ....141...48Y}. 
\label{kinematics}

\subsection{Position-Velocity Diagrams}
We obtained the p-v diagrams by integrating the minor axis over $\pm$10\ac\ (CO) and $\pm$50\ac\ (\HI) from the midplane. In the central region of the CO map, an elongated structure is stretched out from 720 \kms\ to 940 \kms\ and it is in good agreement with  \citet{1999A&A...343..740G}, suggesting that the central   feature might be caused by a bar. Such a bar structure near the centre is also shown in NGC 891 (\citealt{1987PASJ...39...47S}; \citealt{1992A&A...266...21G}; \citealt{1995A&A...299..657G};\citealt{2011AJ....141...48Y}) more distinctively and the radial velocity ($\sim$785 \kms) of the bar feature is  exceeding the fairly flat velocity ($\sim$760 \kms) in the p-v diagram of NGC 891 (Fig. 4 of \citealt{2011AJ....141...48Y}). According to the study by  \citet{1999ApJ...522..699A}, an elongated bar feature with higher radial velocity compared to the outer flat velocity is suggested to be a side-on bar (e.g., NGC 891) while a bar feature with lower velocity than the flat velocity is suggested to be an end-on bar like NGC 4013. In addition, they mentioned that a peanut-shaped bulge is shown when a bar is on the plane of the sky (side-on) wile a box-shaped bulge is shown when a bar is along the line of sight (end-on). The box-shaped bulge can be clearly seen in the box-shaped contours around the bulge of  the 3.6 \um\ image in Fig. \ref{irmaps}.   In contrast, the bar feature near the centre is not present in the p-v diagram of \HI, implying that  no strong \HI\ emission is arising from these small radii.

\subsection{Rotation Curve}
\begin{figure}
\begin{center}
\includegraphics[width=0.45\textwidth,angle=0]{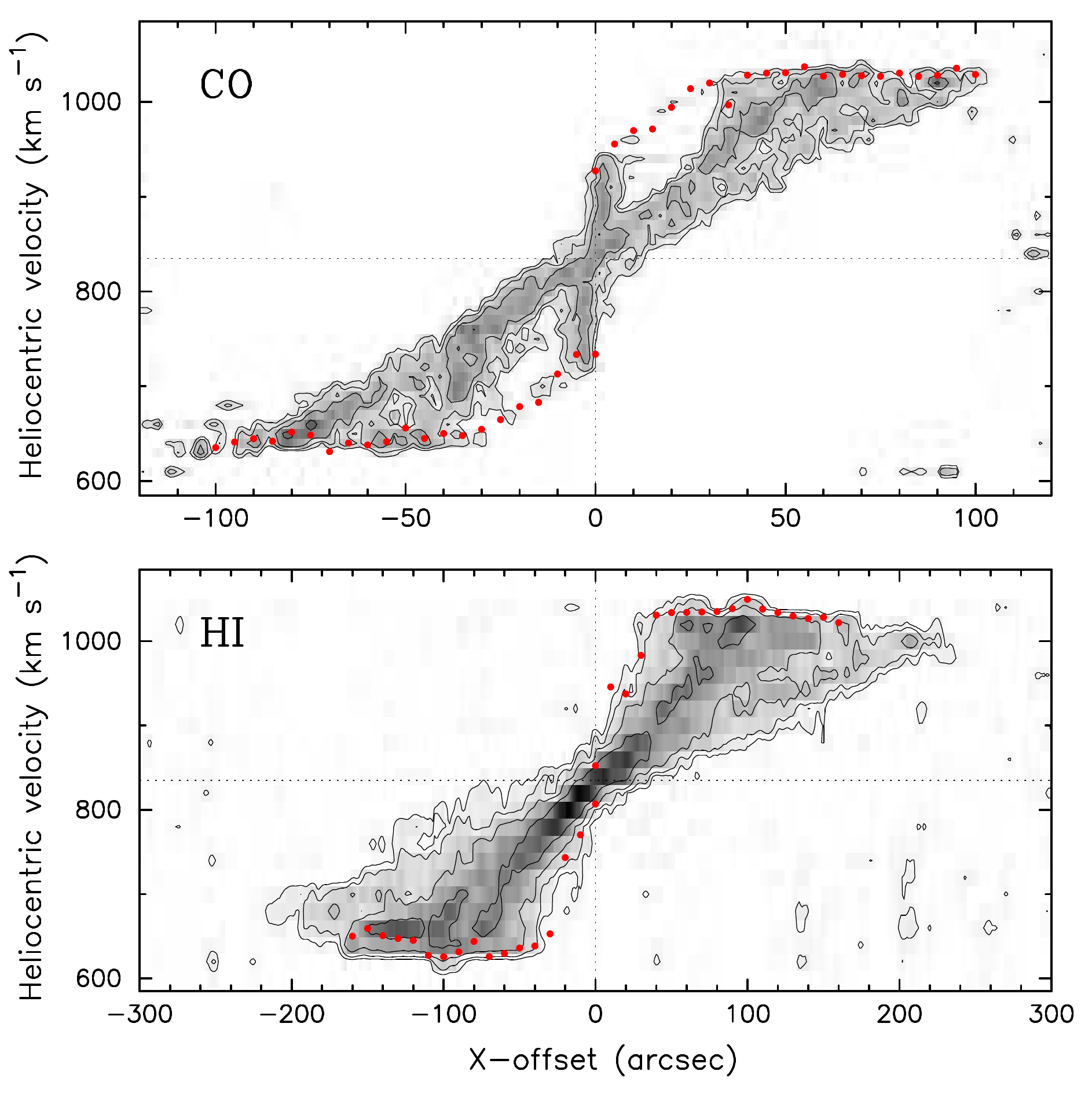}
\caption{Vertically integrated position-velocity diagrams of NGC 4013 CO (top) and \HI\ (bottom) line emission. CO contours are $2.4 \times 1.8^n$ K arcsec, with n=0, 1, 2, 3.
\HI\ contour levels are $90.0 \times 2.1^n$ K arcsec, with n=0, 1, 2, 3.  The lowest contour level is 3$\sigma$. The horizontal dotted lines indicate the heliocentric systemic velocity of 835 \kms. The red circles represent the obtained rotation velocities.
\label{pv4013}}
\end{center}
\end{figure}

\begin{figure}
\begin{center}
\includegraphics[width=0.45\textwidth]{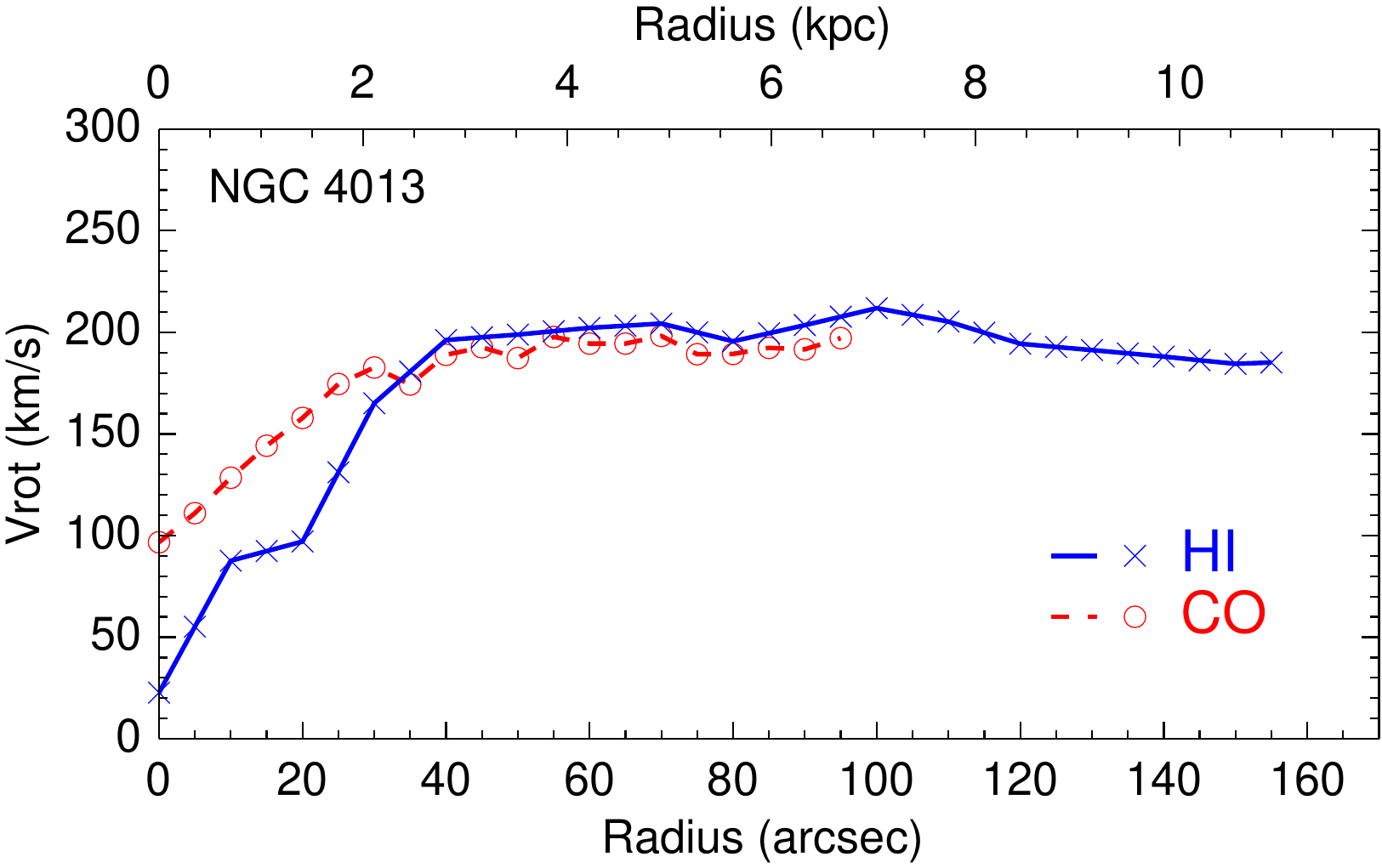}
\caption{Rotation curves of CO (red circle) and \HI\ (blue cross) for NGC 4013. 
\label{vrot}}
\end{center}
\end{figure}

Using the p-v diagram along the midplane, we derived the rotation curve by employing the envelope tracing method \citep{1996ApJ...458..120S}. Note that the p-v diagram is obtained from a slice along the midplane (unlike the vertically integrated p-v diagram for the PVD method) to exclude a lagging halo of \HI\ gas \citep{2015ApJ...808..153Z}.
The envelope tracing method uses the terminal velocity ($V_{\rm ter}$) at each $x$-offset of the p-v diagram. The $x$-offset approaches a radius at the terminal velocity, which is defined to be the highest velocity at the 3$\sigma$ level on the terminal side. Based on the envelope method, the rotation curve is obtained from  the terminal velocity after correcting for the observational velocity resolution ($\sigma_{\rm obs}$) and the gas velocity dispersion ($\sigma_{\rm g}$): 
\begin{equation}
V_{\rm rot} = V_{\rm ter} - \sqrt{\sigma^2_{\rm obs} + \sigma^2_{\rm g}}.
\label{Vrot}
\end{equation}
The observational resolution is 4.3 \kms\ for CO and 8.5 \kms\ for \HI\ based on the channel resolution of 10 \kms\ (CO) and 20 \kms\ (\HI). The velocity dispersion is assumed as 8 \kms\ for \HI\ (\citealt{2006ApJ...650..933B}, and references therein) and 4 \kms\ for CO (\citealt{2011MNRAS.410.1409W}; \citealt{2016AJ....151...15M};  \citealt{2017A&A...607A.106M}). The derived rotation curves for CO and \HI\ are shown as  red filled circles in the p-v diagram (Fig. \ref{pv4013}).  Fig. \ref{vrot} shows the rotation curves as red open circles for CO and blue crosses for \HI\ that are average values of the red-shifted  ($V_{\rm ter} > V_{\rm sys}$) and blue-shifted ($V_{\rm ter} <  V_{\rm sys}$) disc curves. The CO curve starts from 90 \kms\ near the centre due to  non-circular motions induced by the bar, which causes the disagreement between CO and \HI\ curves in the central regions. In addition, the difference of the curves within $\sim$ 35\ac\ may be partially due to the lack of \HI\ emission at small radii.
The CO and \HI\ curves are very consistent each other beyond $r \sim 35$\ac\ where they show roughly flat rotation of $\sim 200$ \kms. 

\section{Radial Density Distribution of NGC 4013}
\label{radialprofile}
We use edge-on galaxies since they are the best targets to study the disc thickness, which enables us to derive the volume density.
Regardless of the advantage, edge-on galaxies were not preferred objects to study the SFL since it is not easy to obtain the radial density distribution due to the projected radial components along the line-of-sight. 
However, we made it easier to derive the gas radial distributions of edge-on galaxies by employing the PVD method \citep{2011AJ....141...48Y}, using the p-v diagram and assuming circular rotation and a flat rotation curve. 
\cite{2011AJ....141...48Y} verified that the PVD method using the assumed flat rotation curve is appropriate for deriving the surface density profile by comparing two different methods and examining how well radial profiles from models recover the profiles from data.

\subsection{Molecular and Atomic Gas}
\begin{figure*}
\begin{center}
\begin{tabular}{c@{\hspace{0.1in}}c@{\hspace{0.1in}}c}
\includegraphics[width=0.32\textwidth]{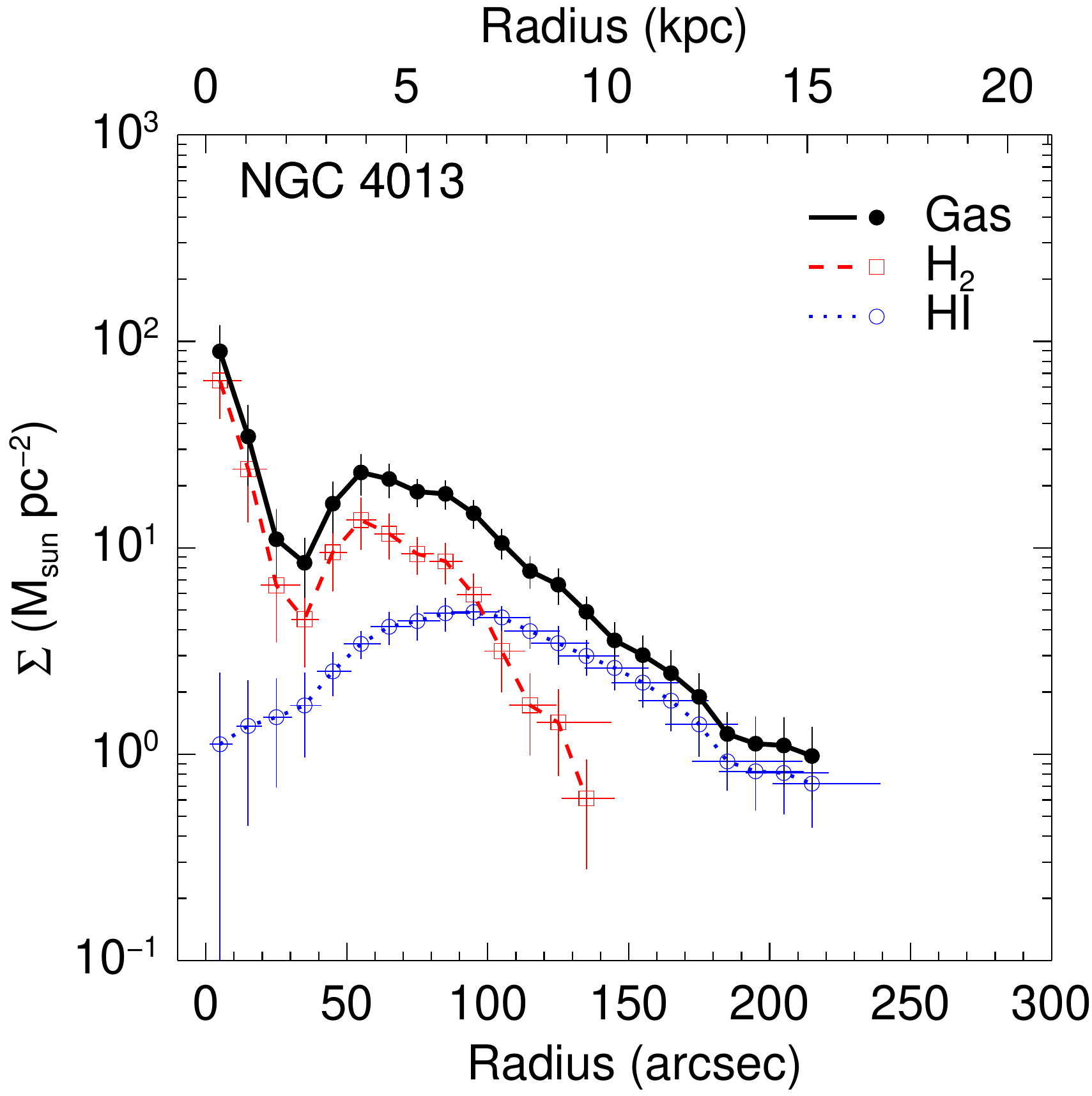}&
\includegraphics[width=0.32\textwidth]{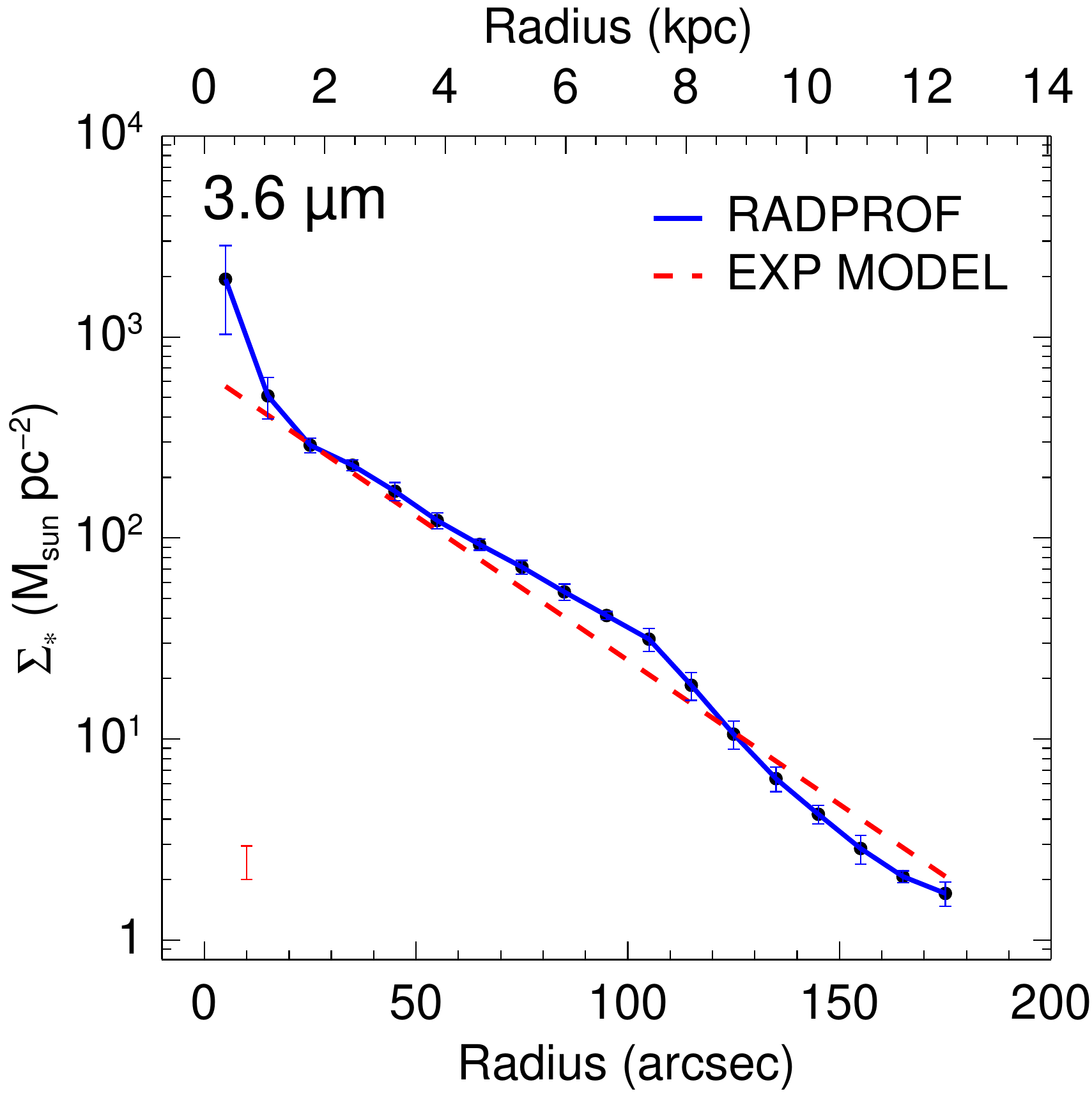}&
\includegraphics[width=0.32\textwidth]{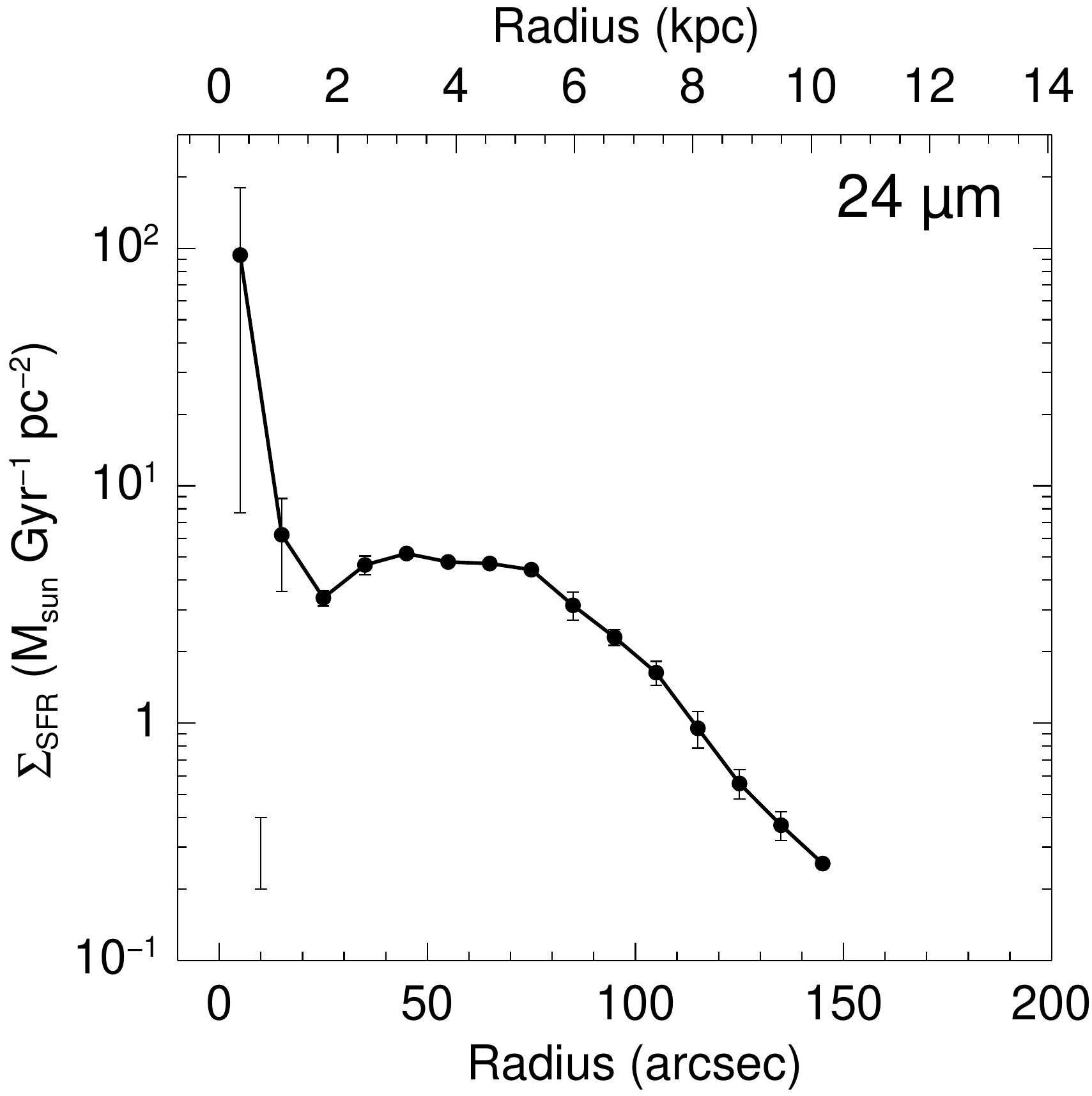}
\end{tabular}
\caption[Radial profiles of \Htwo, \HI, and total gas]{Left: Radial distributions of \sightwo\ (red open squares), \sighi\ (blue open circles), and \siggas\ (solid circles). The vertical error bars are the standard deviation of data points in the radial bins. Middle: stellar surface density profiles obtained from the GIPSY task RADPROF (blue solid line) and the exponential disc model (red dashed line). The error bar on the lower left corner represents the largest difference between the RADPROF and the model profiles except the central regions. 
Right: SFR surface density as a function of radius obtained from $Spitzer$ 24 \um\ imaging using RADPROF.  The vertical error bars on the RADPROF profile show the standard deviation of each point. The vertical error bar on the lower left corner shows a factor of 2 uncertainty obtained by comparing two different radial profiles from the RADPROF and ELLINT tasks. 
\label{radiprof}}
\end{center}
\end{figure*}

First, we convolved the CO data to the \HI\ beam to compare each other and combine them for the total gas. Then we obtained the vertically integrated p-v diagrams (Fig. \ref{pv4013}) for both CO ($\pm 10$\ac) and \HI\ ($\pm 50$\ac).  
We assumed a flat rotation curve of 200 \kms\ based on Fig. \ref{vrot} for the PVD method. The ($x$, $V_{\rm r}$) positions in the p-v diagram are converted to a galactocentric radius (equation \ref{eq_rad}) using the radial velocity ($V_{\rm r}$) of the position and the assumed circular speed ($V_{\rm c}$) of 200 \kms.

\begin{equation}
r =  V_{\rm c} \left< \frac{x}{V_r-V_{\rm sys}} \right> \quad\quad \rm{with}\ \it |V_r-V_{\rm sys}| < V_{\rm c},
\label{eq_rad}
\end{equation}
where the angle brackets represent the mean value of $x/(V_r - V_{\rm sys})$ within a pixel in the p-v diagram. Also, the flux of each pixel is converted into the surface brightness in a face-on disc by taking into account the line-of-sight depth and the flux of one pixel affected by both near and far sides of the edge-on galaxy.
In this procedure, the central regions ($|x| < 40$\ac\ and $|V_{\rm r}-V_{\rm sys}| < 50$ \kms) in the p-v diagrams are excluded due to  a mixture of emissions from many different radii.  Note that there are still many pixels in regions wehre $|x| < 40$\ac\ and $|V_{\rm r}-V_{\rm sys}| > 50$ \kms, assigned to small radii near the center.
Finally, the obtained surface brightnesses are converted to surface mass densities (\sightwo\ and \sighi) using the conversion factors for CO (\citealt{1996A&A...308L..21S}; \citealt{2001ApJ...547..792D}) and \HI: 
\begin{equation}
N(\textrm H_2) \,\rm [cm^{-2}] = 2 \times 10^{20} \,\it I_{\rm CO} \,[\rm K \,\kms],
\label{xco}
\end{equation} 
\begin{equation}
N(\textrm{\HI}) \,\rm [cm^{-2}] = 1.82 \times 10^{18} \,\it I_{\rm HI} \,[\rm K \, \kms].
\label{xh1}
\end{equation}
The face-on surface densities are averaged in a radial bin of 10\ac.
Fig. \ref{radiprof} (left panel) shows the radial surface density profiles of the atomic gas (\sighi) as bule open circles, the molecular gas (\sightwo) as red open squares, and the total gas (\siggas), which is a summation of \sighi\ and \sightwo\ with the inclusion of Helium (a factor of 1.36). Each data point is the average value of data in the radial bin size of 10\ac\ for both CO and \HI\ and the vertical error bar of each point represents the standard deviation (1$\sigma$ uncertainty). In addition, we assume a factor of 2 uncertainty for the CO-to-H$_2$  conversion factor (e.g., \citealt{2013ARA&A..51..207B}) in the molecular density profile. \citet{Sandstrom_2013} found that the average conversion factor is generally flat across the galactic disk of their sample except in the inner region where the mean value is lower by a factor of $\sim$2. We also derived radial profiles using the rotation curve within $r = 40$\ac, where the assumed flat rotation curve for the PVD method does not match to the obtained rotation curve, in order to compare with the derived radial profiles shown in Fig. \ref{radiprof} (left). The different rotation curves within 40\ac\ causes a factor of 1.1--2.6 difference for CO and a factor of 1.5--2.4 difference for \HI\ in the radial profiles, depending on the radius. 
 The horizontal error bar shows obtainable minimum and maximum radii in the radial bin and the minimum and maximum values are obtained by adding or subtracting the angular and velocity resolutions ($\Delta x$ and $\Delta V$) from equation \ref{eq_rad}. 

The molecular gas density profile shows a central concentration following a gap, possibly due to the bar.   This tendency is very similar to the molecular gas distribution of NGC 891 that also has a bar \citep{2011AJ....141...48Y}.  The atomic gas is deficient in the central region and it increases up to 100\ac\ and decreases gradually to the outer region as shown in many other spiral galaxies.

\subsection{Stars and SFR}
\label{starmodel}
The stellar (middle panel) and SFR (right panel) surface density profiles shown in Fig. \ref{radiprof} are obtained from the $Spitzer$ IRAC 3.6 \um\ map (Program ID 215; PI: G. Fazio) and MIPS 24 \um\ map (Program ID 30562; PI: J. Howk) using the Groningen Image Processing System (GIPSY; \citealt{1992ASPC...25..131V}) task RADPROF, which provides a radial density distribution using the Lucy iterative scheme (\citealt{1974AJ.....79..745L}; \citealt{1988A&AS...72..427W})  based on an assumption of axisymmetry. 
In order to calculate the radial distribution, RADPROF requires integrated intensity strips for east and west sides of a galaxy disc. Since RADPROF does not allow an inclination of 90$\degr$, we set INCL (inclination parameter) to be 89$\degr$ for NGC 4013. We have verified that varying the inclination parameter by several degrees do not change the radial distribution. Before employing RADPROF, foreground bright sources on the maps are removed  and filled by neighbour values using the GIPSY tasks BLOT and PATCH, respectively, then the maps are  convolved to the \HI\ resolution.
Using the MIRIAD task IMSPEC, we obtained the vertically integrated ($\pm$ 50\ac) brightness strips for blue-shifted and red-shifted discs and used  the data of strips as inputs to RADPROF.  The RADPROF density profiles in the figure are the average value of both disc profiles.

For comparison, we also obtained an additional stellar density profile by fitting an exponential, locally isothermal, and self-gravitating disc model \citep{1981A&A....95..105V} to the 3.6 \um\ map. When fitting the exponential disc model to the map, the central bulge region is excluded. The disc model is given by 
\begin{equation}
L(r,z) = L_0 \,\textrm{e}^{-r/l} \,\textrm{sech}^2\left(\frac{z}{h_*}\right)\;, \label{model}
\end{equation}
and it is integrated along the line of sight for an edge-on galaxy: 
\begin{equation}
\mu(x,z) = \mu(0,0) \left(\frac{x}{l}\right)  K_1\left(\frac{x}{l}\right) \textrm{sech}^2 \left(\frac{z}{h_*}\right), 
\label{expfit}
\end{equation}
where $L_0$ is the space-luminosity density at the centre, $l$ is the scale length, $\mu(0,0)=2lL_0$, and $K_1$ is the modified Bessel function of the second kind of order one. The obtained scale length ($l$) and scale height ($h_*$) are $\sim$30\ac\ (2100 pc) and $\sim$7\ac\ (490 pc), respectively.   \citet{Comer_n_2011} obtained the scale height of 100 -- 150 pc (thin disk) and 500 -- 620 pc (thick disk) by fitting a profile including two stellar and one gaseous discs. 

We adopted an empirical conversion factor (equation \ref{eq_star}) given by \citet{2008AJ....136.2782L} for converting 
 the 3.6 \um\ intensity ($I\rm_{3.6}$) from RADPROF and the exponential disk model into the stellar surface mass density (\sigstar): 
\begin{equation}
\sigstar\, [\surm] = 280\, (\cos\,i) \,I\rm_{3.6} \,[MJy \,\,sr^{-1}],
\label{eq_star}
\end{equation}
where the inclination $i$ is zero since it is already considered in the disc model and RADPROF. They adopted a $K$-band mass-to-light ratio of 0.5 for the conversion factor and measured a $I\rm_{3.6}$ to $K$-band intensity ratio ($I\rm_{3.6}$/$I_K$) of 0.55. The uncertainty (mainly due to the mass-to-light ratio) is $\sim$30--60$\%$, depending on galaxy colours (\citealt{2008AJ....136.2782L} and references therein). 
Fig. \ref{radiprof} (middle panel) shows the stellar surface densities obtained from the exponential disc model (red dashed line) and RADPROF (blue solid line). The vertical error bars on the RADPROF profile are the standard deviation of the data points, which are the average values in a radial bin of 10\ac.
The exponential profile  is well matched to the RADPROF profile except the central bulge regions. The largest difference between the profiles (except the central regions) is a factor of $\sim$1.5 and it is indicated as an error bar in the lower-left corner of the figure. We will use the exponential distribution as the stellar density (\sigstar) throughout this paper. 

In order to obtain the SFR surface density profile shown in the right panel of Fig. \ref{radiprof}, we used the $Spitzer$ 24 \um\ image with the calibration given by \citet{2007ApJ...666..870C}: 
\begin{equation}
\frac{\sigsfr }{\Msol \,\rm yr^{-1}\, kpc^{-2}} = 1.56 \times 10^{-35} \left(\frac{\it S\rm_{24 \mu m}}{\rm{erg \,s^{-1} \,kpc^{-2}}}\right)^{0.8104},
\label{eqSFR}
\end{equation}
where
\begin{equation}
\frac{S\rm_{24 \mu m}}{\rm{erg \,s^{-1} \,kpc^{-2}}} = 1.5 \times 10^{40} \left(\frac{\it I_{\rm 24}}{\rm MJy \,sr^{-1}}\right).
\end{equation}
The RADPROF solution of 24 \um\ surface brightness is used as $I_{24}$. The SFR profile shape of centrally concentrated density with a gap is very similar to the molecular gas distribution. 
The vertical error bars on the SFR profile are the standard deviation of the data points averaged in a bin size of 10\ac. In the lower-left corner of the figure, we also show an uncertainty of a factor of 2 caused by the RADPROF method, obtained by comparing two different methods (RADPROF and ELLINT) for a sample of face-on galaxies  in the previous study of \citet{2011AJ....141...48Y}. The GIPSY task ELLINT  integrates a map in elliptical annuli to obtain the mean intensity of each ring for the radial density distribution. In the study, the largest difference (a factor of 2) between RADPROF and ELLINT radial profiles  of the galaxy sample   is  used as the uncertainty.

\section{Radial Variation in Vertical Distribution}
\label{verticalprof}
\subsection{Gas Disc Thickness of NGC 4013}

\begin{figure*}
\begin{center}
\begin{tabular}{c@{\hspace{0.1in}}c@{\hspace{0.1in}}c}
\includegraphics[width=0.32\textwidth]{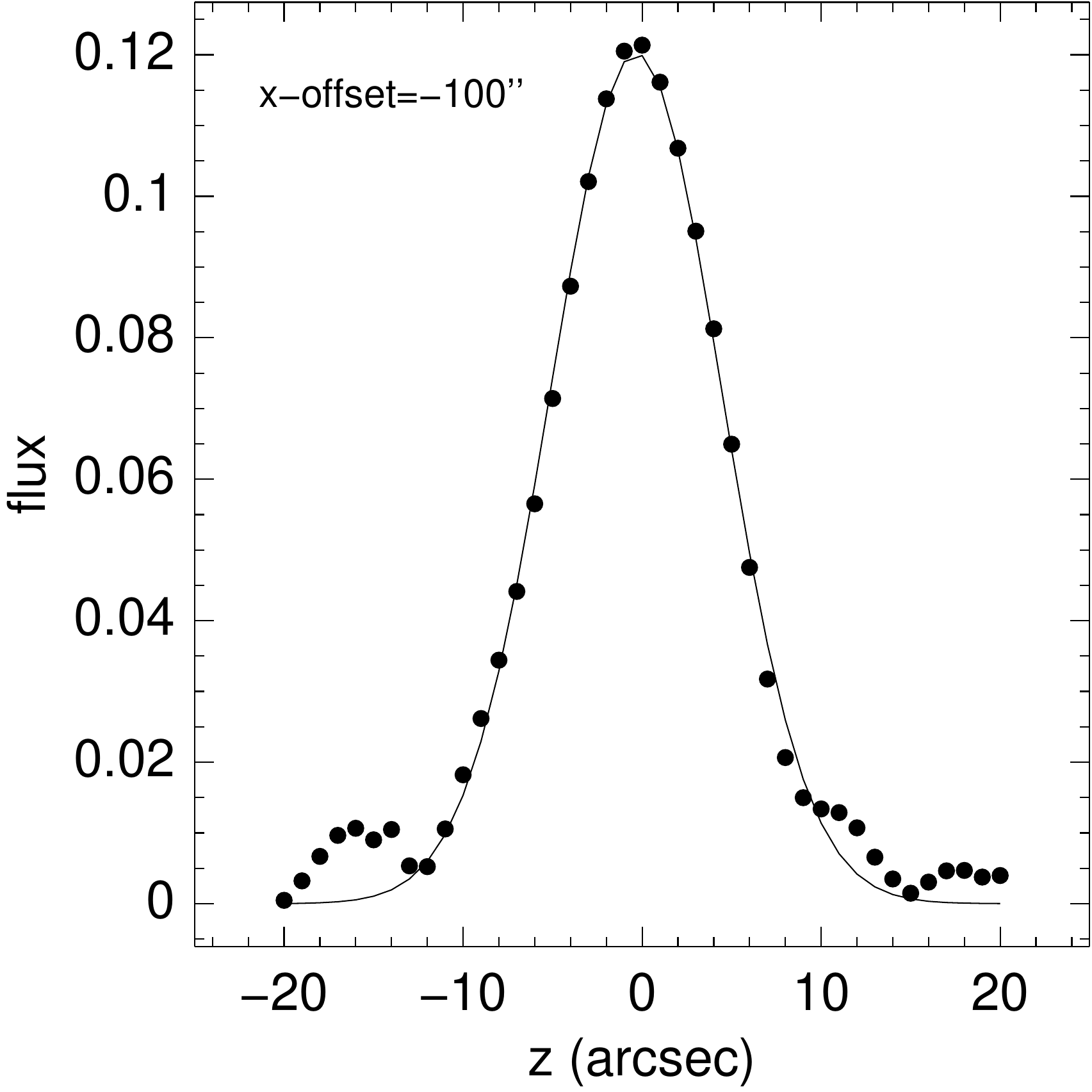}&
\includegraphics[width=0.32\textwidth]{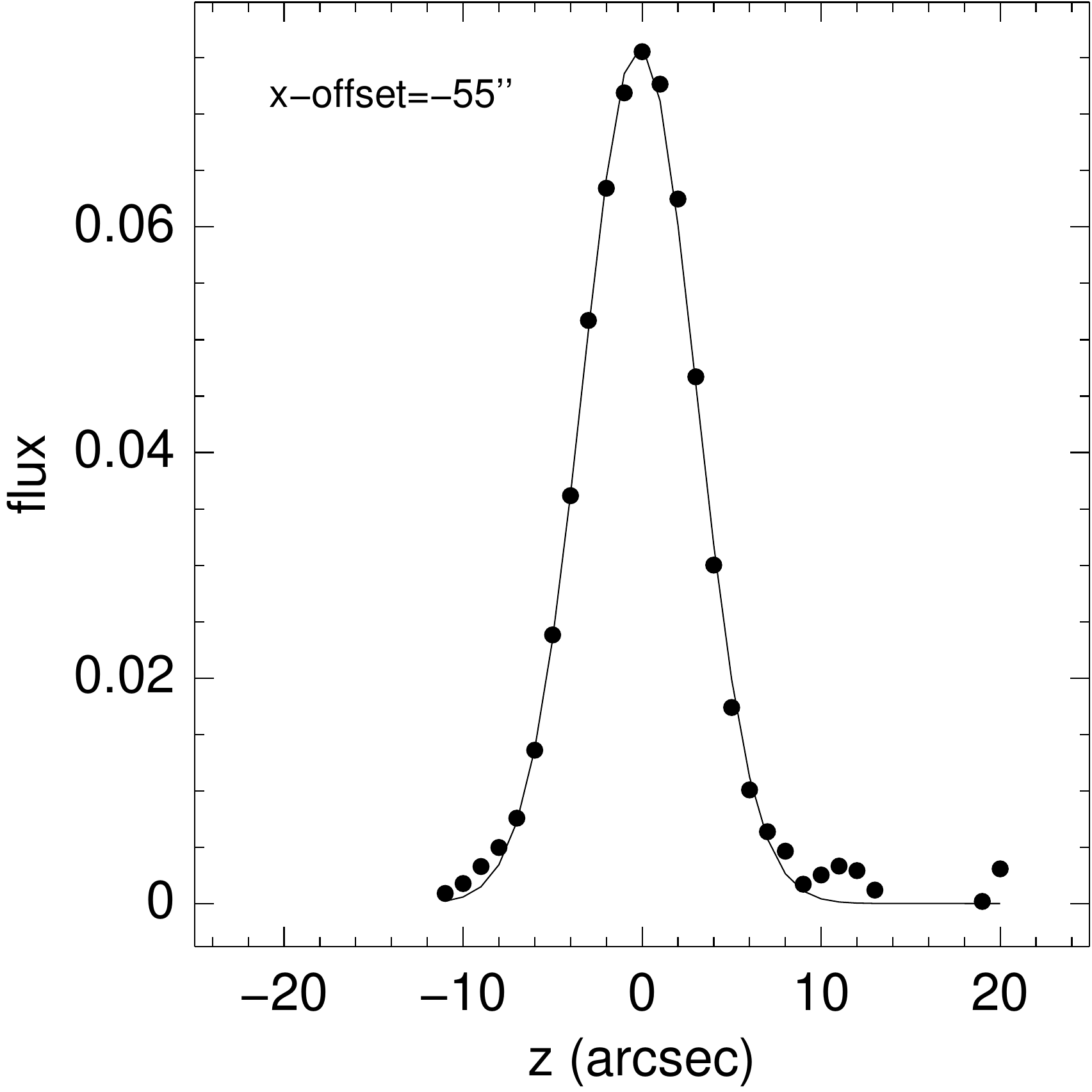}&
\includegraphics[width=0.32\textwidth]{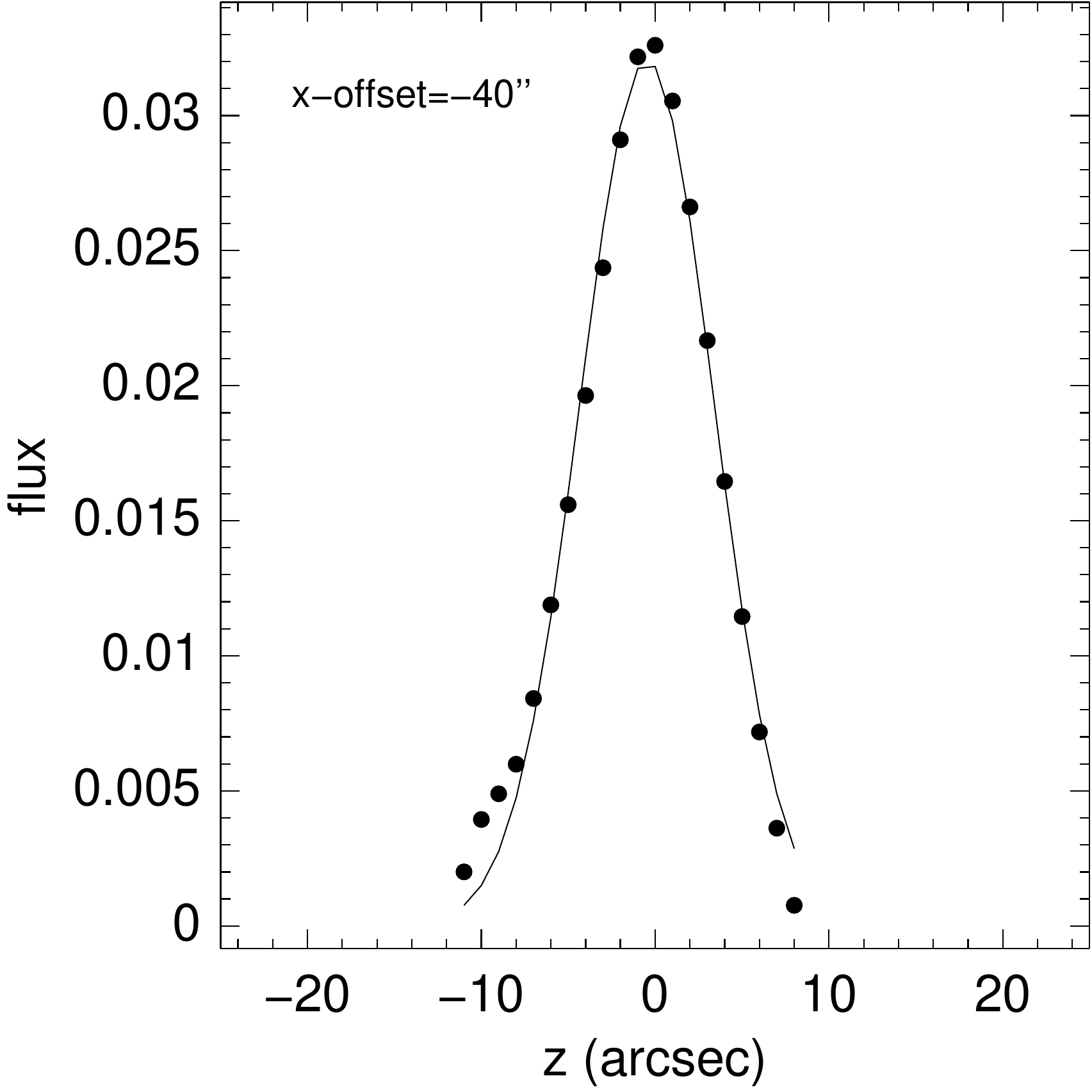}
\end{tabular}
\caption{Gaussian fits (solid line) to the vertical profiles (filled circles) of the terminal-velocity map of NGC 4013 \HI\ at radius -100\ac\ (left), -55\ac\ (middle), and -40\ac\ (right).  
\label{h1verticalfit}}
\end{center}
\end{figure*}

Since the inclination of NGC 4013 is almost 90$\degr$ (\citealt{2013A&A...556A..54V}; \citealt{2015ApJ...808..153Z}), it is relatively easy to measure the scale height. As we have done for the gas disc of NGC 891  \citep{2011AJ....141...48Y}, we simply fitted  a Gaussian function directly to the CO and \HI\ vertical density distributions obtained from terminal-velocity integrated intensity maps,  where the  $x$-offset approaches the actual radius. The maps are masked to reduce the rms noise for more reliable solutions and the most-edge channels of 620--640 \kms\ (red-shifted side) and 1020--1040 \kms\ (blue-shifted side) are used to integrate the CO and \HI\ maps. 
Figure \ref{h1verticalfit} shows Gaussian fitting results at three different $x$ offsets.
Note that the Gaussian width is used as the scale height and the beam size is deconvolved to measure the intrinsic scale height without smearing occurred by the telescope beam. The measured scale heights of CO and \HI\  are shown in the left panel of Fig. \ref{scaleh}. Each data point is an average value in a radial bin of 5\ac\ (350 pc) for CO and 10\ac\ (700 pc) for \HI. The vertical error bars show uncertainties of nonlinear least-squares fitting to the Gaussian function for the scale heights.
The open and solid circles present the blue-shifted  and red-shifted sides of the disc, respectively. 
The red (CO) and blue (\HI) lines are weighted linear least-squares fits to the data points of both discs  and the shaded regions around the lines represent uncertainties of the fits. The functions of the linear best-fit lines shown on the top of the figure will be used as the gas scale heights throughout this paper. We also obtained the best-fit lines for only the blue-shifted and red-shifted sides separately. The differences between the scale heights measured in the two sides separately are not significant: a factor of $\sim$1.2 for \HI\ and $\sim$1.1 for CO. The gradients of the fits in units of pc kpc$^{-1}$ are 6.3 and 23.3 for CO and \HI, respectively. The radial variation of the disc thickness is comparable to that of the other edge-on galaxies of \citet{2014AJ....148..127Y}. 
In this measurement, the \HI\ warp was not included because the noticeable warp is away from the disc midplane and the radial velocity of the warp component is lower than the terminal-velocities that we integrated for the intensity map.  Moreover, the warp starts at about $r=10$ kpc \citep{2015ApJ...808..153Z}, so it should not affect the fitting. In addition, a thick \HI\ disc shown in the bottom panel of Fig. \ref{n4013maps} does not affect the Gaussian width measurement since most thick disc components are not included in the terminal-velocity integrated map.

\subsection{Stellar Disc Thickness of NGC 4013}

\begin{figure*}
\begin{center}
\begin{tabular}{c@{\hspace{0.1in}}c@{\hspace{0.1in}}}
\includegraphics[width=0.45\textwidth]{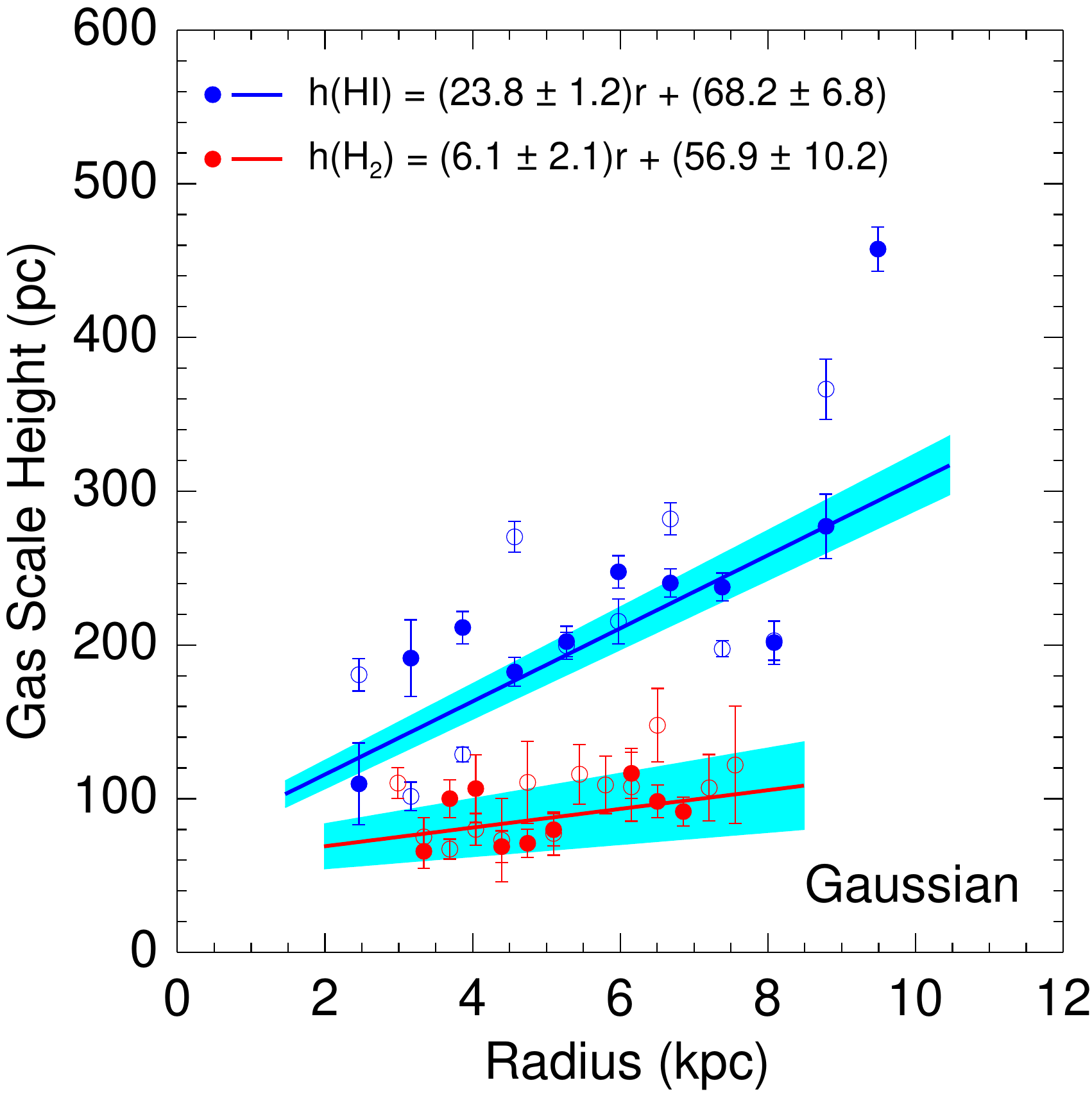}&
\includegraphics[width=0.45\textwidth]{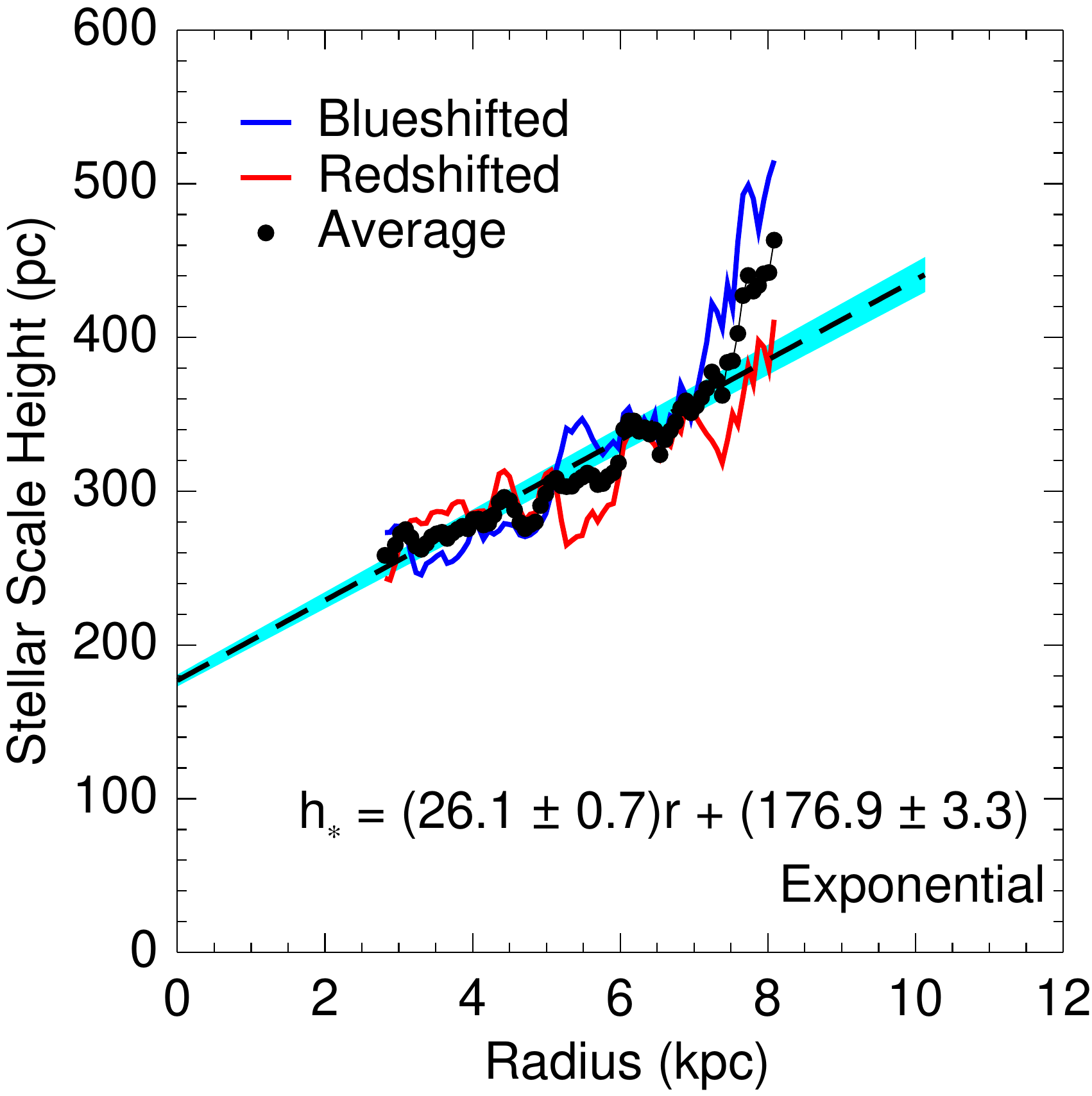}
\end{tabular}
\caption{Left: Molecular (red) and atomic (blue) gas scale heights in NGC 4013 as a function of radius measured by fitting a Gaussian function to the vertical profiles of CO and \HI. The open and filled circles show the blue-shifted and redshifted sides, respectively.  The vertical error bars show   uncertainties of the nonlinear least-squares fit to the Gaussian function for the scale heights. The lines are linear least-squares fits to all the data points and the fitting functions are shown on the top. The shaded regions around the lines represent uncertainties of the best-fits.
Right: Stellar scale height in NGC 4013 as a function of radius obtained by fitting an exponential function to the vertical profiles of 3.6 \um. The red and blue lines show the blue-shifted and redshifted sides, respectively. The filled circles are average values of both sides. The dashed line is the best fit (shown on the bottom) to the average values and the shaded region shows the uncertainty of the fit. 
\label{scaleh}}
\end{center}
\end{figure*}

\begin{figure*}
\begin{center}
\begin{tabular}{c@{\hspace{0.1in}}c@{\hspace{0.1in}}c}
\includegraphics[width=0.32\textwidth]{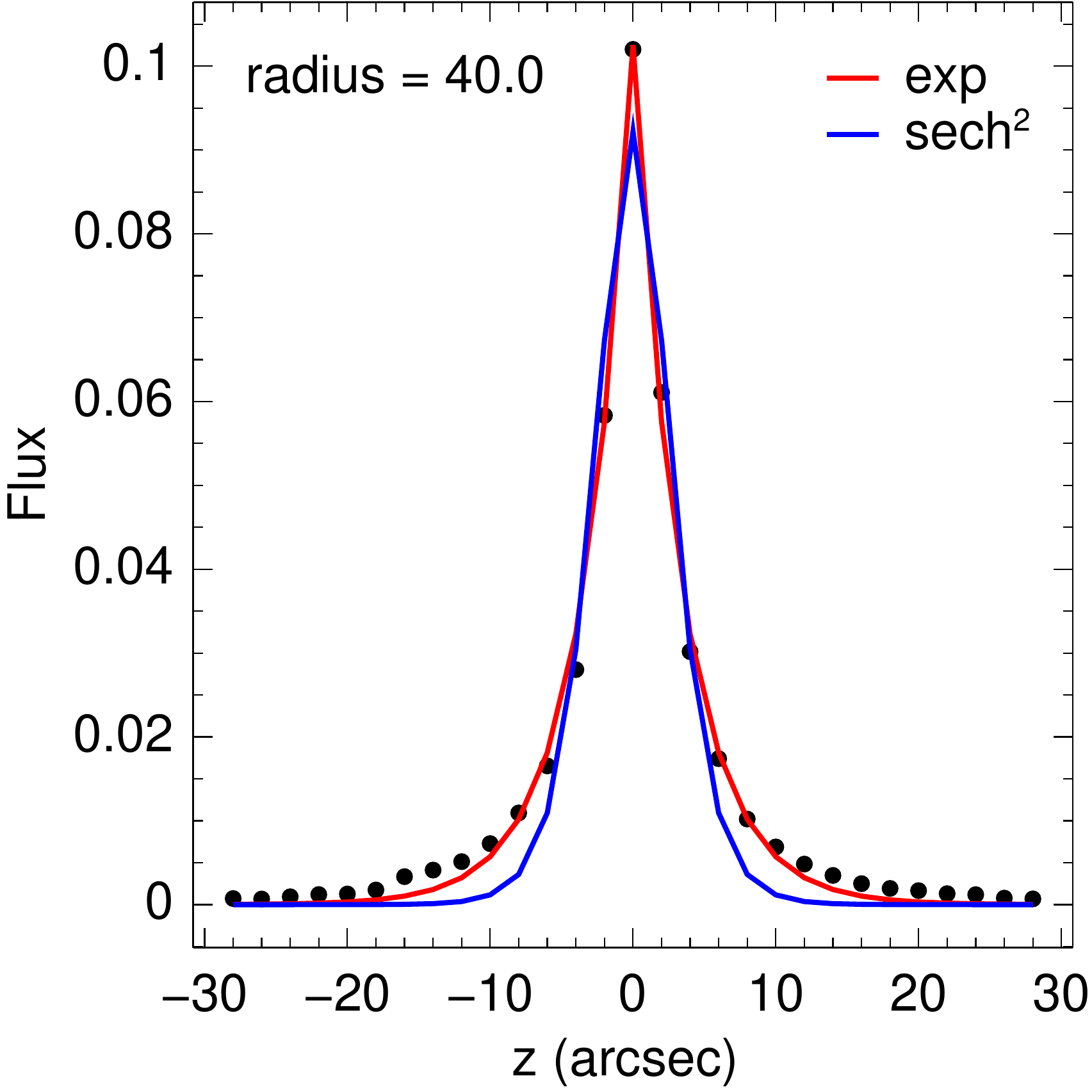}&
\includegraphics[width=0.32\textwidth]{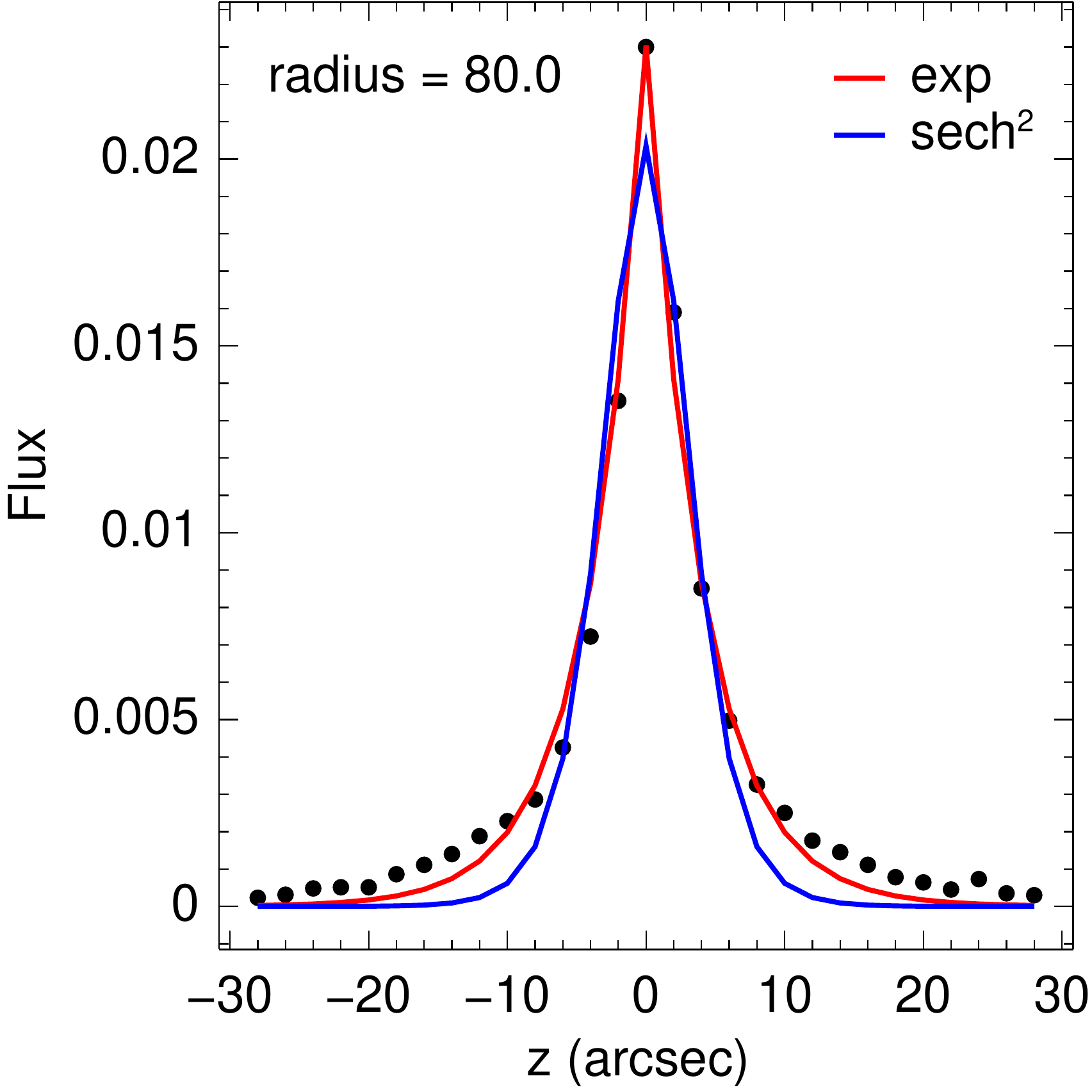}&
\includegraphics[width=0.32\textwidth]{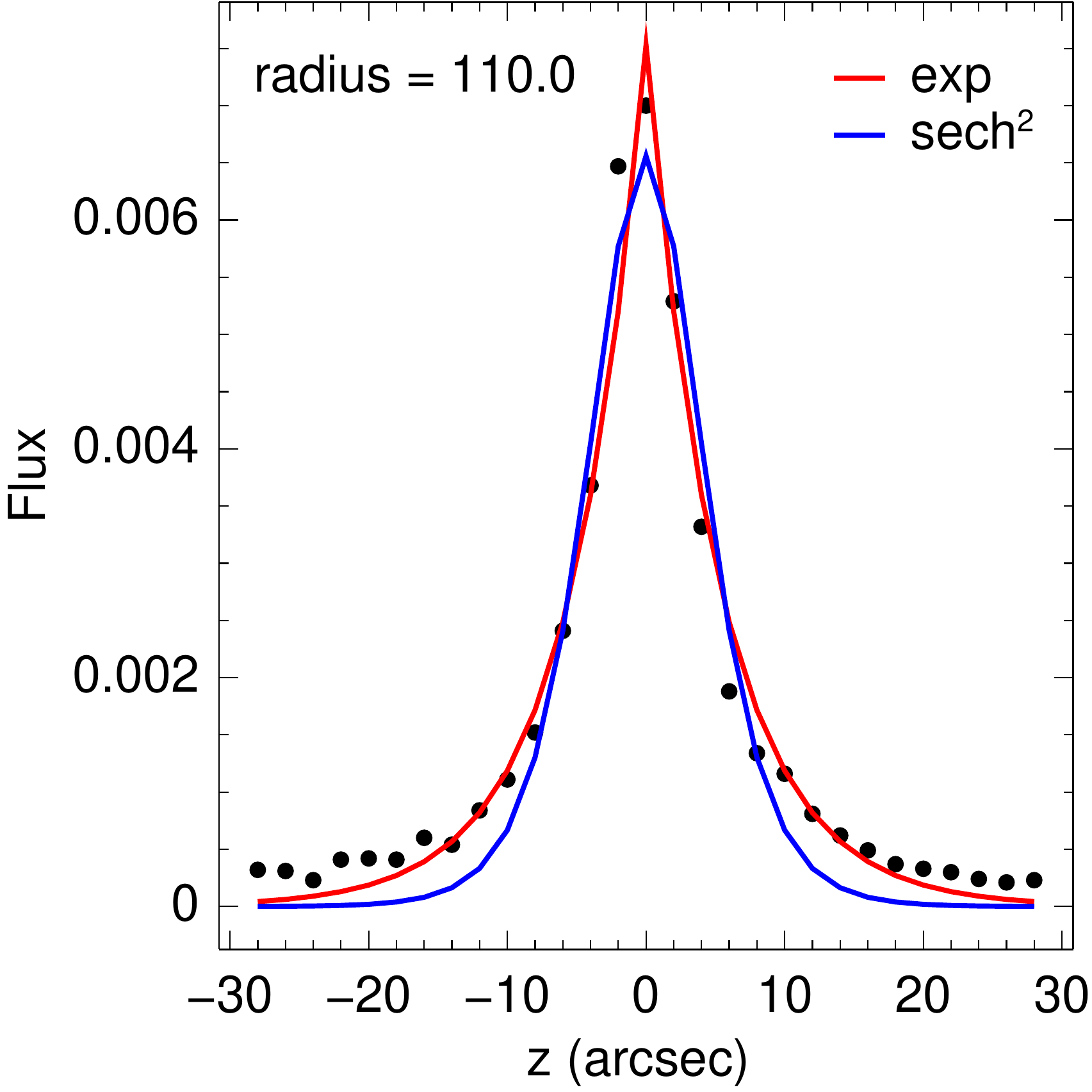}
\end{tabular}
\caption{Exponential (red line) and sech$^2$ (blue line) fitting in NGC 4013 to the vertical profiles (filled circles) of 3.6 \um\ at radius 40\ac\ (left), 80\ac\ (middle), and 110\ac\ (right).  
\label{stellarfit}}
\end{center}
\end{figure*}

The stellar disc thickness is measured by fitting an exponential function to the vertical profile of the 3.6 \um\ map at each radius.
 In order to obtain the vertical profiles at each radius (not the $x$-offset), we ran the task RADPROF at  the vertical distance of $z$ from -30\ac\ to 30\ac\ in a step of 2\ac.  
The radial density profiles at each $z$ obtained from RADPROF are finally used to build up the vertical distribution at each radius (see Fig. \ref{stellarfit}). In the previous work for NGC 891 by  \citet{2011AJ....141...48Y}, we fitted a sech$^2 (z/h_*)$ function to the vertical profiles, however, we use an exp(-$z/h_*$) function to fit the profiles in this work since the exponential fit (red) is better than the sech$^2$ fit (blue) as shown in Fig. \ref{stellarfit}. 
The measured exponential scale heights ($h_*$) from the fits are plotted for blue-shifted side (blue line), red-shifted side (red line), and the average value (filled circles) in the right panel of Fig. \ref{scaleh}. The average values are used to obtain the best fit (dashed line), which is utilized as the stellar scale height in this work.  The best fit function is shown on the bottom of the figure. The central regions ($\pm$ 40\ac) including the bulge part are excluded from the measurement. The shaded region around the dashed line shows uncertainty of the linear best-fit. We also used the sech$^2$ fitting to measure the stellar scale height for comparing with the exponential fitting, but the difference between the scale heights by exponential and sech$^2$ fittings is not noticeable. The stellar scale heights of the less edge-on galaxies were obtained by modeling the 3.6 \um\ data in \citet{2014AJ....148..127Y}. 

\subsection{SFR Disc Thickness of the Sample: NGC 891, 4013, 4157, 4565, and 5907} 
\begin{figure*}
\begin{center}
\begin{tabular}{c@{\hspace{0.1in}}c@{\hspace{0.1in}}}
\includegraphics[width=0.45\textwidth]{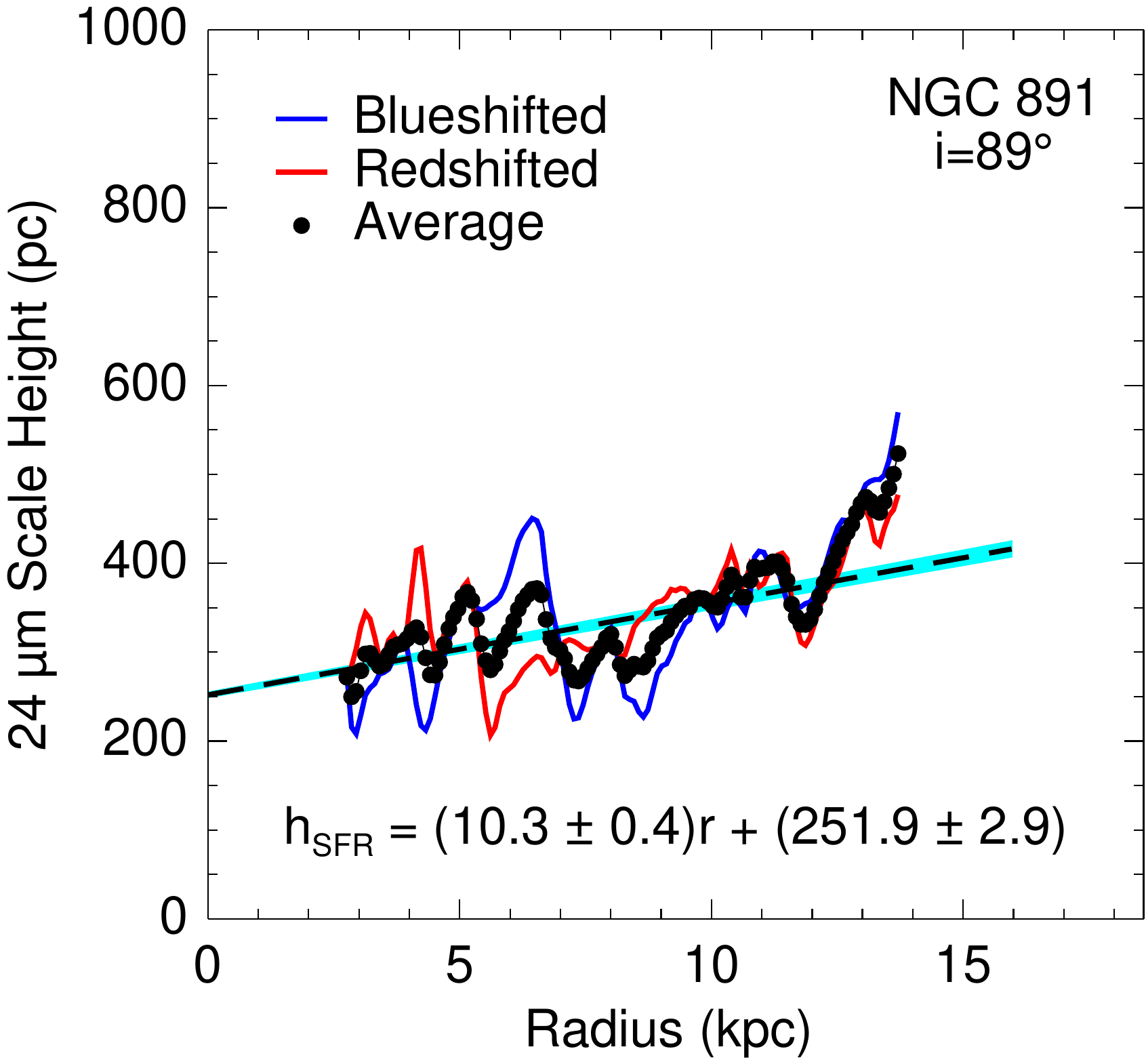}&
\includegraphics[width=0.45\textwidth]{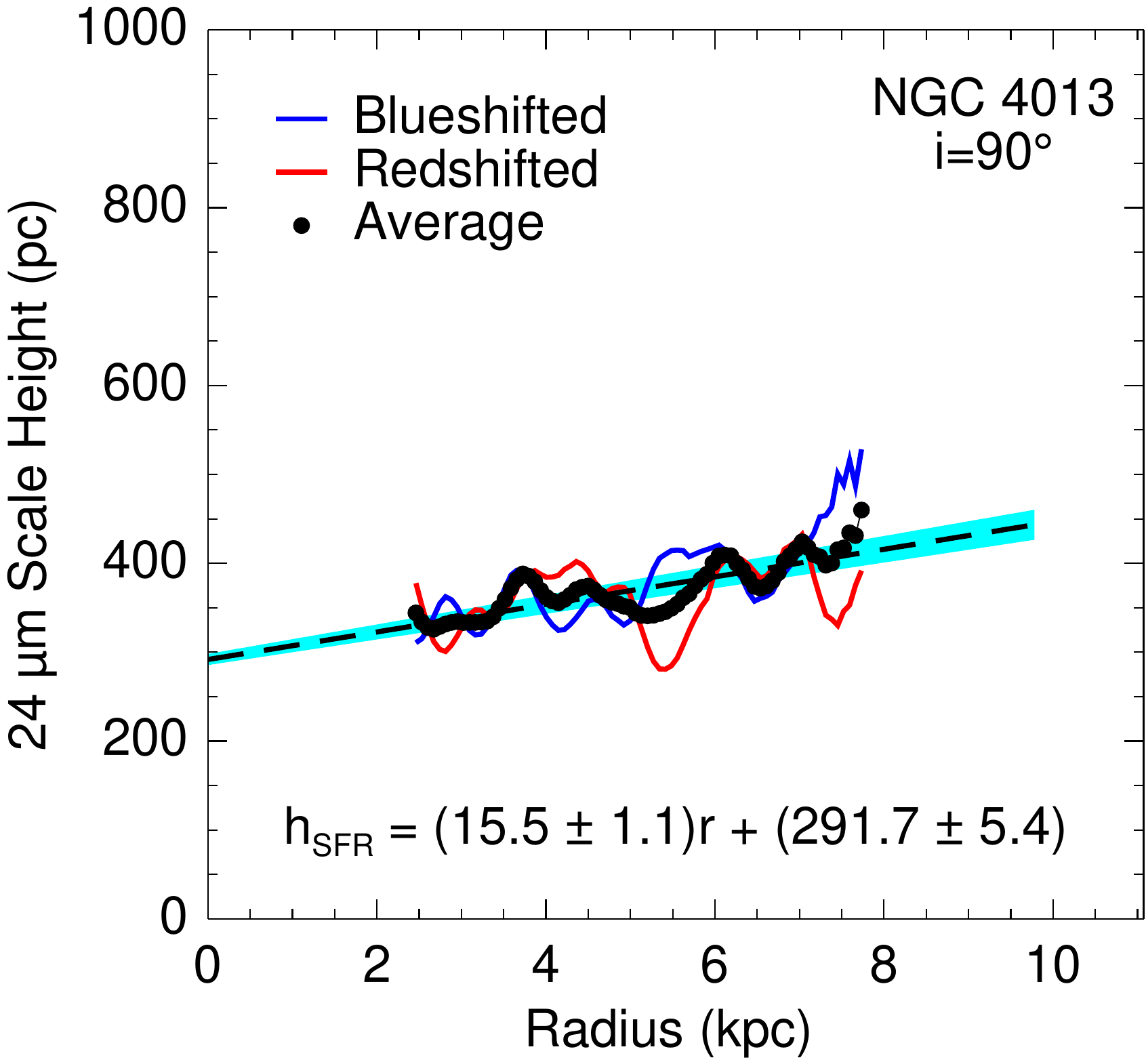}
\end{tabular}
\begin{tabular}{c@{\hspace{0.1in}}c@{\hspace{0.1in}}c@{\hspace{0.1in}}}
\includegraphics[width=0.32\textwidth]{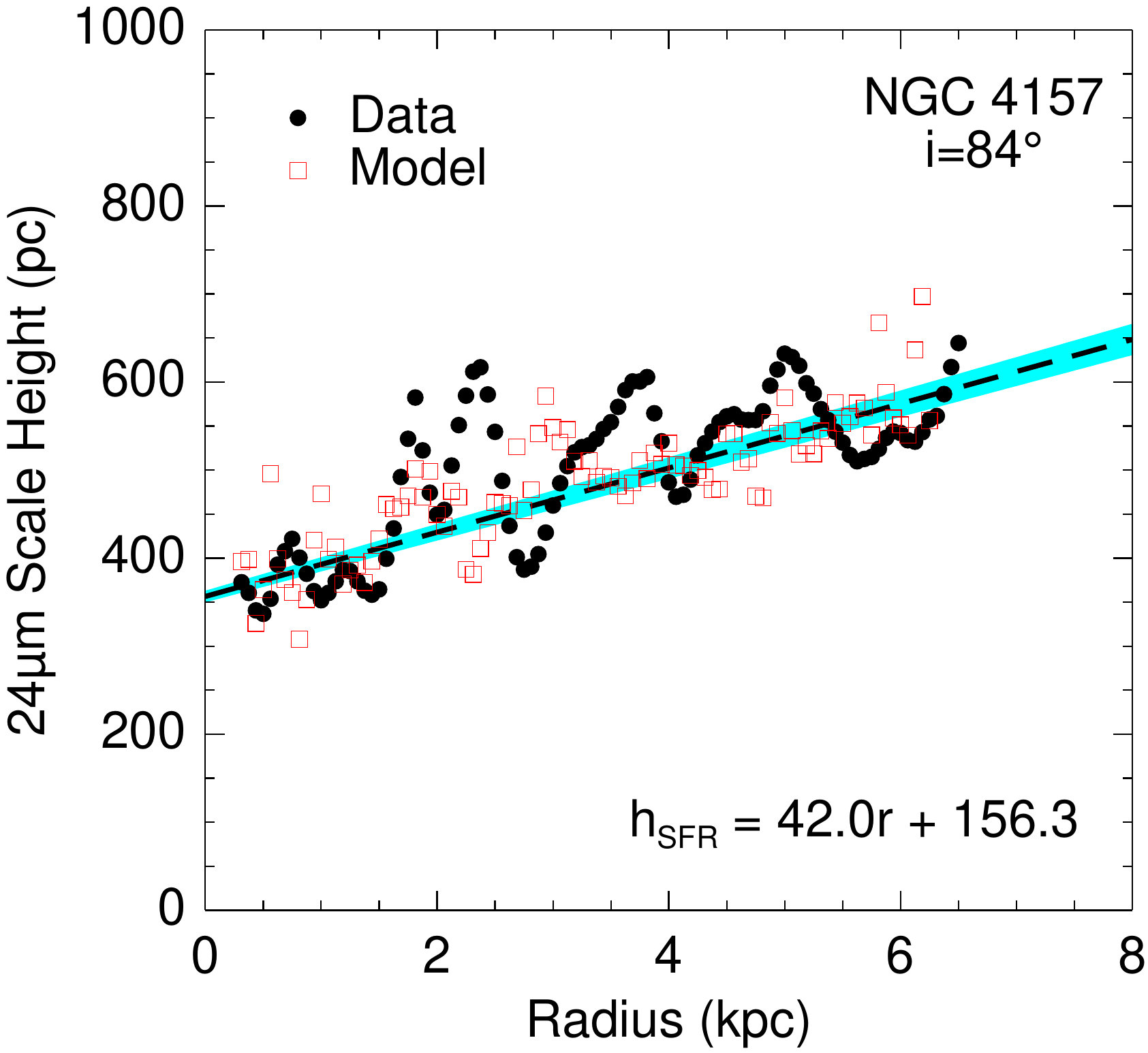}&
\includegraphics[width=0.32\textwidth]{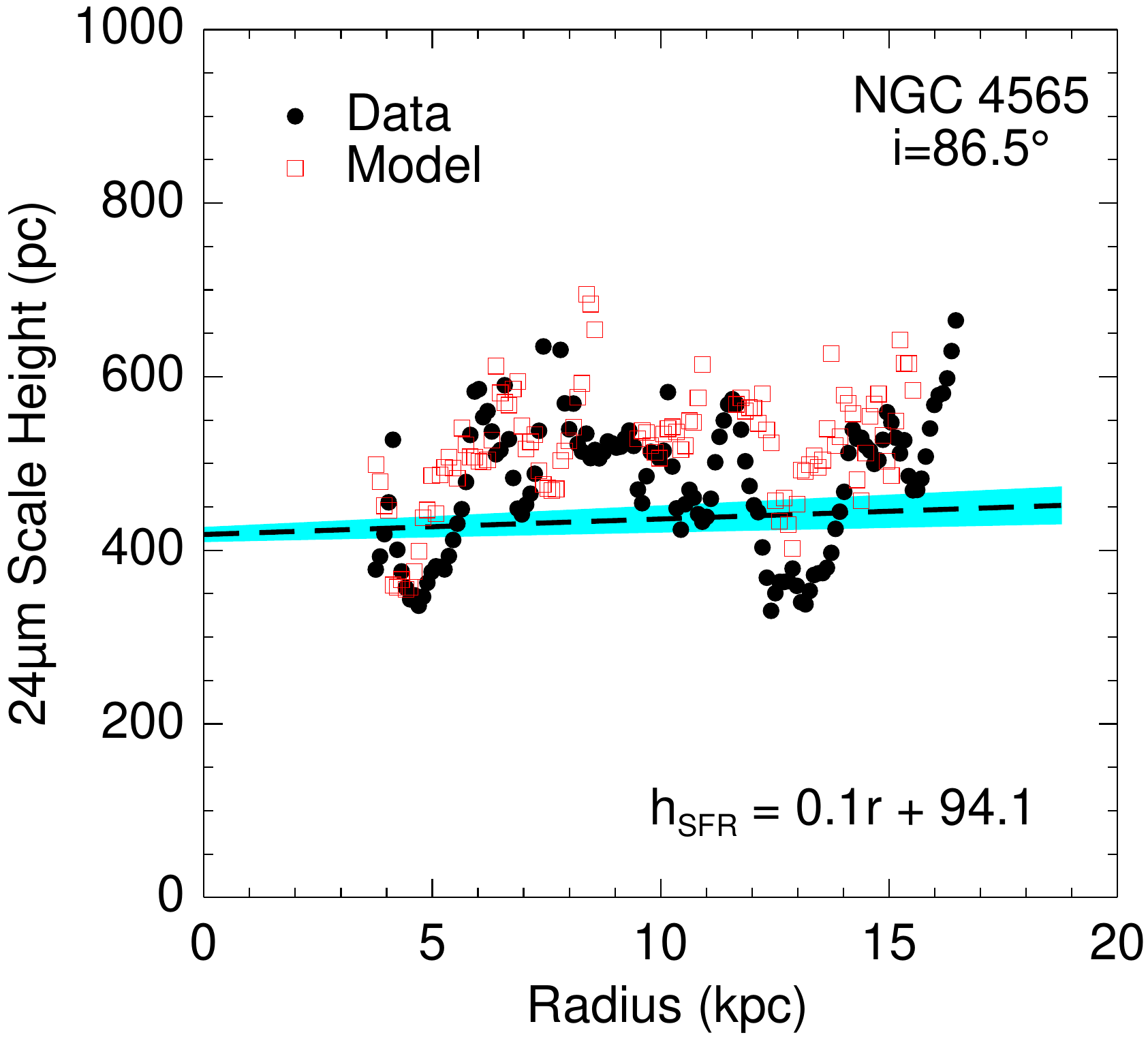}&
\includegraphics[width=0.32\textwidth]{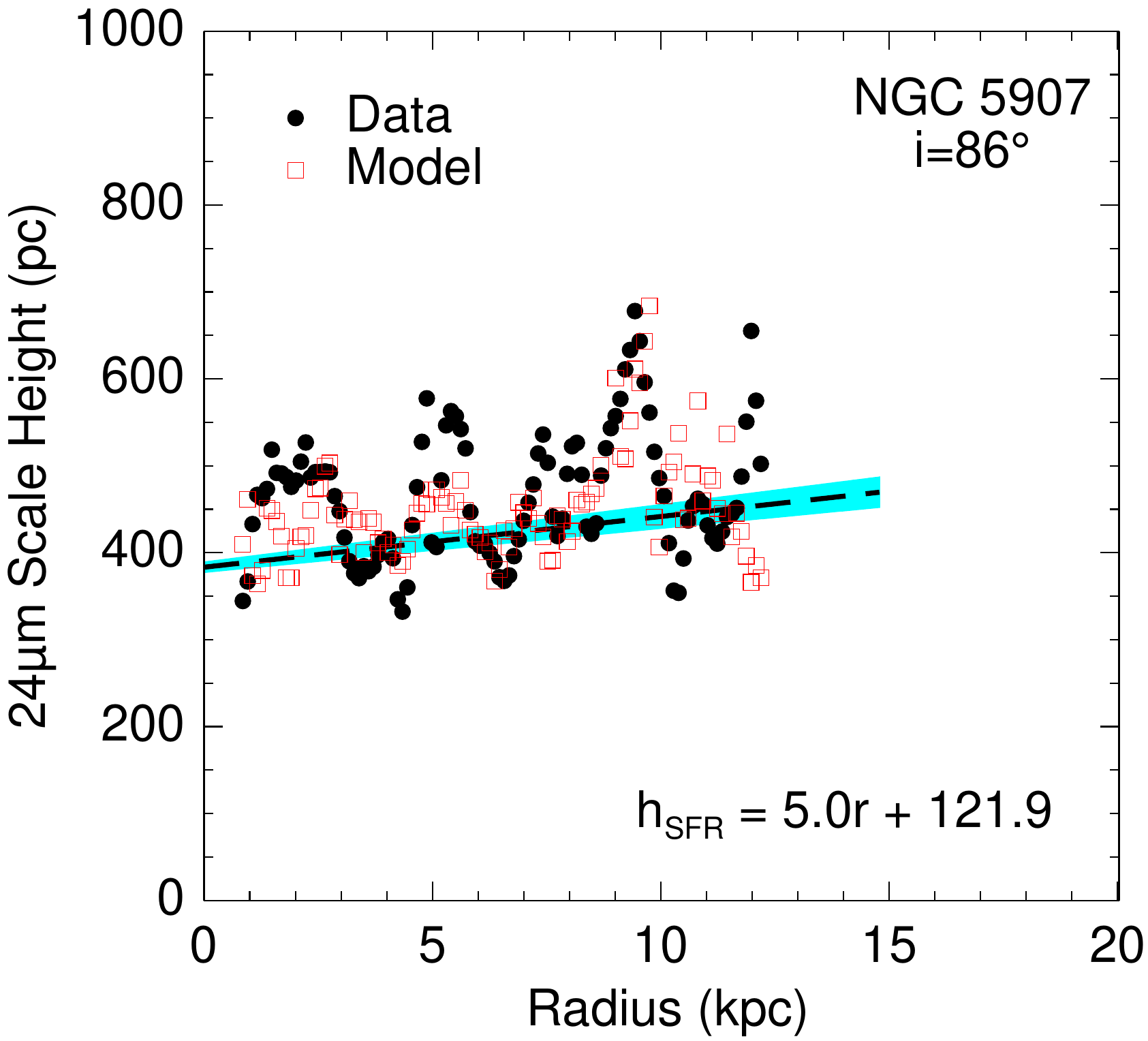}
\end{tabular}
\caption{Top panels: Scale heights of 24 \um\ (SFR) for NGC 891 and NGC 4013. The filled circles are average values of blue-shifted (blue line) and redshifted (red line) sides. The dashed line is the best-fit to the average data points. Each best-fit function is represented in the bottom of panels. The shaded regions around the best-fit line show the uncertainty of the fit. Bottom panels: Projected scale heights measured from the 24 \um\ maps (filled circles) and the best model (open squares) for the less edge-on galaxies of NGC 4157, NGC 4565, and NGC 5907. The dashed line is the best-fit to the data. The intrinsic scale height used to generate the best model is indicated as a function in the bottom of each panel. 
\label{sfr_sh}}
\end{center}
\end{figure*}

In addition to the gas and stellar scale heights, we measured the scale height of the 24 \um\ data for the SFR volume density to investigate the volumetric SFL. The same  analysis (based on RADPROF) that we used to obtain the stellar scale height is applied to the 24 \um\ (SFR) scale height for almost edge-on galaxies. 
Fig. \ref{sfr_sh} (top panels) shows the  SFR scale heights for NGC 891 (left) and NGC 4013 (right). The best fit  to the average scale height of the blue-shifted and red-shifted scale heights will be used when deriving the SFR volume density; a function of the best-fit is indicated in the bottom. 
For the less edge-on galaxies,  we used the modeling method applied to the stellar scale heights of NGC 4157, NGC 4565, and NGC 5907 \citep{2014AJ....148..127Y}. In summary, we generated a 24 \um\ model galaxy at the inclination of the galaxies (Table \ref{galprop}) with a guess scale height using the GIPSY task GALMOD. In this model, the derived 24 \um\ surface density profile is used.  Then, we measured the line-of-sight projected scale height of the generated model galaxy using the same method employed for the stellar and SFR scale heights of NGC 891 and NGC 4013. The projected scale height of the model was compared with that of the 24 \um\ data. If the scale heights are not consistent each other, then we tried a new scale height for a next model.  Finally, we found the best model that reproduces the scale height of the data within several trials. In Fig.  \ref{sfr_sh} (bottom panels), we compared the projected scale heights of the data (filled circles) and the best model (open squares) for the less edge-on galaxies; they agree reasonably well each other.   The input scale height used to generate the best model is indicated as a function in the bottom of each panel and will be used to estimate the SFR volume density. The input scale heights are quite lower than the projected scale heights of the data, but their slopes are not much different from each other. In the case of NGC 891 and NGC 4013, the input scale heights, which are obtained from the data, are in good agreement with the measured projected scale heights of their model. The scale height of NGC 4565 is almost constant like its CO scale height, while the scale heights of other galaxies increase with radius, slowly or moderately. Note that the measurements of the SFR scale heights are pretty uncertain. First, the MIPS 24 \um\ image is at resolution of 5.9\ac\ ($\sim 300$ pc, on average of the sample), so the emission is not well-resolved vertically. Second, the dust may be heated by young stars near the midplane, whose UV radiation escapes to large heights above the disk.

\begin{table}
\begin{center}
\caption{Galaxy Properties}
\label{galprop}
\begin{tabular}{lccccc}
\hline
Galaxy  &Distance &PA& Inclination &Physical Scale\\
&(Mpc)&(\degr)&(\degr)&(pc arcsec$^{-1}$)\\
&(1)&(2)&(3)&(4)\\
\hline
NGC 891&9.5&23&89.0&46\\
NGC 4013&14.5&65&90.0&70\\
NGC 4157&12.9&63&84.0&62 \\
NGC 4565&9.7&135&86.5&47\\
NGC 5907&11.0&115&86.0&53\\
\hline
\end{tabular}\\
(1) Distance from the literature: \citet{1981A&A....95..116V} for NGC 891, \cite{1999AJ....117.2102I} for NGC 4157, \cite{2005A&A...432..475D} for NGC 4565, and  \cite{2006A&A...459..703J} for NGC 5907. The distance of NGC 4013 is based on the velocity with respect to the 3K CMB  \citep{1996ApJ...473..576F} and $H_0$ = 73 \kms\ Mpc$^{-1}$. (2) Position angle obtained from the 3.6 \um\ image using the MIRIAD task IMFIT. 
(3) Inclination from the literature: \citet{2007AJ....134.1019O} for NGC 891,  \citet{2015ApJ...808..153Z} for NGC 4013, \citet{2014AJ....148..127Y} for NGC 4157, 4565, and 5907. (4) Physical scale in units of pc per arcsec. 
\end{center}
\end{table}

\subsection{Resolved H$\small\textsc{I}$ Disc Thickness of NGC 4157 and NGC 5907}

\begin{figure}
\includegraphics[width=0.45\textwidth]{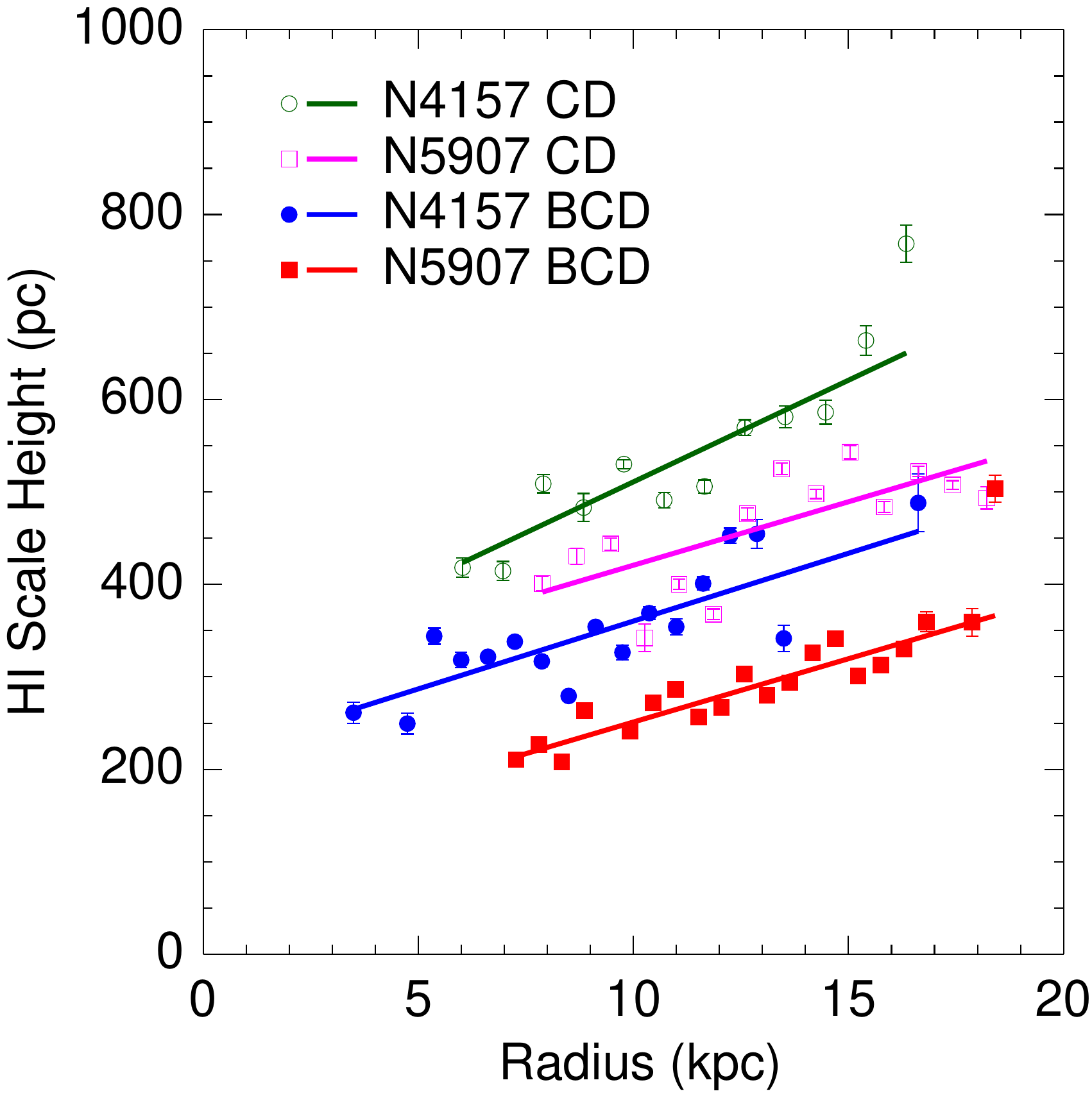}
\caption{\HI\ scale heights of NGC 4157 (circles) and NGC 5907 (squares) as a function of radius. The open symbols represent the old scale heights from the VLA C plus D arrays and the solid symbols represent the new scale heights from the combined maps of B, C, and D arrays. 
\label{scaleh_h1}}
\end{figure}

Using the new \HI\ maps (B, C, and D arrays) of NGC 4157 and 5907, we reproduced the \HI\ scale heights and compared them to the old scale heights (from our previous maps of C plus D arrays) presented in \citet{2014AJ....148..127Y}. Since NGC 4157 and 5907 are less edge-on galaxies, Olling's method \citep{1996AJ....112..457O} was employed to measure the scale heights, considering the projection effects as we have done in the previous study. The Olling method, based on observational data and a face-on view of model galaxy, calculates an inclination and use it to measure the disk thickness (see the Appendix of \citealt{2014AJ....148..127Y} for more details).
The comparisons between the old (CD) and new (BCD) scale heights are shown in Fig. \ref{scaleh_h1} as circles for NGC 4157 and squares for NGC 5907. The most noticeable difference between the scale heights is that the BCD scale heights are lower than the CD scale heights by a factor of $\sim$1.5.  It seems that the difference may be caused by the lower resolution ($\sim$ 15\ac) of the CD data that do not well resolve the disc thickness. We checked that the BCD data convolved to the CD resolution reproduce similar scale heights with the CD data. 
The differences between the gradients of CD and BCD are within a factor of 1.5.

\subsection{Vertical Velocity Dispersions of NGC 4013}
\label{sec_vd}
\begin{figure}
\includegraphics[width=0.45\textwidth]{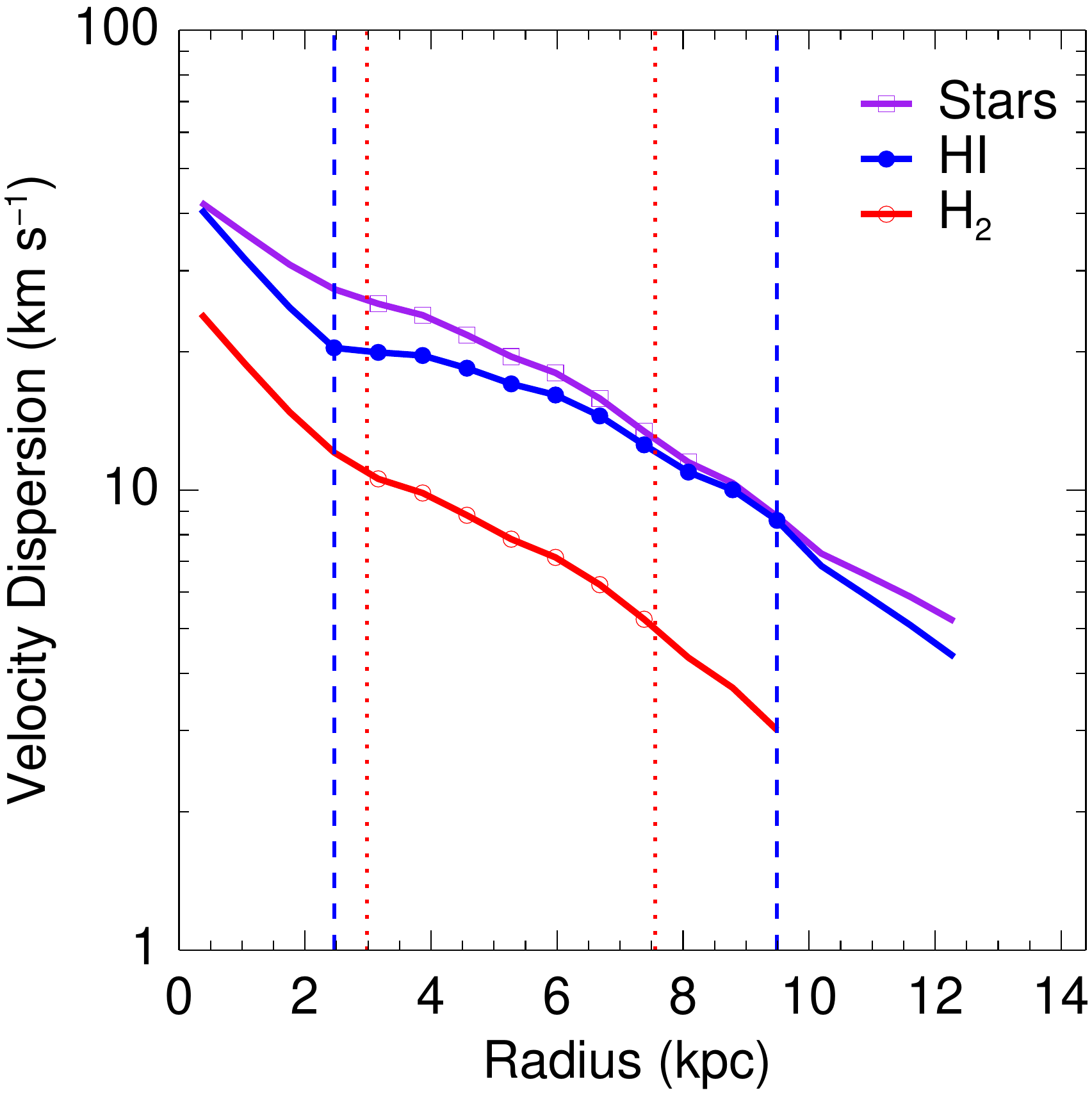}
\caption{Inferred vertical velocity dispersions for CO (red open circle), \HI\ (blue filled circle), and stars (purple open square) assuming hydrostatic equilibrium. The vertical lines enclose the regions where the data points of CO (dotted) and \HI\ (dashed) scale heights are available as shown in Fig. \ref{scaleh}. 
\label{vdisp}}
\end{figure}

The direct measurement of the vertical velocity dispersion is not possible for an edge-on galaxy, since the velocity direction is perpendicular to the line of sight. Instead, we inferred the dispersion by solving the Poisson equation assuming hydrostatic equilibrium given by \citet{2002A&A...394...89N}:
\begin{equation}
\sigma_i^2 = \frac{4\pi G\rho_{\rm 0,tot}\rho_{0i}}{- (d^2\rho_i/dz^2)_{z=0}},
\label{poisson}
\end{equation}
where  $\rho_i = \rho_{0i}$ and $d\rho_i/dz = 0$ at $z=0$ (midplane). The subscript $i$ is for the gas (H$_2$ and \HI) or stars ($_*$). The total midplane volume density ($\rho_{\rm 0,tot}$) is a summation of the total gas  and stars. We ignore the dark matter in the total density since it is not an important factor for the vertical distribution although the flat rotation curve shows the dark matter contribution .  The scale heights  used for  the total density profiles are up to $\sim$100 pc (H$_2$), $\sim$300 pc (\HI), and $\sim$400 pc (stars), where the dark matter density and its gradient along $z$ are negligible. \citet{2002A&A...394...89N} also found that a fraction of the dark matter is less than 6\% within $z \leq 1$kpc at a radius of 12 kpc where the \HI\ and H$_2$ scale heights are largest, suggesting that the dark matter is not important factor in the total density for the scale height. 
We also verified that the total density including the dark matter does not affect to results. For this reason and to be consistent with our previous works (\citealt{2011AJ....141...48Y}; \citeyear{2014AJ....148..127Y}), we assume that the dark matter density is insignificant compared to the stellar density in the disc. On the other hand, \citet{Bacchini_2019} showed that the dark matter has a strong impact on \HI\ flaring by including or excluding the dark matter halo in the potential of NGC 2403. However, this is tested only in NGC 2403, which is less massive than our sample of galaxies. In addition, the dark matter does not play a crucial role in changing the scale height within the optical radius ($\sim$9 kpc) of NGC 2403. 
Since our study is focused on the inner region of the galaxies within the optical radius where the 24 \um\ emission (SFR tracer) is available,  the dark matter would not be important in this study.  
We obtained the volume densities by integrating a  Gaussian distribution for gas and an exponential function for stars along the vertical direction: 
\begin{equation}
\rho_{\rm 0g} = \frac{\siggas}{h_{\rm g}\sqrt{2\pi}},  \qquad 
\rho_{0*} =  \frac{\sigstar}{2 h_*}.
\label{eq_vol}
\end{equation}
The numerical solutions to equation (\ref{poisson}) for gas and stars are:
\begin{equation}
\sigma_{\rm g} = \sqrt{4\pi G h_{\rm g}^2 \rho_{\rm 0,tot}}, \qquad 
\sigma_* = \sqrt{2\pi G h_*^2 \rho_{\rm 0,tot}}.
\label{eq_vdisp}
\end{equation}
Fig. \ref{vdisp} shows  the inferred vertical velocity dispersions from the measured disc thicknesses and volume densities. 
The vertical lines show the regions where the CO and \HI\ data points are available in the scale heights of Fig. \ref{scaleh}: red dotted lines for CO and blue dashed line for \HI. 
The velocity dispersions outside the regions are derived using the scale heights at the nearest  lines. 
All the dispersions decrease with radius and the values are comparable to those of other galaxies we found in our previous studies ( \citealt{2011AJ....141...48Y}, \citeyear{2014AJ....148..127Y}) and other studies (e.g., \citealt{Boomsma_2008}; \citealt{2009AJ....137.4424T}). We also derived the velocity dispersions using the total density including the dark matter. The differences between the velocity dispersions with and without the dark matter are a factor of 0.8$\sim$1.1 for \HI\ and 1.1$\sim$1.2 for CO and stars. The velocity dispersions (Fig. \ref{vdisp}) will be used when we derive the interstellar gas pressure in the next section.

\subsection{Galaxy Modeling for NGC 4013, 4157, and 5907}
\begin{figure*}
\begin{center}
\begin{tabular}{c@{\hspace{0.1in}}c@{\hspace{0.1in}}}
\includegraphics[width=0.5\textwidth,angle=0]{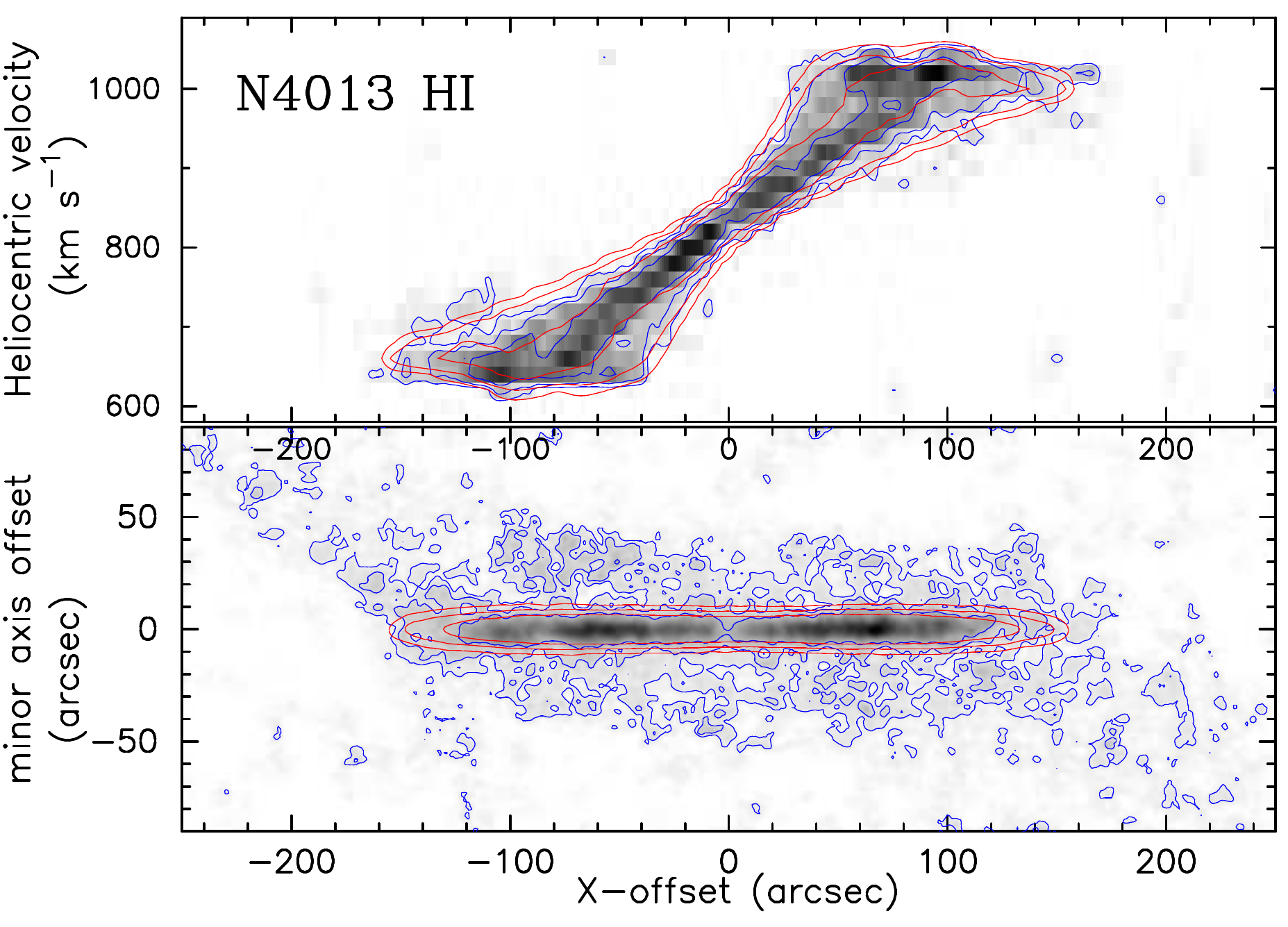}&
\includegraphics[width=0.5\textwidth,angle=0]{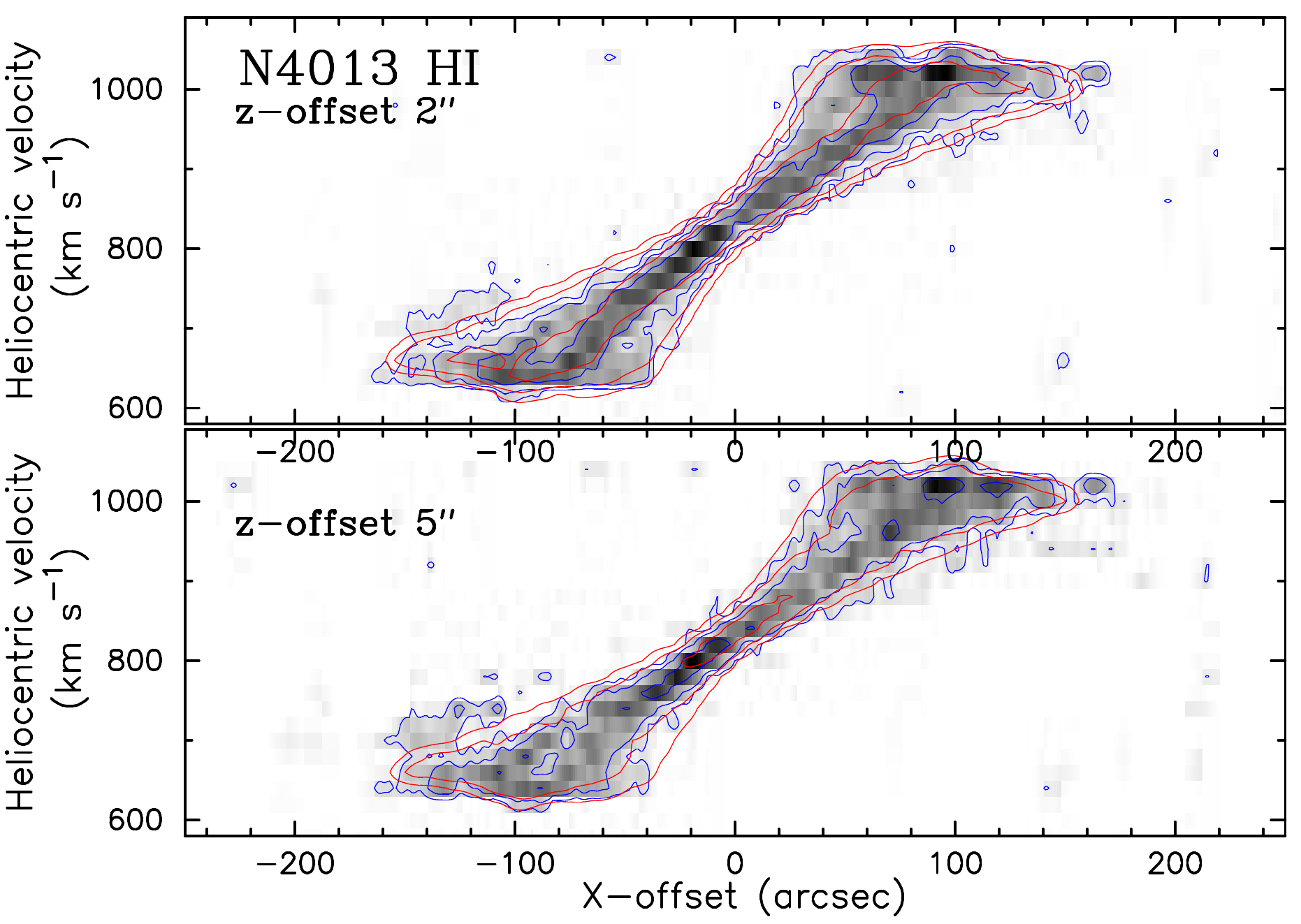}\\
\includegraphics[width=0.5\textwidth,angle=0]{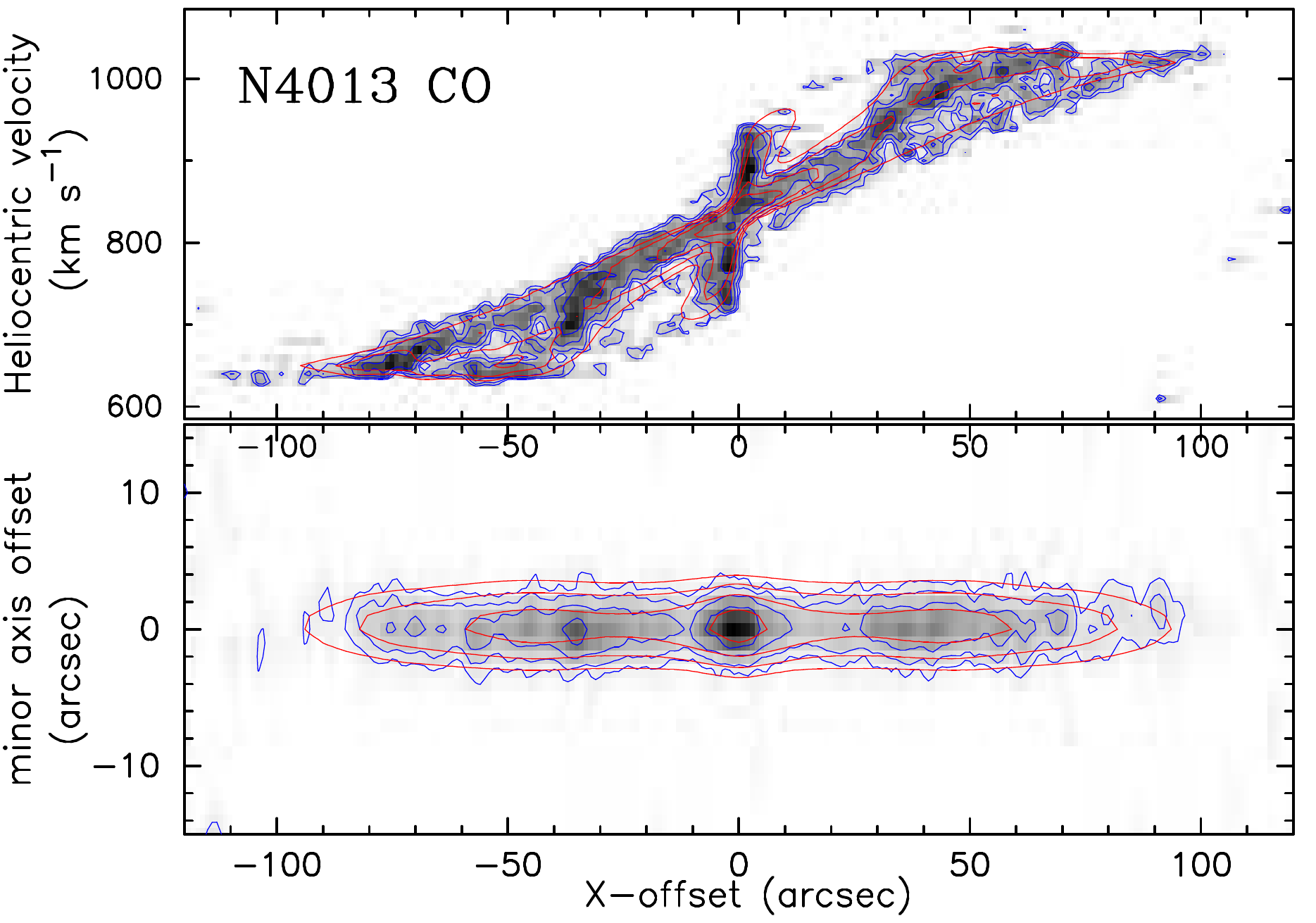}&
\includegraphics[width=0.5\textwidth,angle=0]{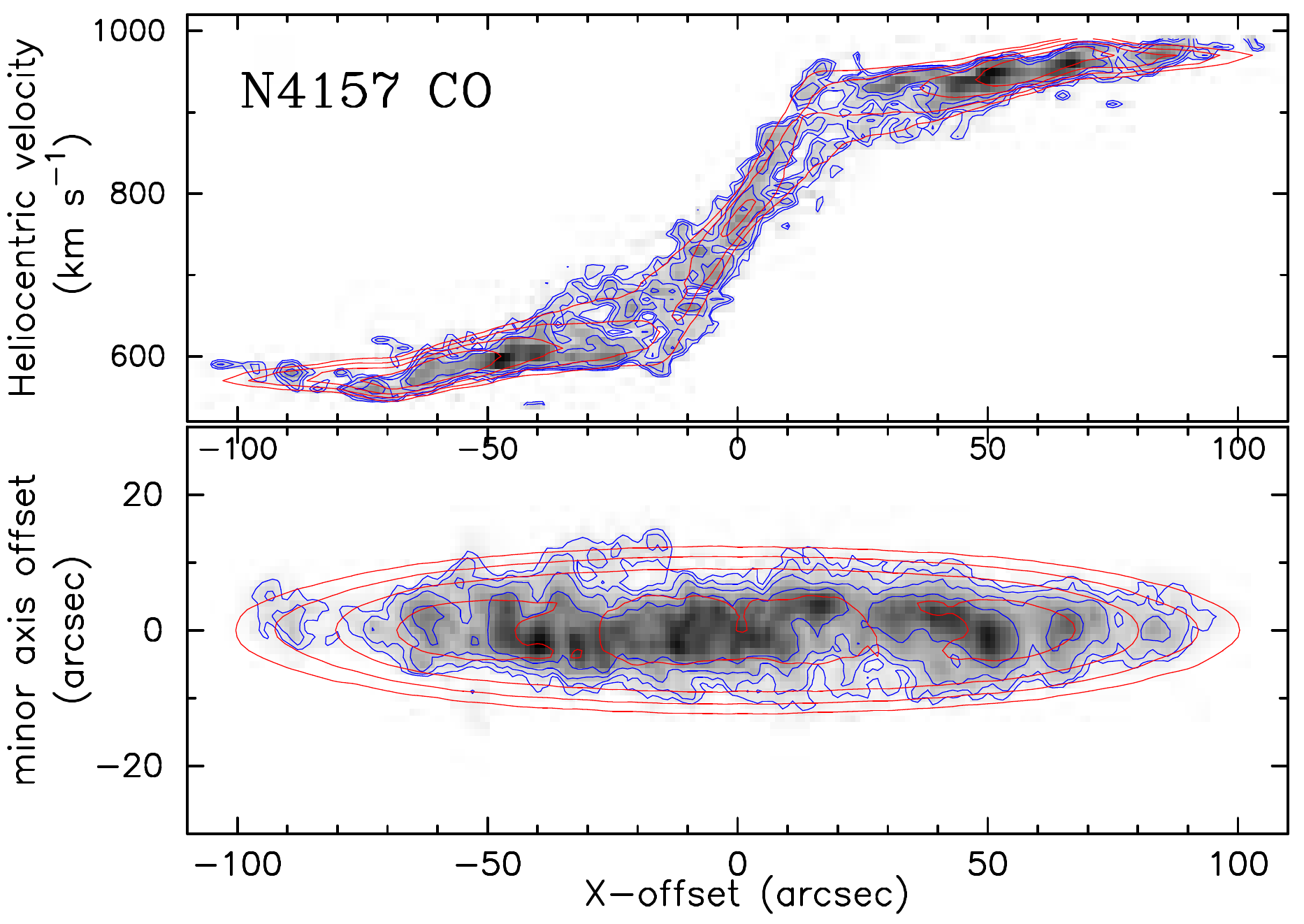}\\
\includegraphics[width=0.5\textwidth,angle=0]{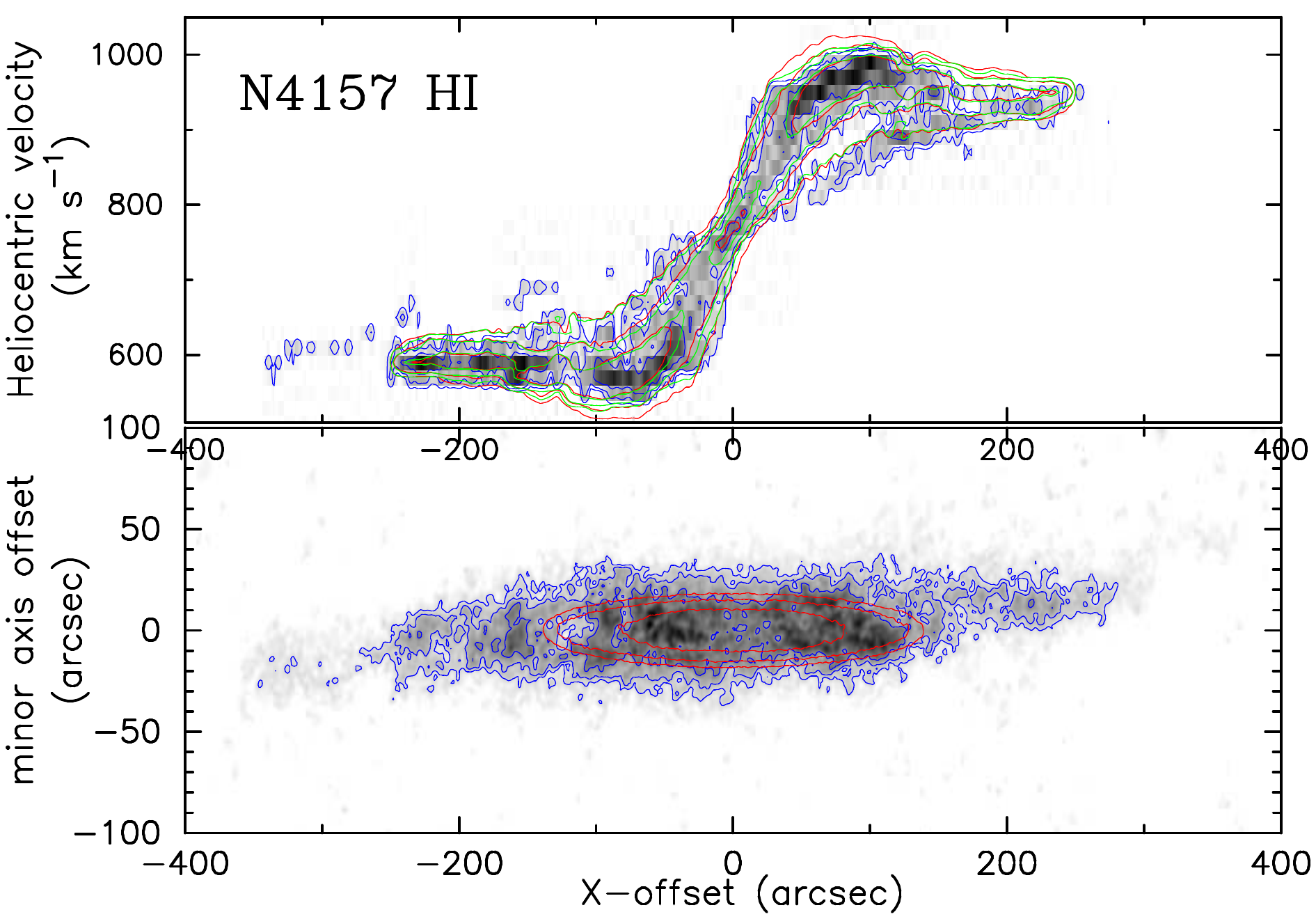}&
\includegraphics[width=0.5\textwidth,angle=0]{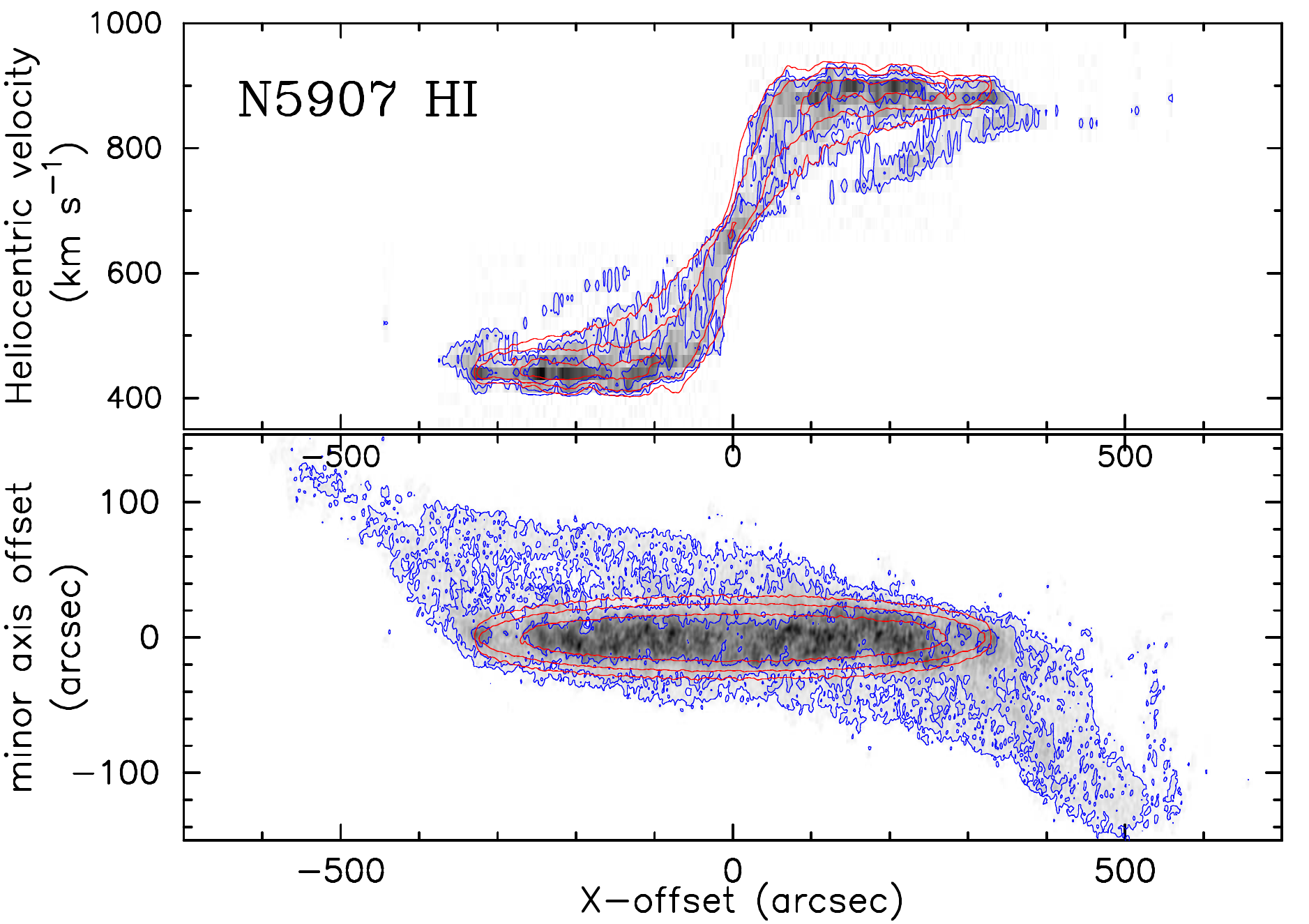}\\
\end{tabular}
\caption{\HI\ and CO position-velocity diagrams along the midplane and integrated intensity maps of NGC 4013, 4157, and 5907 with TiRiFiC model galaxy contours (red) overlaid on the maps (blue contours). The lowest level of contours in the maps is  $3\sigma$. The green contours on the  \HI\ p-v diagram of NGC 4157 show another \HI\ model generated by using a different vertical velocity dispersion (decreased by a factor of 2). Refer to the text for details.  
\label{fig_tirific}}
\end{center}
\end{figure*}

In order to examine whether a model galaxy with the derived scale height and velocity dispersion reproduces the \HI\ or CO emission, we built models of the galaxies  using the Tilted Ring Fitting Code (TiRiFiC) by \citet{2007A&A...468..731J}. We used the derived rotation curve (VROT), surface brightness (SBR), scale height (Z0), and  velocity dispersion (SDIS) with radius as main input parameters for the TiRiFiC modeling, allowing VROT to vary. We have checked that the input (data) and output (model) rotation curves are in good agreement. 
The radially dependent velocity dispersion SDIS is an optional parameter whose default is zero, while a global dispersion (CONDISP) is not dependent on radius and has no default value. We used CONDISP of 8 \kms\ for \HI\ and 4 \kms\ for CO. Comparisons between the models (red contours) and the data (blue contours) on the midplane p-v diagrams  and the integrated intensity maps are shown in Fig. \ref{fig_tirific}. 
In addition, we show more \HI\ p-v diagrams of NGC 4013 at different heights from the midplane in the figure (top-right).  
Overall, the model galaxies are in good agreement with the data in those maps except for the \HI\ thick disk with warps (in the intensity maps) that are excluded in the measurement of scale heights. In the case of NGC 4157 HI (bottom-left), the outermost red contour  in the p-v diagram seems to suggest that the derived vertical velocity dispersion in the region is overestimated. In order to check whether the velocity dispersion is overestimated, we generated a new model with a decreased velocity dispersion by a factor of 2. The p-v diagram of the new model  (green contours) is overlaid on the blue (data) and red (model) contours, indicating that the outermost blue contour on the terminal velocity appears to match with the green contour rather than with the red contour. 
Note that the CO model of NGC 5907 is presented  in Fig. 17 of \citet{2014AJ....148..127Y}.
Generally, these comparisons suggest that the derived scale height and vertical velocity dispersion agree reasonably well with the data. 

\begin{figure*}
\begin{center}
\begin{tabular}{c@{\hspace{0.1in}}c@{\hspace{0.1in}}}
\includegraphics[width=0.45\textwidth]{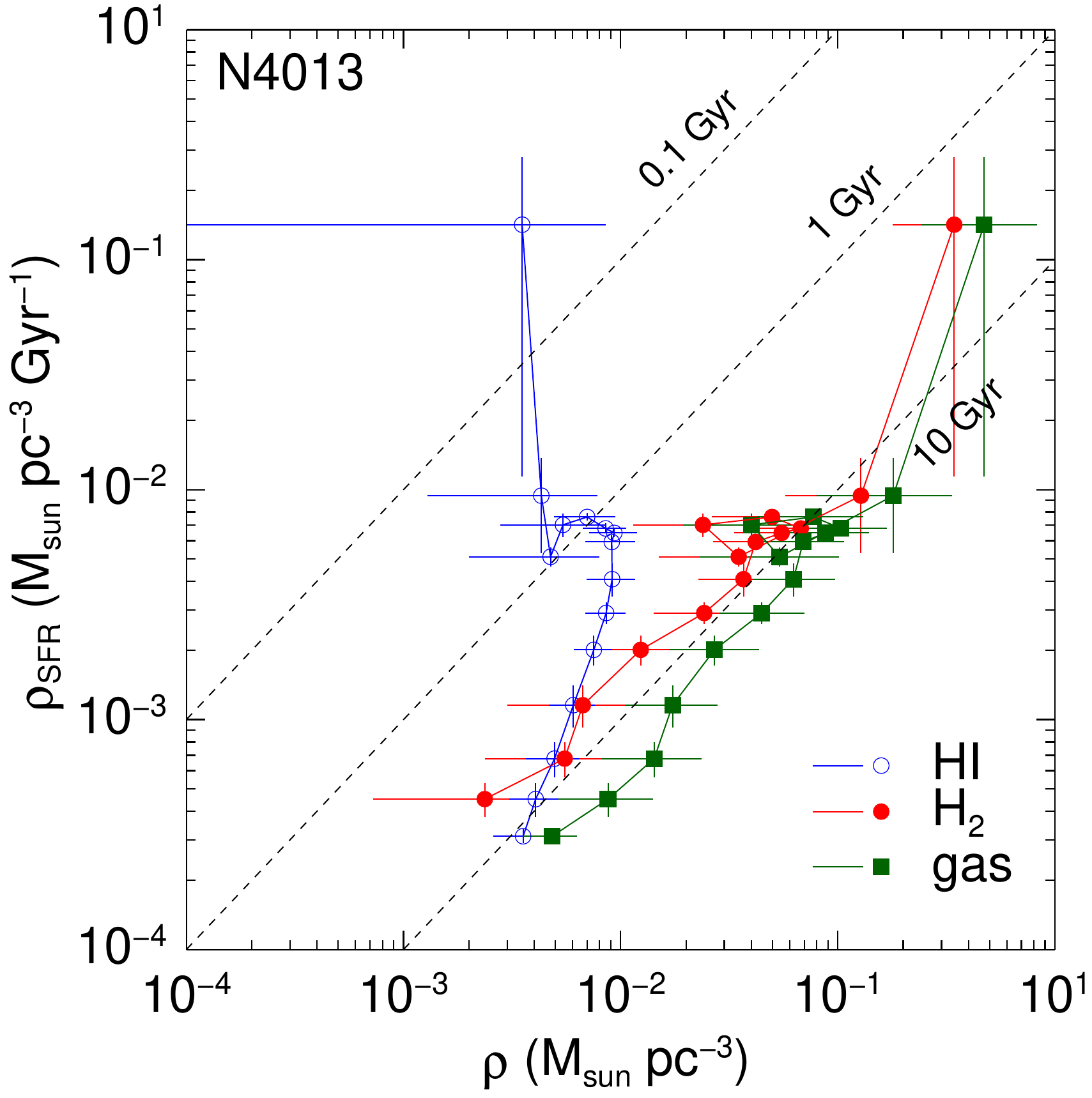}&
\includegraphics[width=0.45\textwidth]{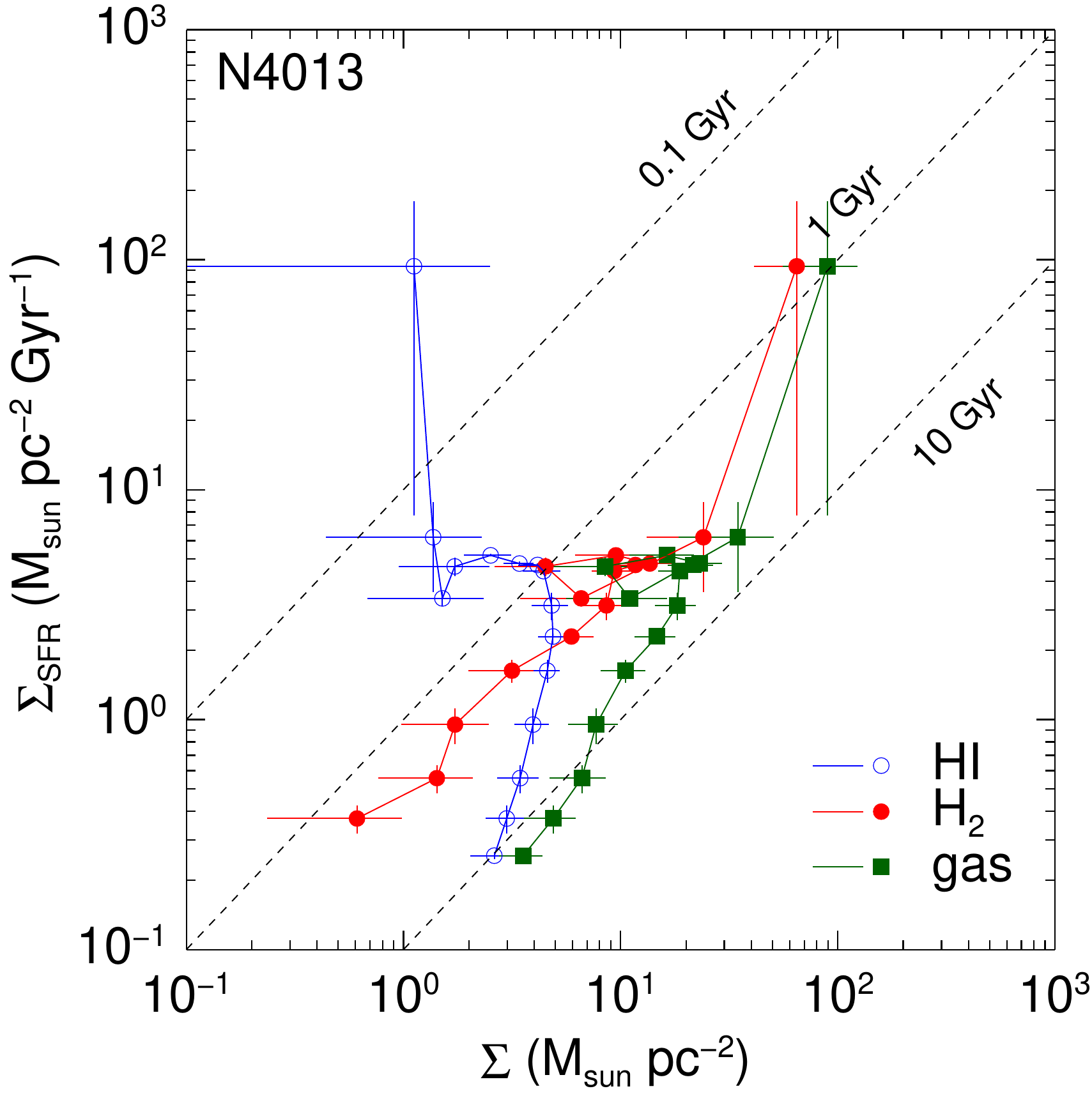}
\end{tabular}
\caption{Left: SFR volume density against H$_2$ (red filled circle), \HI\ (blue open circle), and total gas (green filled square) volume densities of NGC 4013. The vertical and horizontal error bars represent the uncertainties of the volume densities. Right: \sigsfr\ as a function of \sightwo, \sighi, and \siggas\ for NGC 4013.  The dashed lines show constant SFE and the gas depletion time (SFE$^{-1}$) is labelled for the lines. The vertical and horizontal error bars represent the uncertainties of the surface densities. 
\label{sfl}}
\end{center}
\end{figure*}

\begin{figure*}
\begin{center}
\begin{tabular}{c@{\hspace{0.1in}}c@{\hspace{0.1in}}} 
\Large Volume & \Large Surface\\
\includegraphics[width=0.4\textwidth]{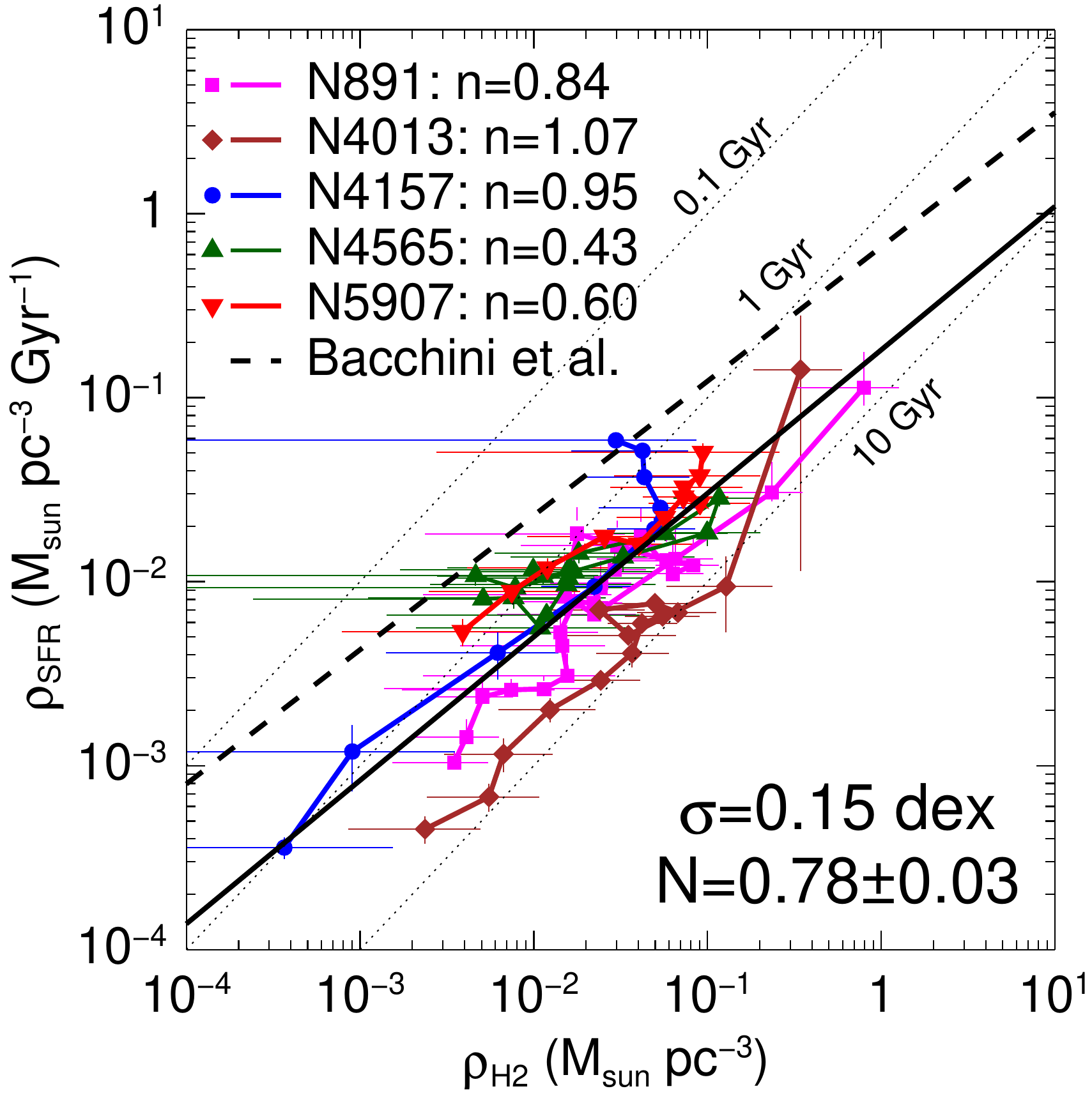}&
\includegraphics[width=0.4\textwidth]{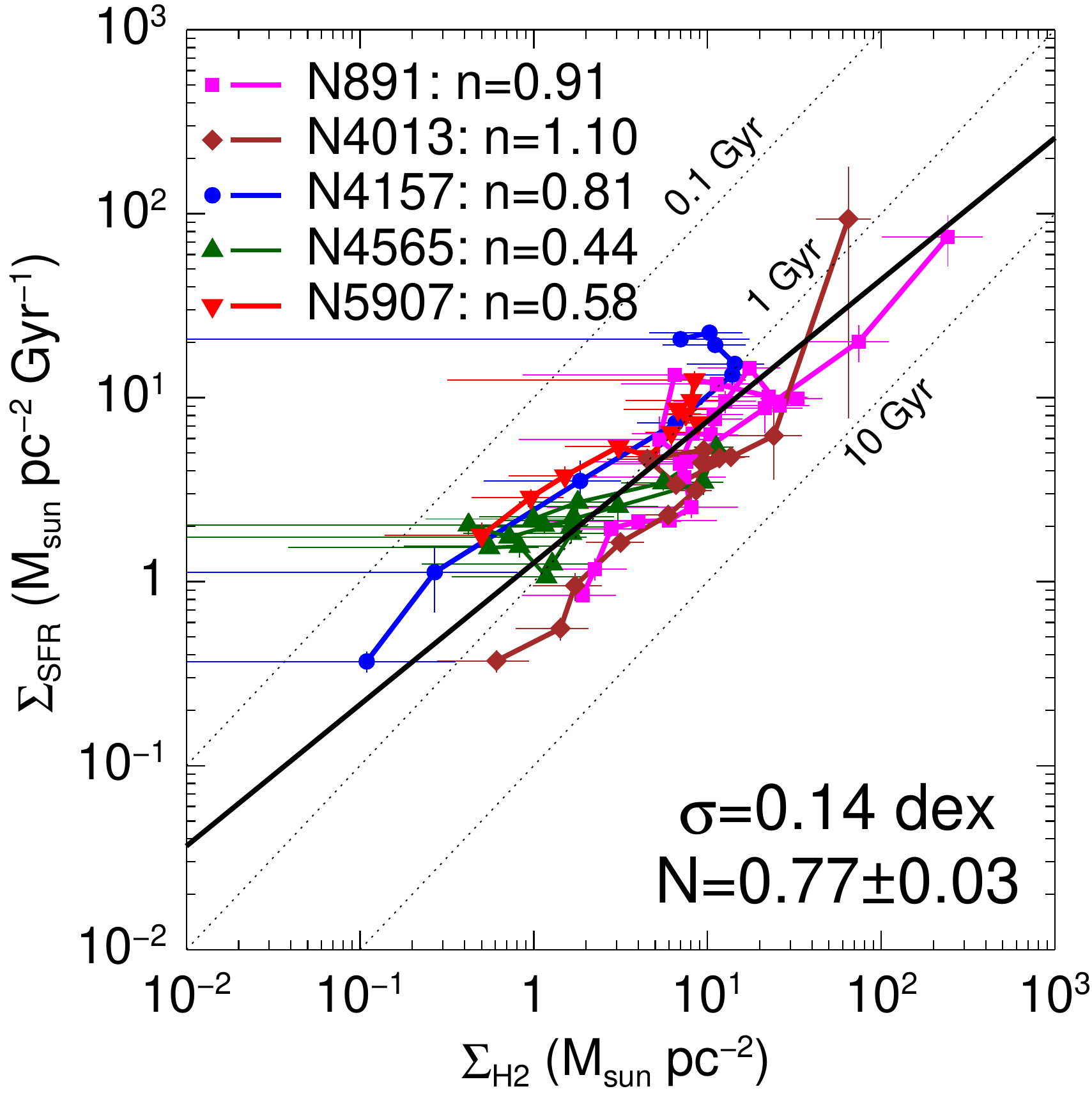}\\
\includegraphics[width=0.4\textwidth]{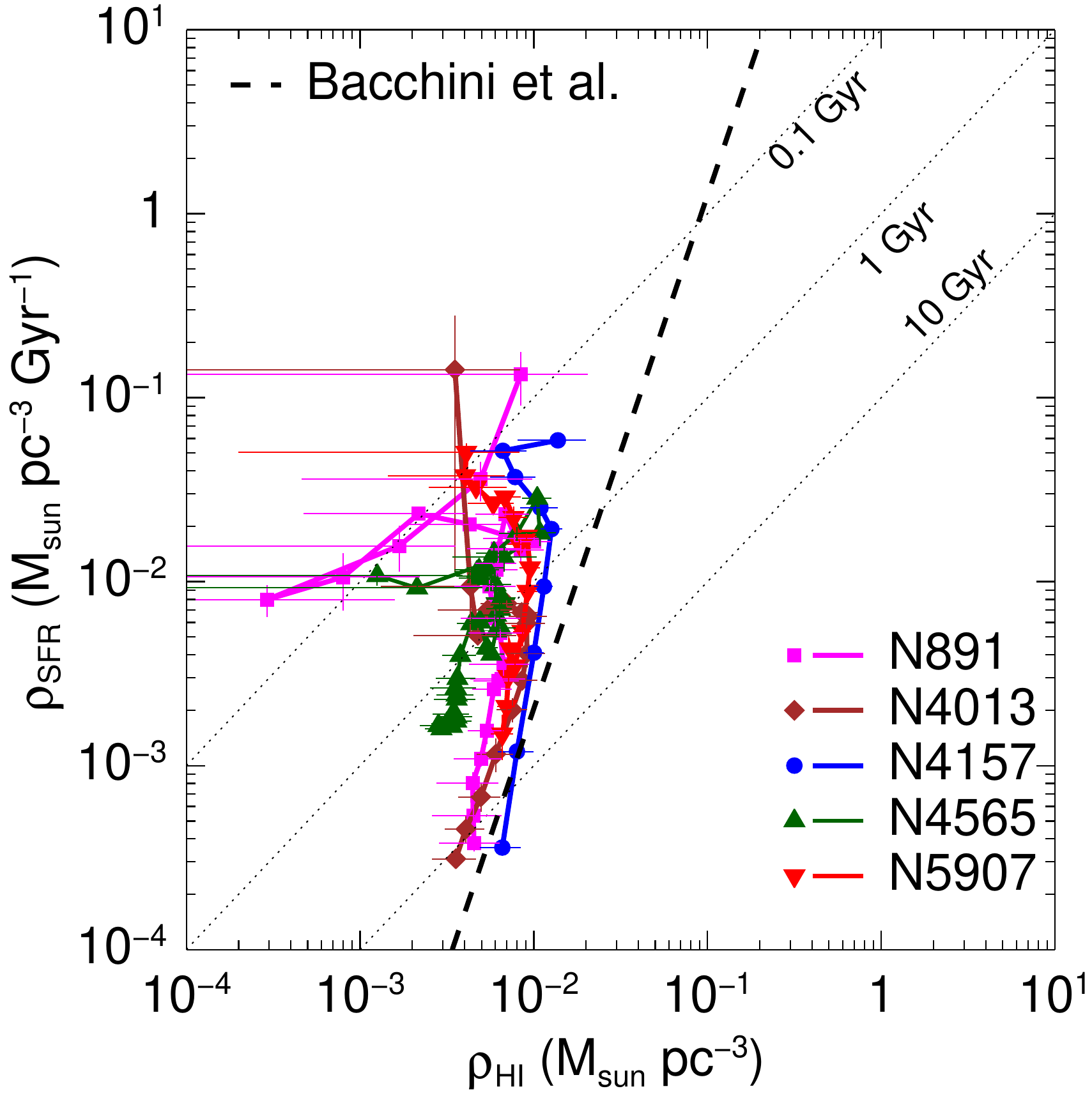}&
\includegraphics[width=0.4\textwidth]{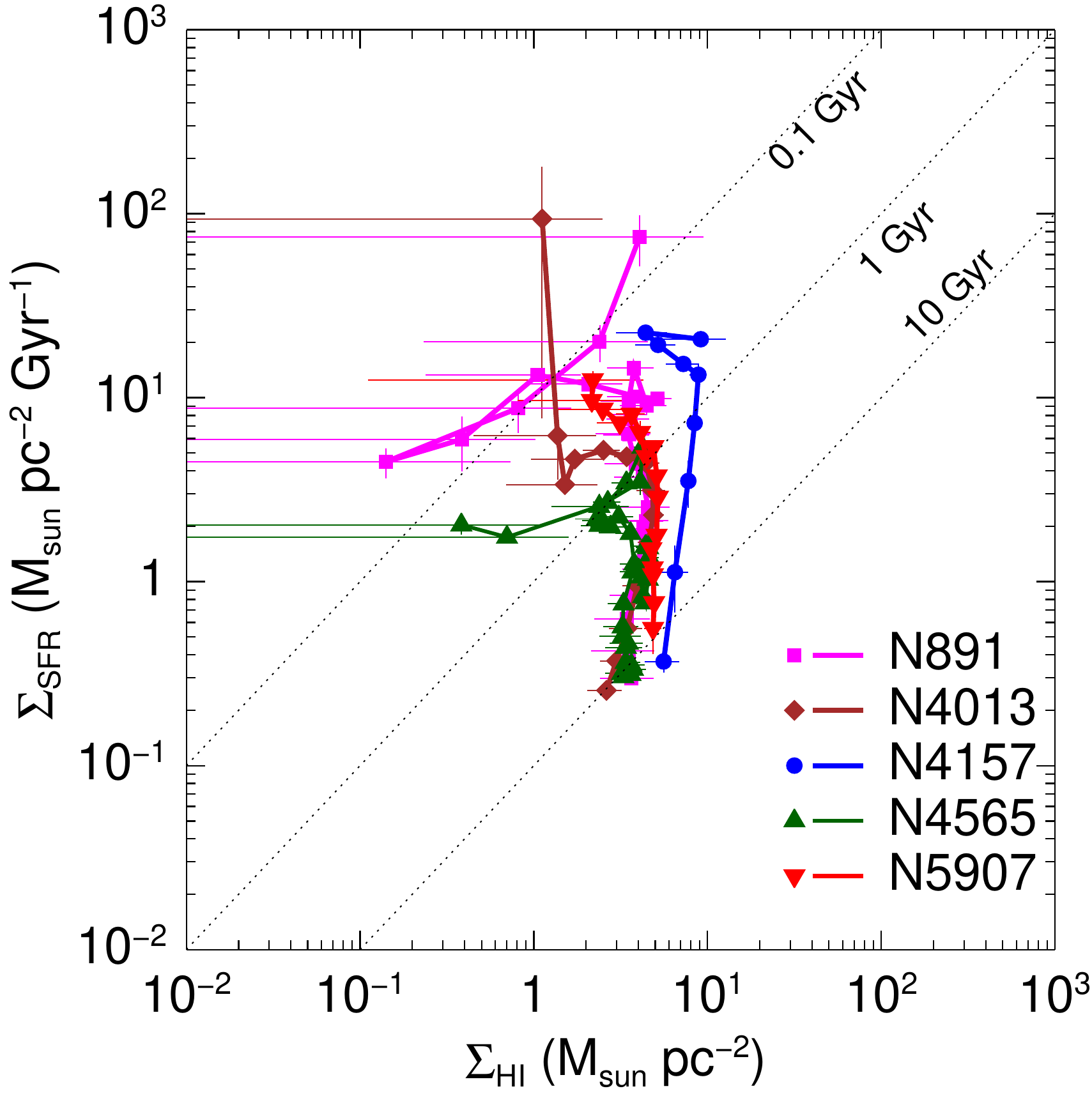}\\
\includegraphics[width=0.4\textwidth]{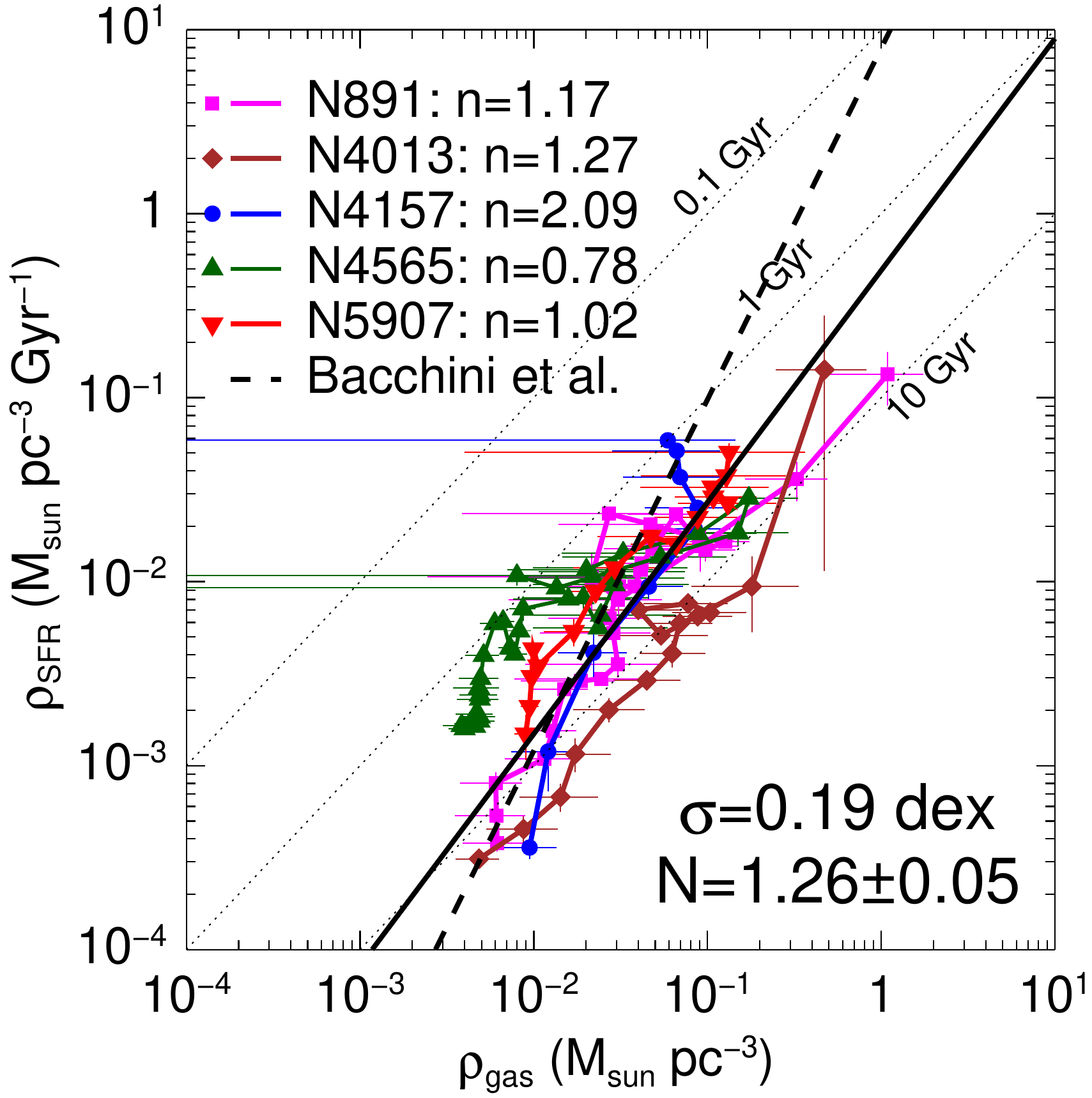}&
\includegraphics[width=0.4\textwidth]{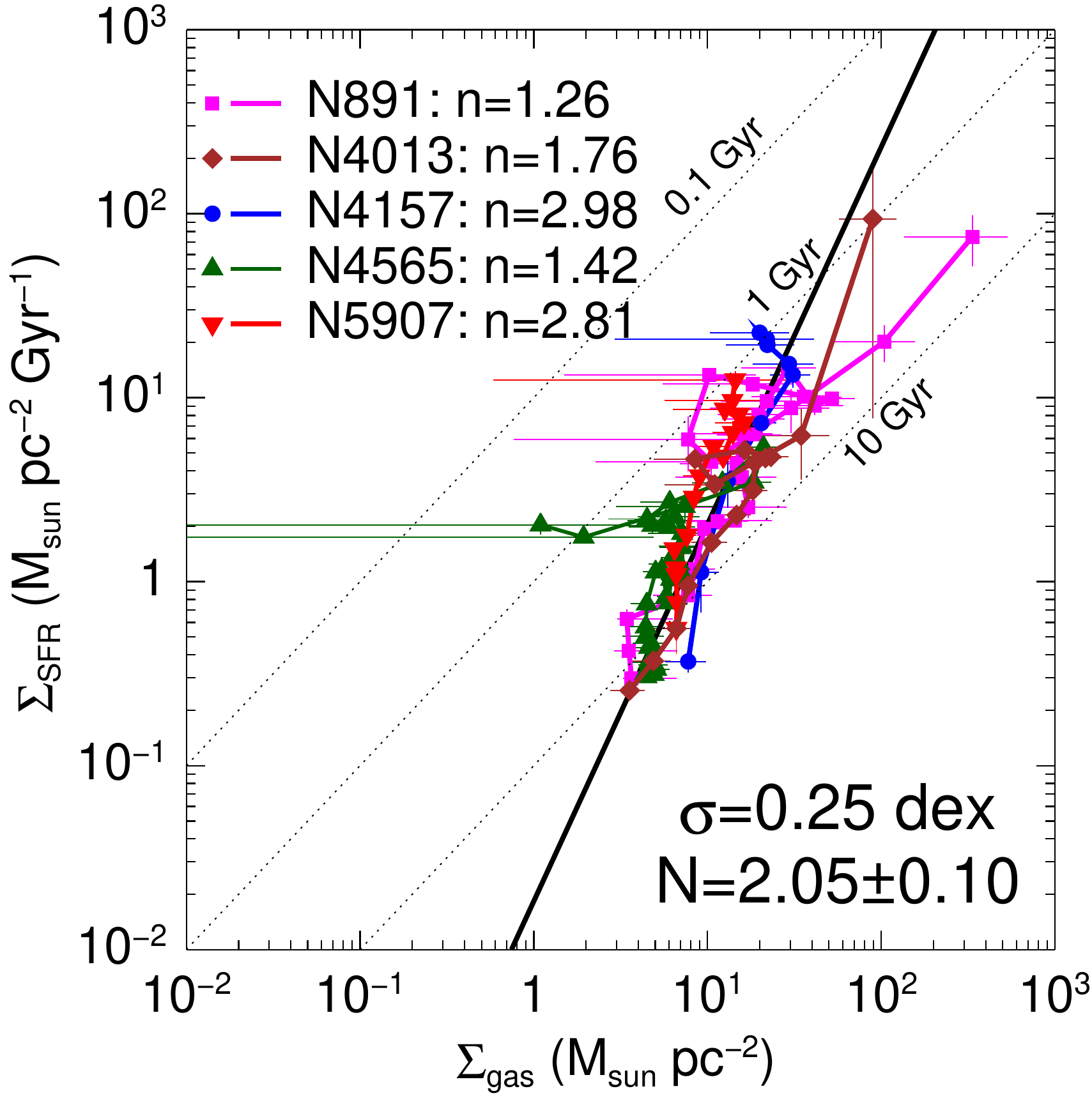}
\end{tabular}
\caption{ Relationship between SFR and H$_2$ (top), \HI\ (middle), and total gas (bottom) based on volume (let-hand panels) and surface (right-hand panels) densities for our galaxy sample. Each power-law index is indicated in the upper-left corner and an average index (black solid line) and rms scatter are presented in the lower-right corner (except for \HI).  The dashed lines in the volumetric relations are the best fits of \citet{Bacchini_2019}.  The dotted lines show constant SFE and the gas depletion time (SFE$^{-1}$) is labelled for these lines. 
\label{sfl_5gals}}
\end{center}
\end{figure*}

\section{Star Formation Prescriptions}
\label{SFprescriptions}
In this section, we examine the volumetric SFL and the relation between the volumetric SFR and the interstellar gas pressure and then compare them with the relations based on surface densities. 
\subsection{SFR versus Gas}

Very recently, \citet{Bacchini_2019} showed a tight correlation  between the SFR and the gas (\HI, H$_2$, \HI+H$_2$) based on volume densities  using the scale heights calculated from the equation of hydrostatic equilibrium. 
We estimated the volume densities (equation \ref{eq_vol}) of the total gas ($\rho_{\rm gas}=$ \siggas/$h_{\rm g}\sqrt{2\pi}$) and the SFR ($\rho_{\rm SFR}=$ \sigsfr/2$h_{\rm SFR}$) using the measured scale heights and surface densities as functions of radius and plotted them for NGC 4013 in Fig. \ref{sfl} (left panel) to investigate whether the volumetric SFL exists and  
examine how much the volumetric power-law relation ($\rho_{\rm SFR} \propto \rho_{\rm gas}^{\rm N}$) is stronger than the relation based on surface densities (right panel). 
The vertical and horizontal error bars of the surface densities (right panel) are the uncertainties of the radial profiles in Fig. \ref{radiprof}. The error bars of the volume densities (left panel) are obtained using the uncertainties of the surface density profiles and the uncertainties of the scale heights. In addition, the factor of 2 uncertainty for the CO-to-H$_2$ conversion factor would affect the volume density of the total gas by a factor of 60\%. 
The comparison between the volumetric and surface SFLs shows that both relations for the molecular gas are not much different while the relations for the total gas are somewhat different: similar slopes in the molecular correlation and different slopes in the total gas correlation.
 We used the ordinary least-squares (OLS) bisector \citep{1990ApJ...364..104I} to fit the power-law relation on a log-log scale. 
The power-law  index of NGC 4013 for the ``molecular'' volumetric SFL (1.07) is similar to the index  for the molecular surface SFL (1.10), but the total gas index for the volumetric SFL (1.27)  is smaller than the index for the surface SFL (1.76). This tendency is clearly shown in Fig. \ref{sfl_5gals}, where we compared the volumetric SFL (left panels) with the surface SFL (right panels) for H$_2$ (top), \HI\ (middle), and the total gas (bottom) of the sample. The best-fit line to  the whole sample represents an average value of the best-fit slopes of five galaxies and the average index is indicated in the lower-right corner.  Each power-law index is presented in the upper-left corner of the panels. As we mentioned, the molecular SFL does not show a difference between volume and surface densities. The molecular power-law indices for volume and surface densities are very similar to each other with similar rms scatters:  0.78 with $\sigma = 0.15$ dex for volume density and 0.77 with $\sigma = 0.14$ dex for surface density, on average. On the other hand, the ``total gas'' SFL (bottom panels) based on  volume (left) and surface (right) densities shows a big difference in the power-law slopes: 1.26 for volume density and 2.05 for surface density. The average rms scatter is smaller in the volumetric SFL (0.19 dex) than in the surface SFL (0.25 dex) for the total gas.
The total gas trends for the volumetric SFL are similar to the molecular gas trends and they have  flatter slopes with smaller scatters compared to the total gas surface SFL. 
We found no difference between the volumetric and surface SFLs for the atomic gas (middle panels of Fig. \ref{sfl_5gals}) in contrast to a study by \citet{Bacchini_2019} who found a tight correlation in the volumetric SFL for \HI\ (dashed lines in the figure) by determining the volume densities assuming the  hydrostatic equilibrium to obtain the scale heights.

\begin{figure*}
\begin{center}
\begin{tabular}{c@{\hspace{0.1in}}c@{\hspace{0.1in}}}
\includegraphics[width=0.45\textwidth]{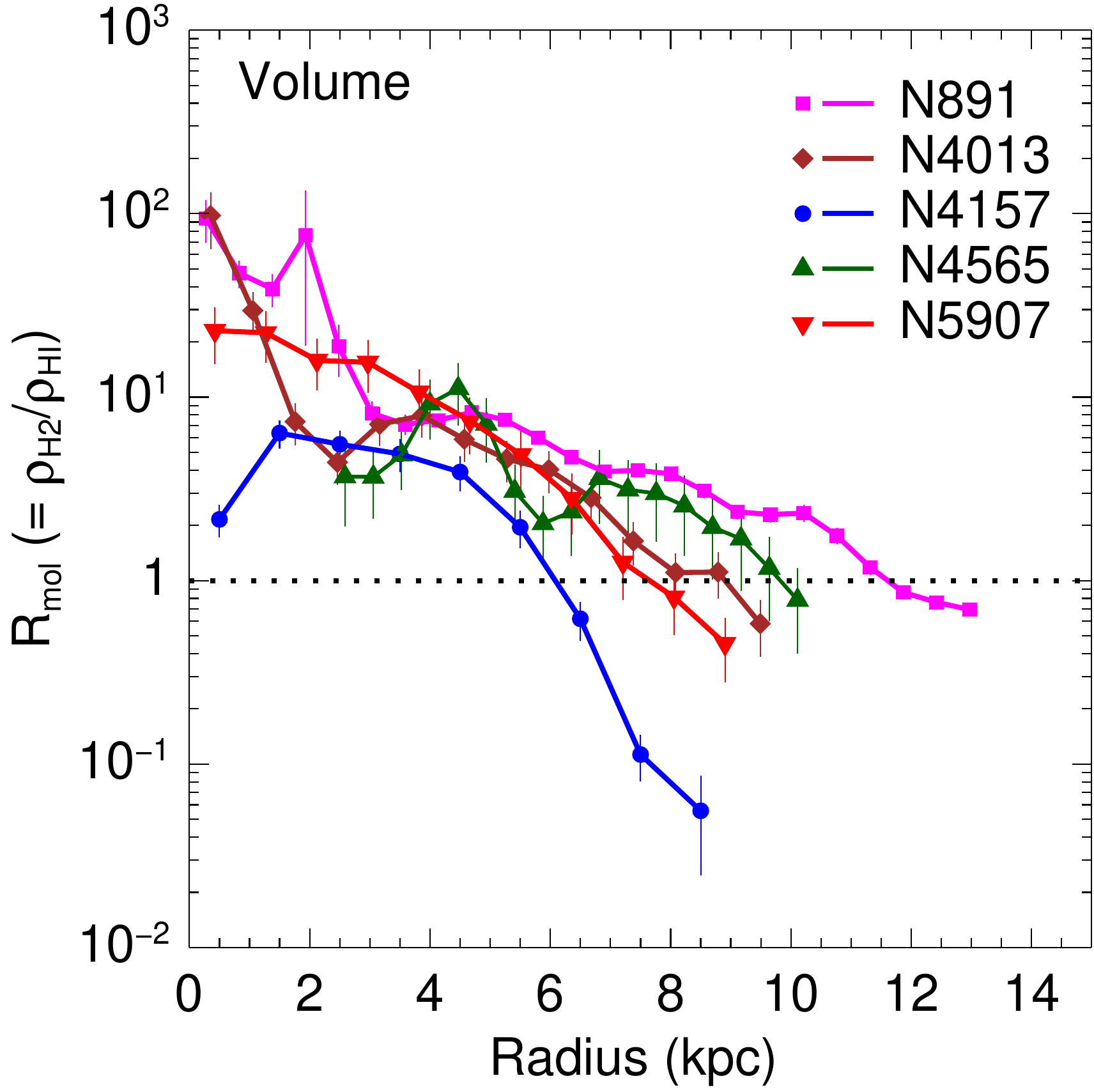}&
\includegraphics[width=0.45\textwidth]{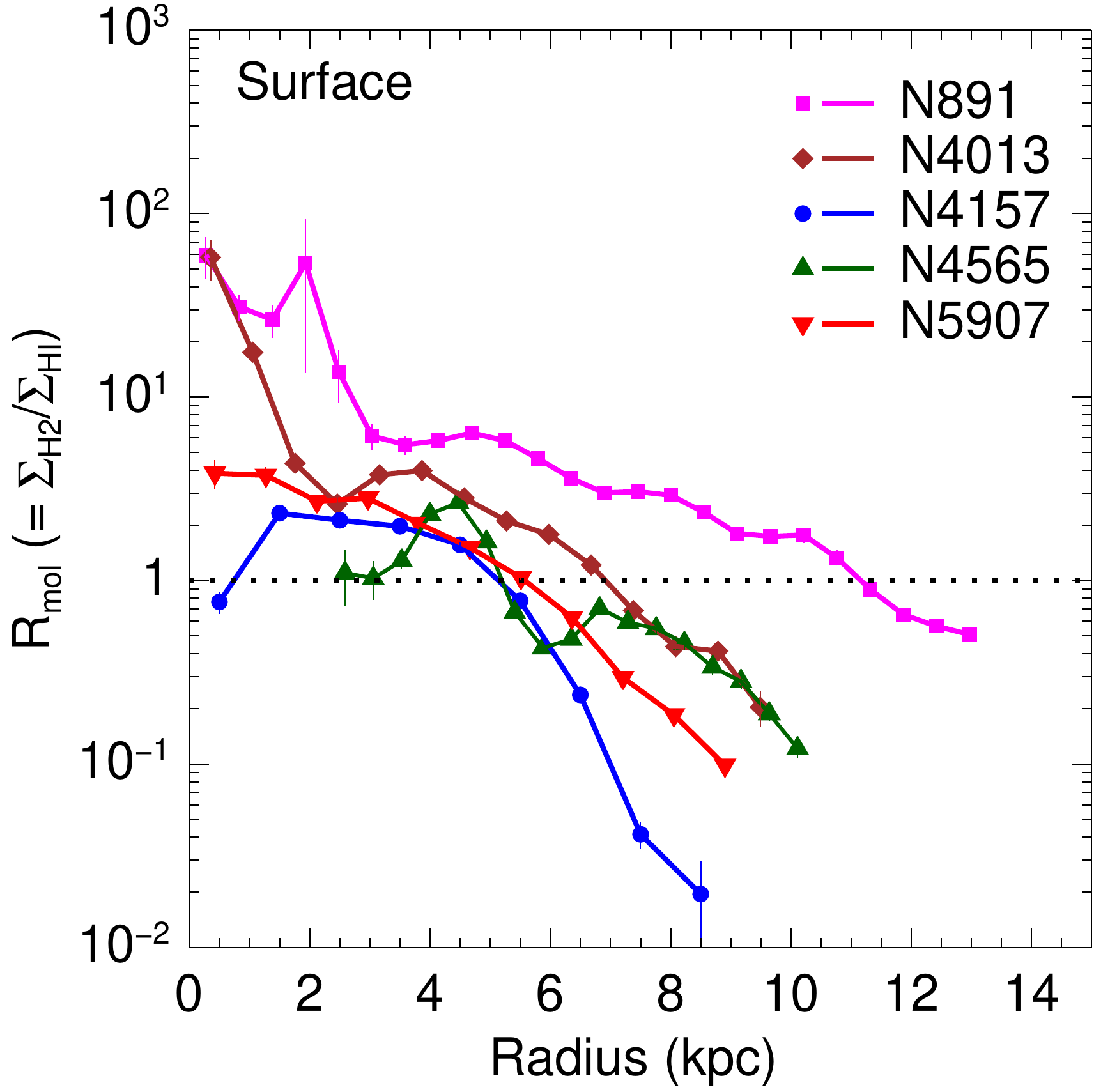}
\end{tabular}
\caption{The ratio of molecular to atomic gas as a function of radius for volume (left) and surface (right) densities. The horizontal dotted line indicates $R_{\rm mol} = 1$.  The vertical error bars on the data points represent the standard error of the mean.
\label{rmol}}
\end{center}
\end{figure*}

Even though the volumetric and surface SFLs are similar to each other in terms of the molecular  and atomic gas individually, the volumetric and surface SFLs for the total gas are different: the surface SFL of the total gas has a steeper slope with a larger scatter than the volumetric SFL. 
The reason for the difference is that the atomic volume density contributes little to the total gas volume density, accordingly the volumetric SFL, while the atomic surface density is high enough to steepen the power-law slope of the surface SFL.
In addition, the transition radius, where the ratio of molecular to atomic gas density is  equal to unity, is larger in the volume density than in the surface density. Fig. \ref{rmol} shows the ratio of molecular to atomic gas ($R_{\rm mol}$) as a function of radius and each transition radius of the sample is listed in Table \ref{table_rt}. This demonstrates that most transitions occur farther  away from the centre in volume densities than in surface densities, implying that the molecular gas is dominant in the volume density farther than in the surface density due to the radial variation in scale height. 

\begin{table}
\begin{center}
\caption{Transition Radius in units of kpc \label{table_rt}}
\vspace*{0.3cm}
\begin{tabular}{cccc}
\hline
Galaxy &Surface&Volume\\
\hline
NGC 891& 11.2 & 11.6 \\
NGC 4013& 7.0 & 8.4 \\
NGC 4157& 5.2 & 6.2\\
NGC 4565& 6.2 & 9.8\\
NGC 5907& 5.6 & 7.7\\
\hline
\end{tabular}
\end{center}
\end{table}

\begin{figure*}
\begin{center}
\begin{tabular}{c@{\hspace{0.1in}}c@{\hspace{0.1in}}}
\includegraphics[width=0.4\textwidth]{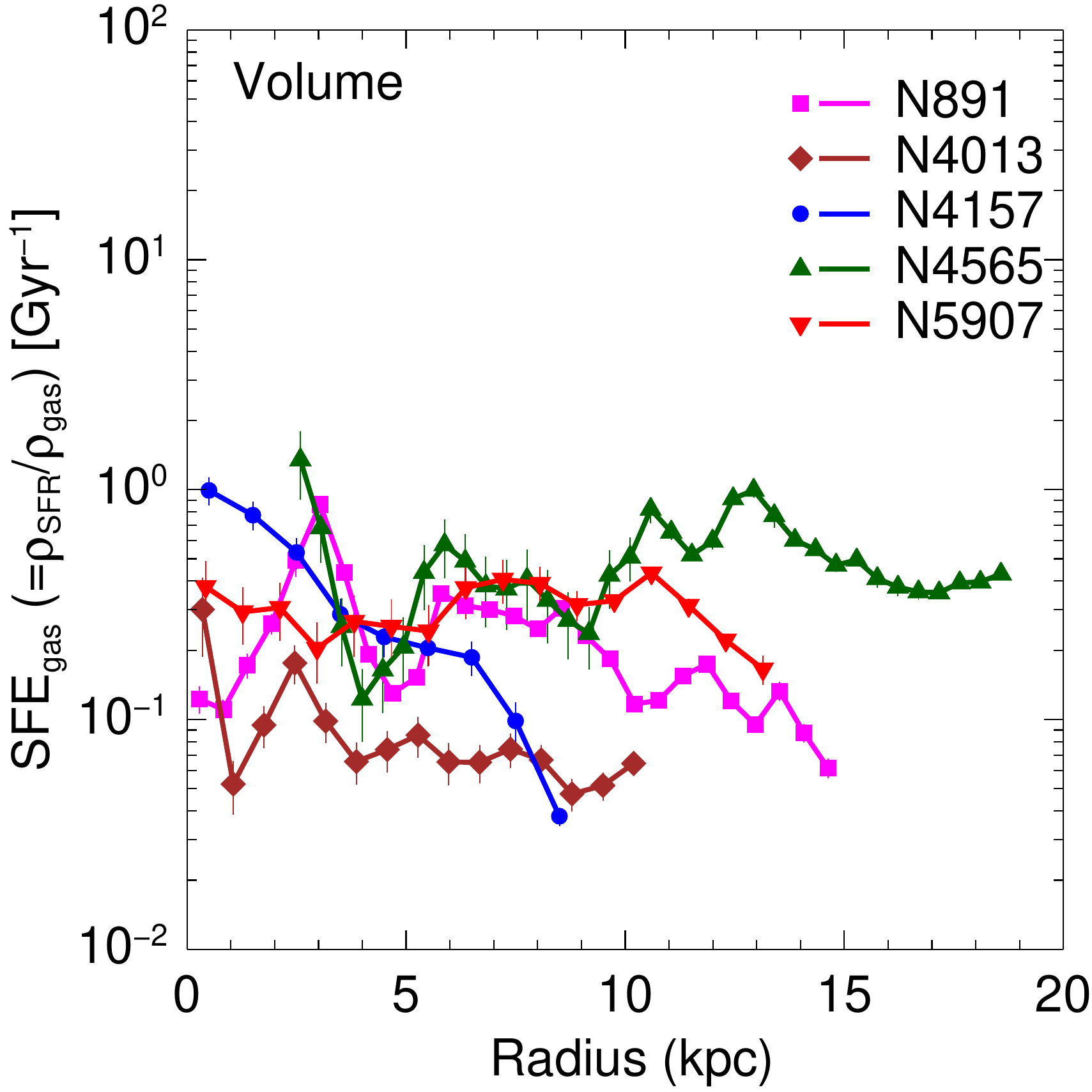}&
\includegraphics[width=0.4\textwidth]{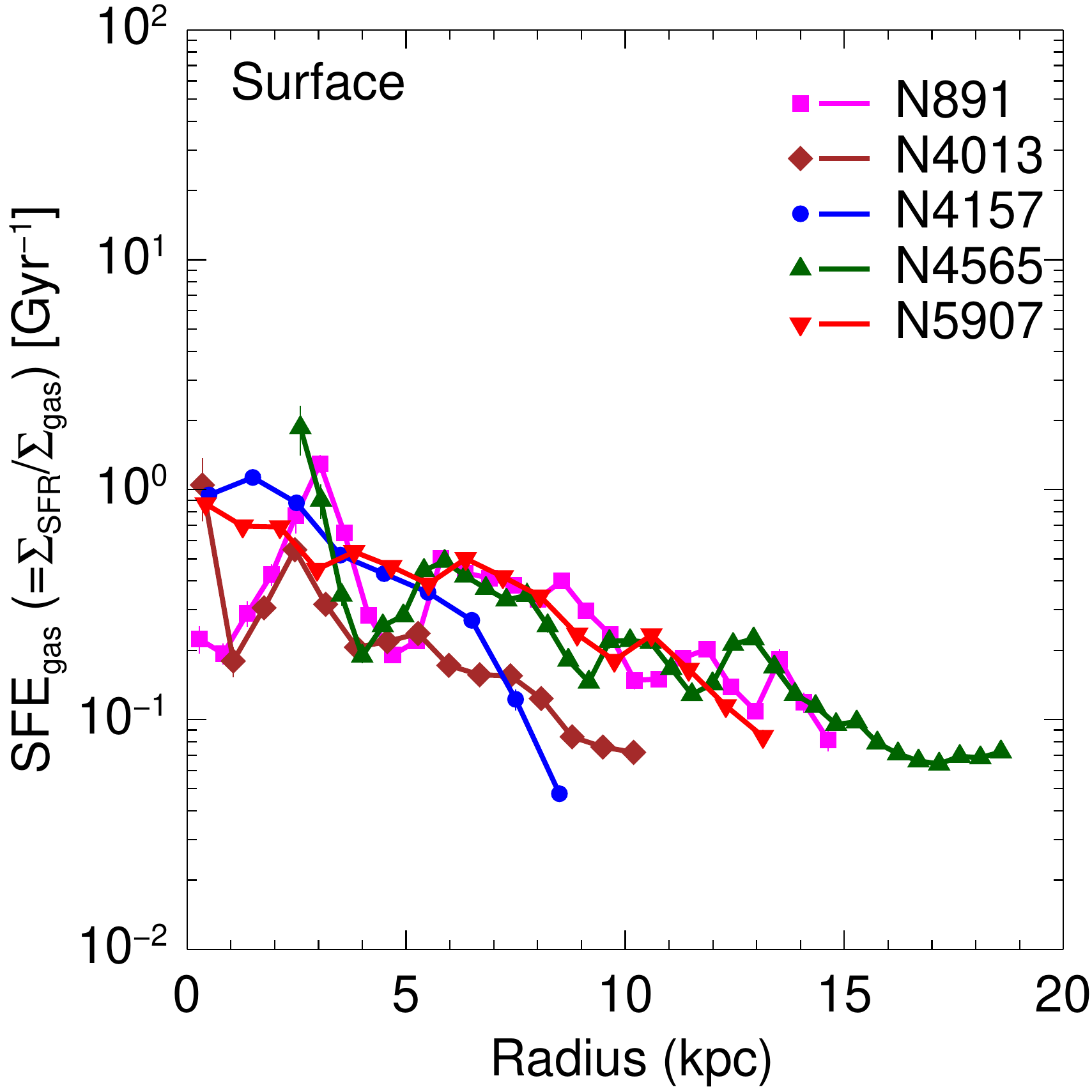}\\
\includegraphics[width=0.4\textwidth]{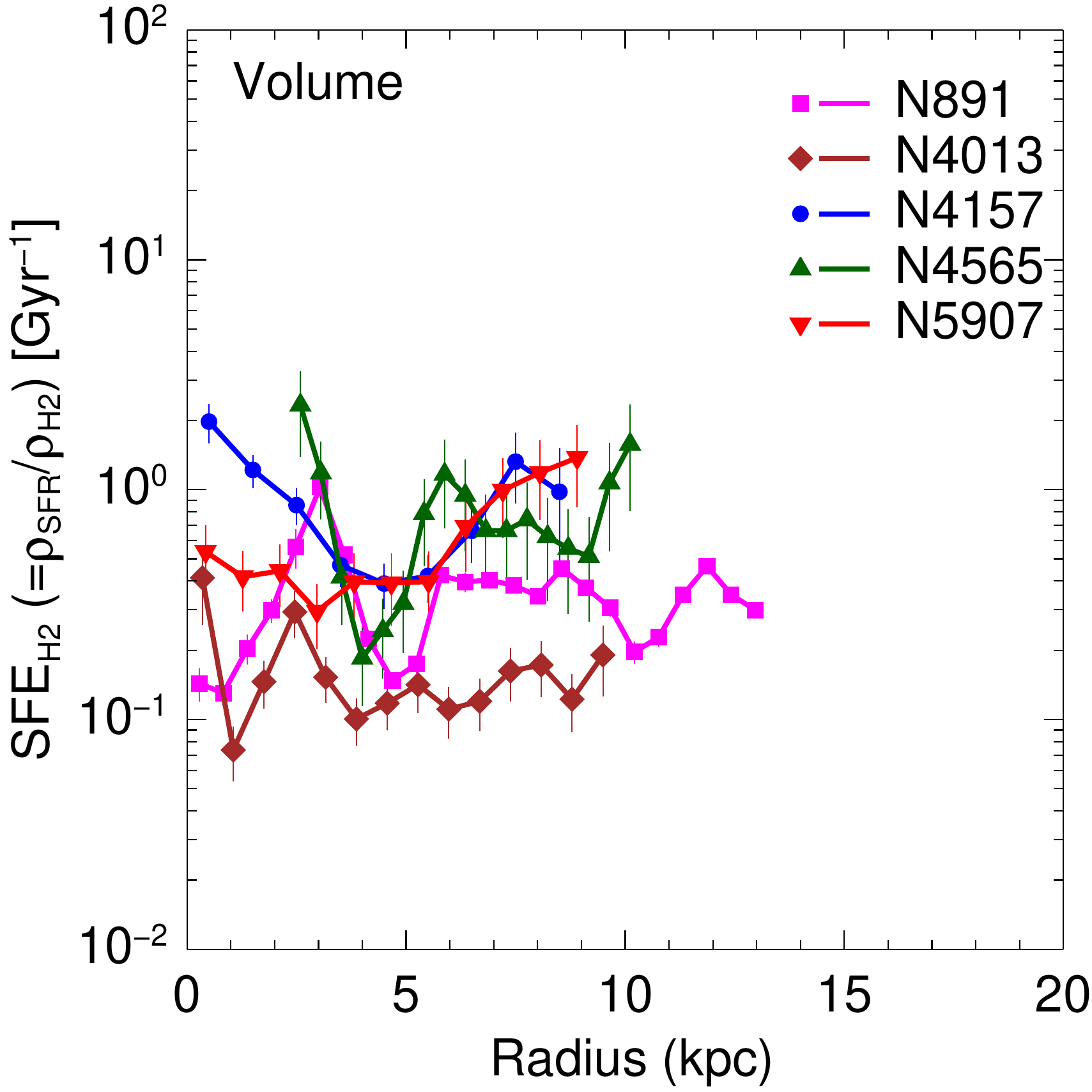}&
\includegraphics[width=0.4\textwidth]{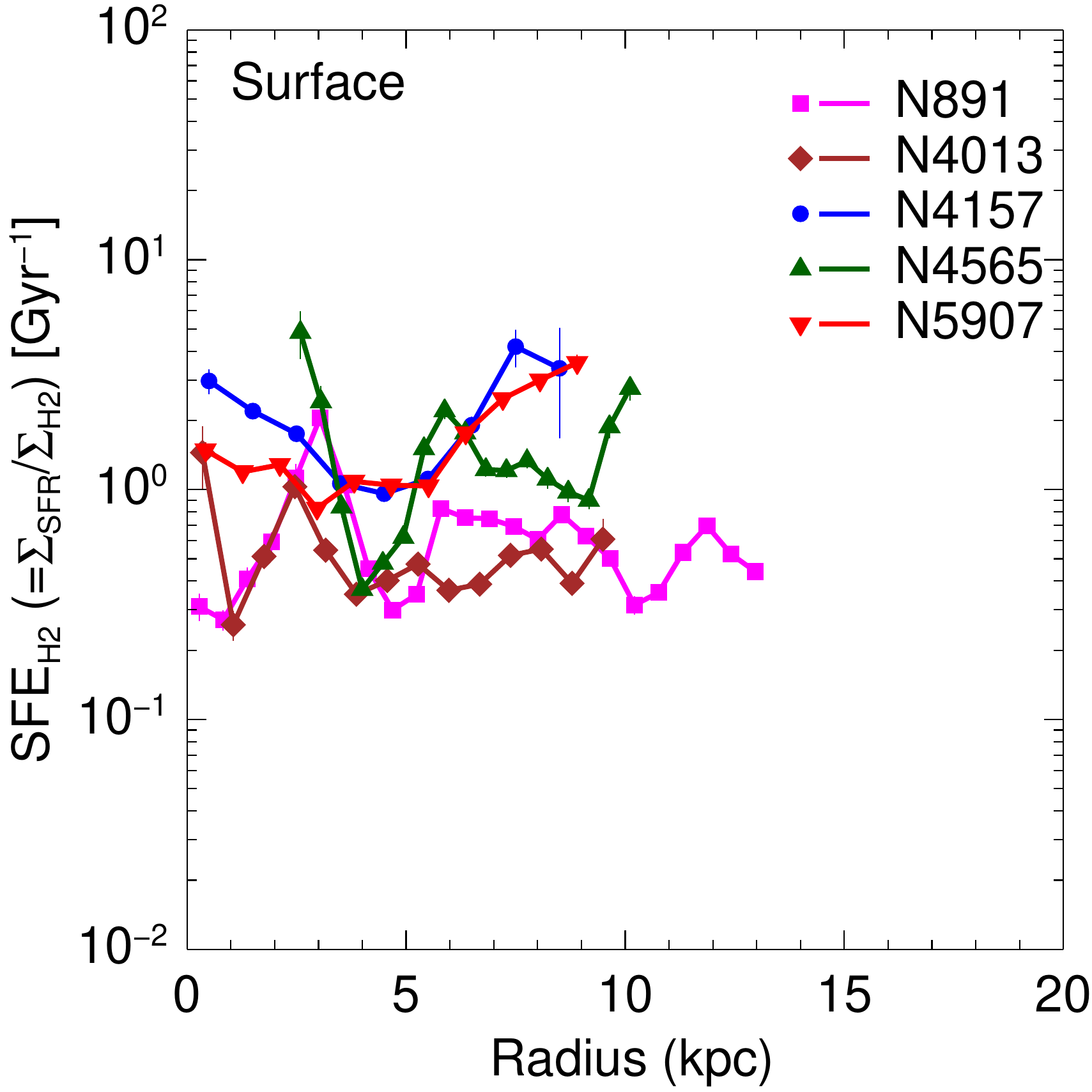}\\
\includegraphics[width=0.4\textwidth]{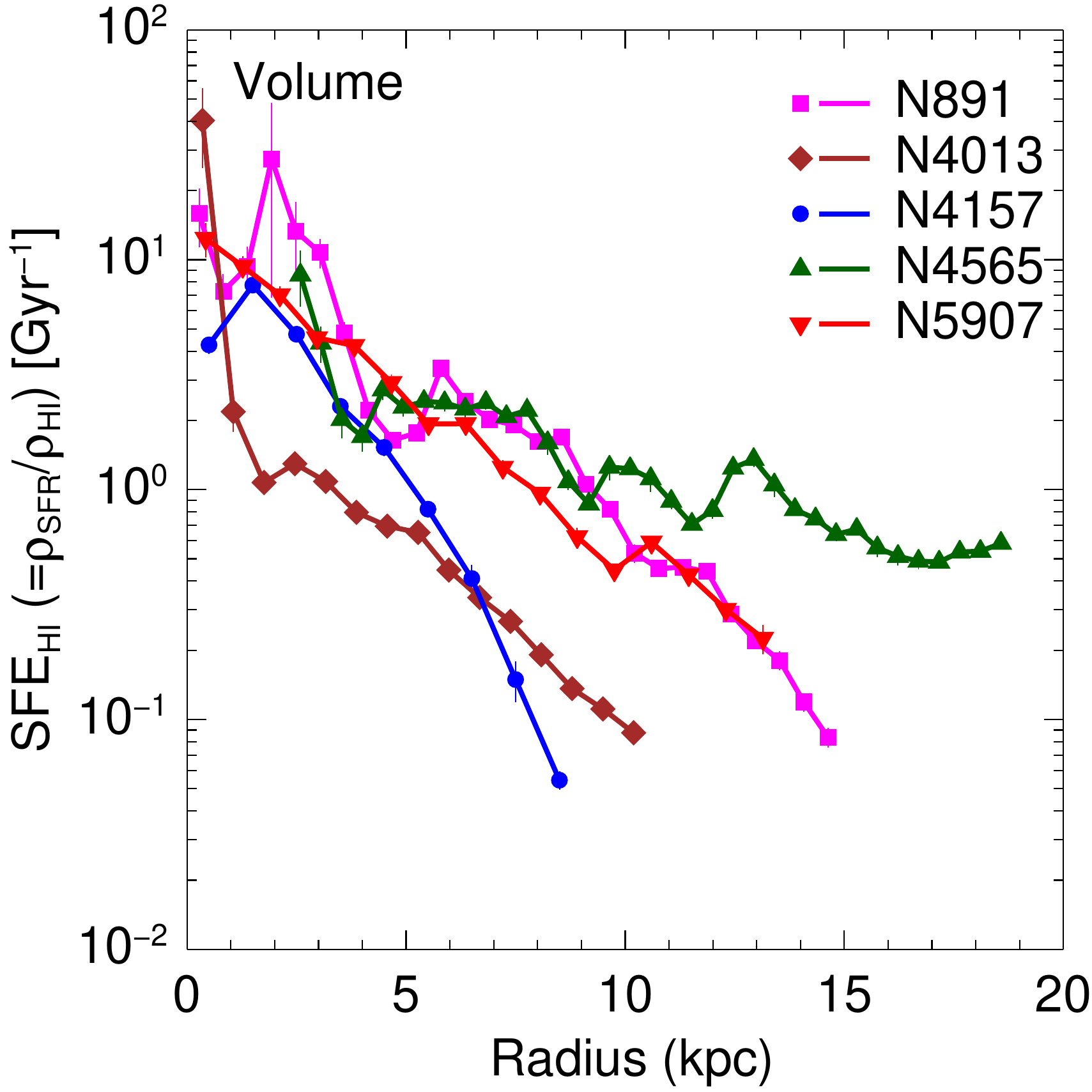}&
\includegraphics[width=0.4\textwidth]{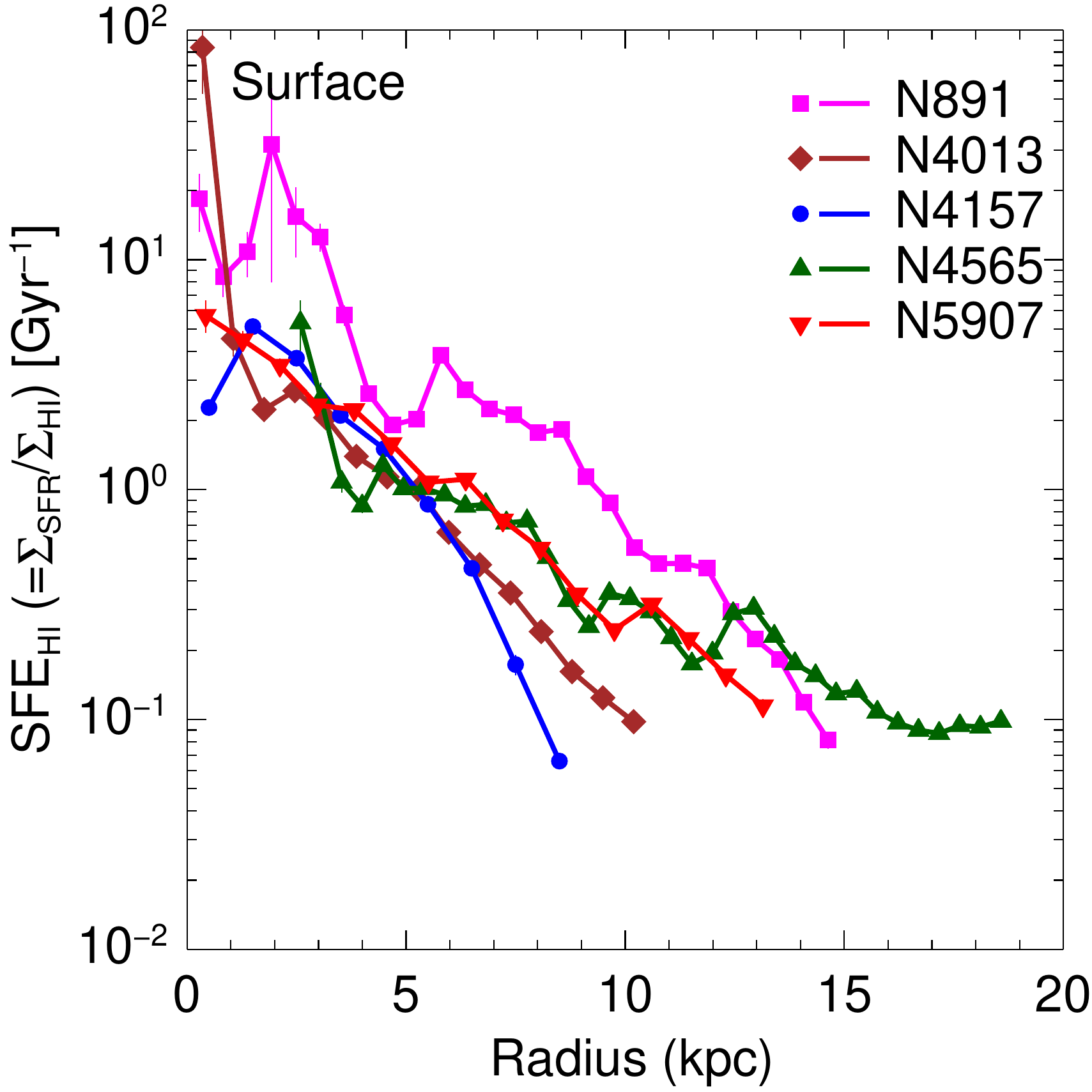}\\
\end{tabular}
\caption{Star formation efficiency as a function of radius in terms of the total gas (top panels), the molecular gas (middle panels), and the atomic gas (bottom panels). The left-hand panels are SFE based on volume density and the right-hand panels are SFE based on surface density.  The vertical error bars on the data points represent the standard error of the mean.
\label{sfe}}
\end{center}
\end{figure*}

In the star formation efficiency (SFE) in terms of the total gas, we also found a noticeable difference between volume and surface densities.   Fig. \ref{sfe} shows the SFE$_{\rm gas}$  profiles (top panels) based on the volume ($\rho_{\rm SFR}/\rho_{\rm gas}$) and surface (\sigsfr/\siggas) densities. The surface SFE$_{\rm gas}$ (right panel) decreases with radius like the previous studies (e.g., \citealt{2008AJ....136.2782L}; \citealt{2016MNRAS.463.2092Y}). 
On the other hand, the volumetric SFE$_{\rm gas}$ (left panel) appears to be roughly constant. 
The difference between the volumetric and surface SFEs indicates that a scale height variation significantly affects the SFE$_{\rm gas}$. 
In terms of the molecular gas, the constancy of SFE$_{\rm H_2}$ (= \sigsfr/\sightwo) is observed in many studies (e.g.,  \citealt{1999AJ....118..670R}; \citealt{2008AJ....136.2846B}; 
 \citealt{Leroy_2013}; \citealt{2016MNRAS.463.2092Y}). Fig. \ref{sfe} (middle panels) also verifies the constant SFE$_{\rm H_2}$ for both volume and surface densities; no difference is observed between them.  However, the gas depletion time is apparently different: $\sim$ 0.8 Gyr for the surface SFE$_{\rm H_2}$ and $\sim$ 1.7 Gyr for the volumetric SFE$_{\rm H_2}$, on average. The difference is because the 24 \um\ scale height is generally larger than the \Htwo\ scale height. This will increase the the SFE and reduce the depletion time in terms of surface density because additional SFR is being integrated at large $z$ without additional \Htwo. Our guess is that either there is propagation of FUV far from star-forming regions to heat the dust, or some of the star formation is occurring in \HI$-$dominated regions.
In the case of the SFE for the atomic gas (SFE$_{\rm HI}$), both volumetric and surface SFEs decrease exponentially with radius as shown in the bottom panels of Fig. \ref{sfe}.

\subsection{SFR versus Interstellar Gas Pressure}
\begin{figure*}
\begin{center}
\begin{tabular}{c@{\hspace{0.1in}}c@{\hspace{0.1in}}}
\includegraphics[width=0.45\textwidth]{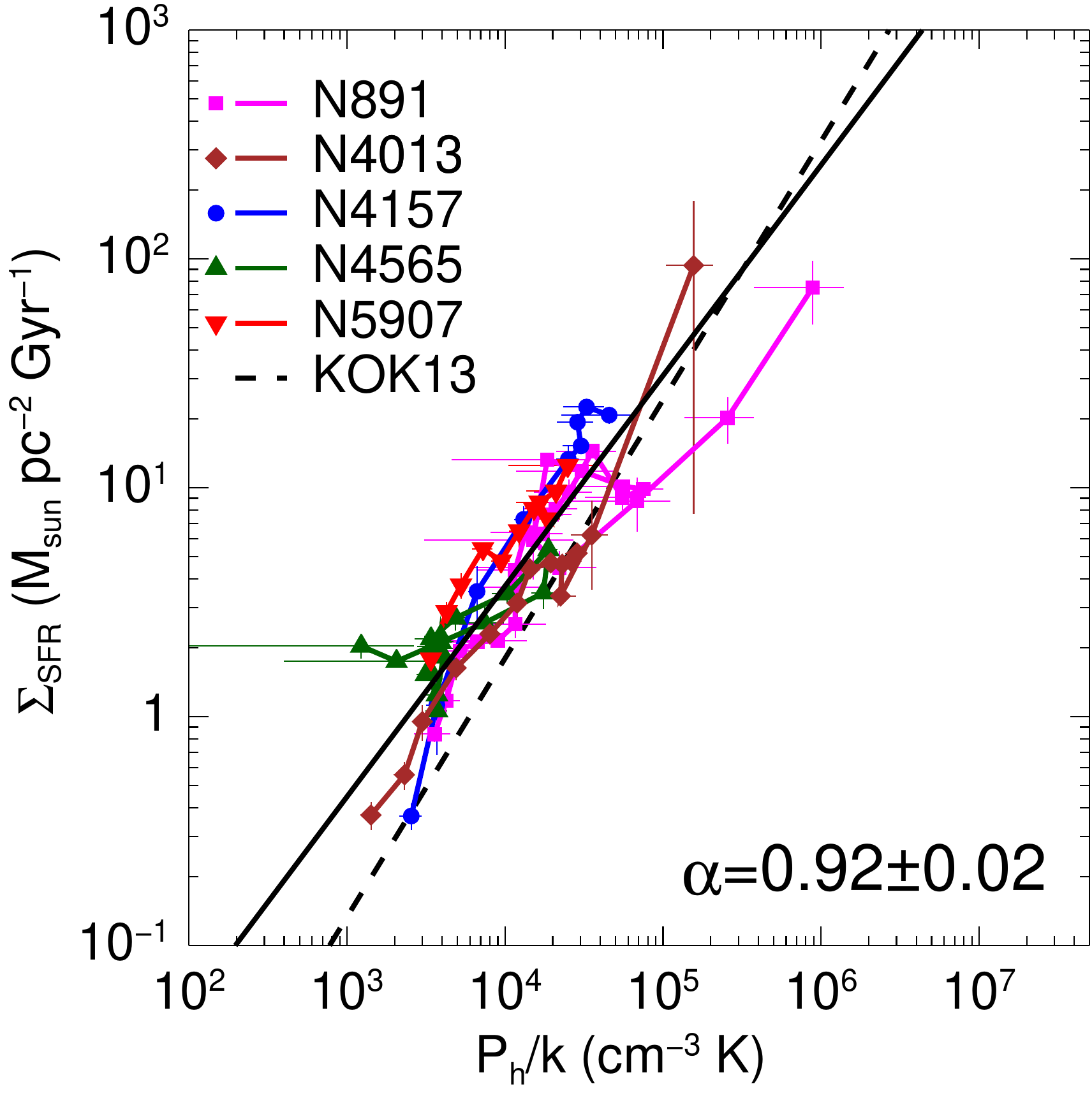}&
\includegraphics[width=0.45\textwidth]{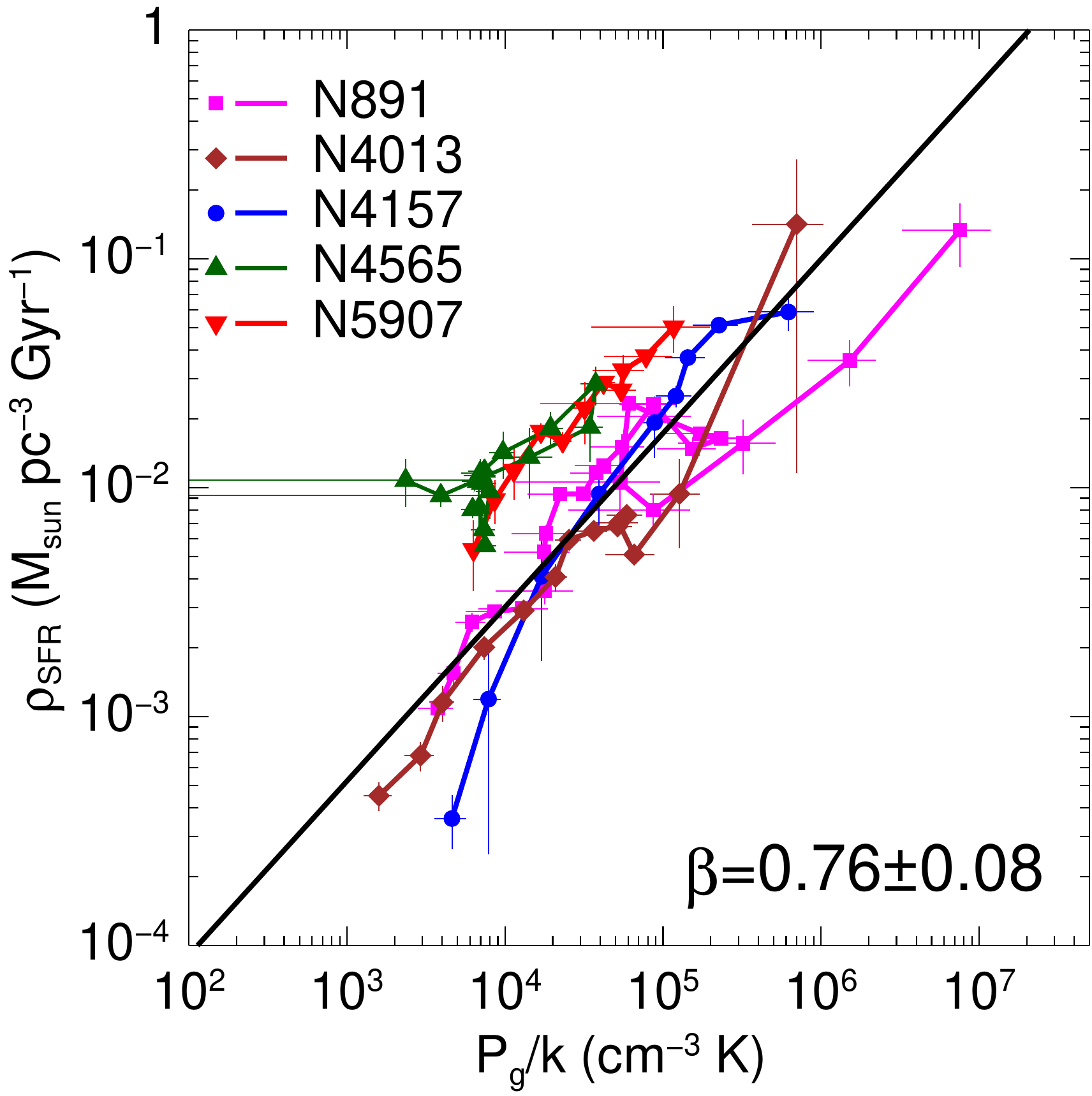}
\end{tabular}
\caption{\sigsfr\ as a function of the hydrostatic midplane pressure (left) and  $\rho_{\rm SFR}$  as a function of the interstellar gas pressure (right). The solid line represents an average slope of the five galaxies and the average slope is shown in the lower-right corner. 
The dashed line in the left panel is the best-fit to simulation data given by \citet{2013ApJ...776....1K} and the power-law slope is 1.13. \label{SFRvsP}}
\end{center}
\end{figure*}

Many previous studies (e.g., \citealt{2002ApJ...569..157W}; \citealt{2006ApJ...650..933B}; \citealt{2014AJ....148..127Y}) found a tight power-law correlation between $R_{\rm mol}$ and the interstellar gas pressure. The tight correlation suggests a  close relationship between the SFR and the pressure since the SFR is strongly correlated with the molecular gas and high pressure increases the gas density. In this section, we directly examine how the pressure is related to SFR.  
In Fig. \ref{SFRvsP} (left), we plot \sigsfr\ as a function of the hydrostatic midplane pressure calculated from equation (\ref{eq_Ph}) given by \citet{2011AJ....141...48Y}:
\begin{equation}
P_{h} = 0.89 (G\sigstar)^{0.5} \siggas \frac{\sigma_{\rm g}}{z_*^{0.5}}\;,
\label{eq_Ph}
\end{equation}
where the gas velocity dispersion $\sigma_g$ is assumed to be 8 \kms \citep{2004ApJ...612L..29B} and the stellar scale height ($z_*$) is obtained from the exponential fitting to 3.6 \um\ map (see Section \ref{starmodel}). 
This figure shows a well-defined power-law correlation as expected from the relationship between $R_{\rm mol}$ and the hydrostatic midplane pressure. The solid line represents the best-fit power law of the galaxy sample and the dashed line shows the best-fit 
relation from a dynamical equilibrium model provided by  \citet{2013ApJ...776....1K} based on three-dimensional numerical hydrodynamic simulations:   
\begin{equation}
\frac{\sigsfr}{\Msol \,\rm yr^{-1}\, kpc^{-2}} = 1.8 \times 10^{-3} \left(\frac{P/k_{\rm B}}{10^4\, \rm cm^{-3}\, K} \right)^{1.13}.
\label{eq_kok13}
\end{equation}
The best-fit slope of \citet{2013ApJ...776....1K} is steeper than our best-fit slope of 0.92, on average. 
Recently, \citet{2020arXiv200208964S} also reported a power-law relation between them with a slope of 0.84 based on 28 galaxies from Atacama Large Millimeter/submillimeter Array (ALMA) observations.
In the relationship between \sigsfr\ and $P_h$, we used the assumed constant values for the gas velocity dispersion and the stellar scale height,  contrary to  the radial variation in the scale height and the velocity dispersion that we measured. As we presented in Section \ref{verticalprof}, the scale height increases with the galactocentric radius and the vertical velocity dispersion decreases with the radius. 
Therefore, it is necessary to examine the relationship between the SFR and the pressure using the observed quantities. Using the volume densities and the vertical velocity dispersions of \HI\ and H$_2$, we obtained the interstellar gas pressure by summing the atomic and molecular gas pressures at the midplane:
\begin{equation}
P_{\rm g}(r) \approx \rho_{\rm H_2}(r)\,\sigma_{\rm H_2}^2(r) +  \rho_{\rm HI}(r)\,\sigma_{\rm HI}^2(r).
\label{eq_Pg}
\end{equation}
In Fig. \ref{SFRvsP} (right), we compared the pressure  with the SFR volume density and found a tight power-law correlation ($\rho_{\rm SFR} \propto P_g^\beta$) for each galaxy with an average slope of $\beta=0.76$, ranging from 0.584 to 1.094. 
Even though the power-law slope of the volume density basis is flatter than that of the surface density basis, we do not see any significant difference in the tightness of the power-law relation. The rms scatters around the best-fits of the galaxies are also similar to each other. However, the correlation between $\rho_{\rm SFR}$ and $P_g$ is more appropriate and realistic than the correlation between \sigsfr\ and P$_h$ since the observed scale height and velocity dispersion vary with radius.

\section{Summary and Conclusions}
\label{sum}
We derived surface mass densities and scale heights of CO, \HI, 3.6 \um\ (stars), and/or 24 \um\ (SFR) for our edge-on or highly inclined galaxy sample (NGC 891, 4013, 4157, 4565, and 5907) to estimate the midplane volume densities as a function of radius. We also inferred the vertical velocity dispersions from the scale heights and the volume densities. Using the velocity dispersions and the volume densities of our galaxy sample, we investigated prescriptions for the SFR.  First, we examined how the gas volume density is correlated with the SFR volume density and how the volumetric SFL is different from the surface SFL. We also compared the volumetric and surface star formation efficiencies (SFEs) in terms of the total, molecular, and atomic gas. Next, we showed how the interstellar gas pressure regulates the SFR and how the relationship between the pressure and the SFR using  the volume densities and the varying velocity dispersions  differs from the relationship using the surface densities and assumed contant values for the scale height and the velocity dispersion. The results and conclusions are summarized as follows. 

1. We measured the scale heights of CO, \HI, stars, and the SFR for NGC 4013 by fitting the Gaussian (CO and \HI) and the exponential (stars and the SFR) functions to their vertical distributions. All the scale heights increase with radius as we already presented in \citet{2014AJ....148..127Y} for the other galaxies of our sample.  The scale height of the SFR is a new property that we measured in this paper for the first time. 
The SFR scale heights of our galaxy sample increase with radius, but the gradient of NGC 4565 is not significant like its CO scale height. 
We reproduced the \HI\ scale heights of NGC 4157 and 5907 with the VLA BCD images that have a higher angular resolution ($\sim$ 4.5\ac) than the old CD images ($\sim$ 15\ac). The BCD scale heights are lower than the CD scale heights by a factor of $\sim$1.5. 

2. We estimated the volume densities of the gas and the SFR using the scale heights and compared the volumetric SFL with the surface SFL for the galaxy sample. We found that both the volumetric and surface SFLs in terms of the molecular gas have similar slopes and tightness of the correlation: 0.78$\pm$0.03 with $\sigma=$ 0.15 dex for volume density and 0.77$\pm$0.03 with $\sigma=$ 0.14 dex for surface density, on average. 
However, the SFLs in terms of the total gas show a significant difference; the power-law slope of the volumetric SFL (1.26$\pm$0.05) is quite lower than the slope of the surface SFL (2.05$\pm$0.10). In addition, the rms scatter of the volumetric SFL (0.19 dex) is smaller than the rms scatter of the surface SFL (0.25 dex).
On the other hand, we found no significant difference between the volumetric and surface SFLs in terms of the atomic gas. They both do not show a good power-law correlation.   

3. We compared the ratio of molecular to atomic gas ($R_{\rm mol}$) in terms of the volume ($\rho_{\rm H_2}$/$\rho_{HI}$) and surface (\sightwo/\sighi) densities and found that the transition radius, where $R_{\rm mol} = 1$, of the volume density is larger than that of the surface density. 

4. In the comparison of the total gas SFEs based on the volume ($\rho_{\rm SFR}$/$\rho_{\rm gas}$) and  surface (\sigsfr/\siggas) densities, we found that the volumetric SFE$_{\rm gas}$ is roughly constant while the surface SFE$_{\rm gas}$ decreases as a function of radius. In the case of the molecular gas, both the volumetric SFE$_{\rm H2}$ (= $\rho_{\rm SFR}$/$\rho_{\rm H2}$) and the surface SFE$_{\rm H2}$ (= \sigsfr/\sightwo) are roughly constant, though the gas depletion time is different ($\sim 0.8$ Gyr for surface density and $\sim 1.7$ Gyr for volume density). The atomic gas SFE$_{\rm HI}$ clearly decreases with radius within the optical radius. The difference between the volumetric and surface SFE$_{\rm gas}$ reflects the radial variation in scale height. 

5. We derived the hydrostatic midplane pressure ($P_h$) using assumed constant values of the stellar scale height and the gas velocity dispersion, as well as the interstellar gas pressure ($P_g$) using the varying scale heights and the velocity dispersions, in order to examine the relation with the SFR. Both correlations of \sigsfr\ versus $P_h$ and $\rho_{\rm SFR}$ versus $P_g$ show a tight power-law with indices of 0.92 and 0.76, respectively. There is no significant difference in the tightness of the relationships. The tight correlation indicates that the  interstellar gas pressure plays a key role in regulating the SFR.

\section*{Acknowledgements}
We thank the anonymous referee for useful suggestions and comments that improved this paper. 
The National Radio Astronomy Observatory is a facility of the National Science Foundation operated under cooperative agreement by Associated Universities, Inc.




\bibliographystyle{mnras}
\bibliography{refer} 





\bsp	
\label{lastpage}
\end{document}